\documentclass[12pt,reqno]{article}
\usepackage[utf8]{inputenc}
\usepackage[T1]{fontenc}
\usepackage{amsmath, amsfonts, amssymb, amsthm,mathrsfs,booktabs, bm,bbm, mathtools, thmtools, thm-restate}
\usepackage{setspace,cancel, pgf,tikz,sectsty,xcolor}
\usepackage{epigraph}
\usepackage{palatino}
\usepackage[title]{appendix}
\usepackage[authoryear]{natbib}
\usepackage[colorlinks=true]{hyperref}
\hypersetup{colorlinks=true, linkcolor=blue, citecolor=blue}
\usetikzlibrary{arrows}
\allowdisplaybreaks

\newtheorem{theorem}{\textsc{Theorem}}[]

\newtheorem{lemma}{\textsc{Lemma}}[]
\newtheorem{corollary}{\textsc{Corollary}}[]
\newtheorem{proposition}{\textsc{Proposition}}[]

\DeclareMathOperator*{\argmax}{arg\,max}
\DeclareMathOperator*{\argmin}{arg\,min}
\DeclareMathOperator*{\supp}{supp}

\DeclareMathOperator*{\E}{\mathbb{E}}
\DeclareMathOperator*{\R}{\mathbb{R}}
\usepackage[nameinlink]{cleveref}
\crefname{manualasm}{assumption}{assumptions}
\crefname{cor}{corollary}{corollaries}
\crefalias{prop}{proposition}
\crefname{claim}{claim}{claims}
\crefname{ex}{example}{examples}
\crefname{defn}{definition}{definitions}
\crefname{rmk}{remark}{remarks}
\crefname{alg}{algorithm}{algorithms}
\newtheorem{observation}{\textsc{Observation}}[]

\usepackage[left=3.4cm,right=3.4cm,top=3.5cm,bottom=3.5cm]{geometry}

\title{Bundling against Learning\thanks{We thank the editor and three anonymous referees, as well as Alex Bloedel, Nina Bobkova, Ben Brooks, Piotr Dworczak, Alex Frankel, Nima Haghpanah, Kevin He, Annie Liang, Andrew McClellan, Phil Reny, and various seminar and conference audiences at the Becker Friedman Institute, the Chicago Market Design Conference, the SED Conference, the IO Theory Conference, the European Workshop on Market Design, ASSA 2026, Warwick, FSU, BU, BC, WashU Olin, Northwestern, Georgetown, UW Madison, Harvard-MIT, University of Bonn, University of Naples Federico II, and University of Tokyo for helpful comments and suggestions. This work is supported by the Biehler Junior Faculty Fellowship at the University of Chicago Booth School of Business (Pernoud).}
}
\author{Agathe Pernoud\thanks{Booth School of Business, University of Chicago. \href{mailto:agathe.pernoud@chicagobooth.edu}{agathe.pernoud@chicagobooth.edu}.} \hspace{2.5cm}   Frank Yang\thanks{Department of Economics, Stanford University.  \href{mailto:shuny@stanford.edu}{shuny@stanford.edu}.}}
\date{\vspace{0.7cm}September 2026}

\begin{document}
\setlength{\abovedisplayskip}{6.5pt plus 2pt minus 2pt}
\setlength{\belowdisplayskip}{6.5pt plus 2pt minus 2pt}
\setlength{\abovedisplayshortskip}{4.5pt plus 2pt minus 1pt}
\setlength{\belowdisplayshortskip}{6.5pt plus 2pt minus 2pt}

\maketitle
\begin{abstract}
A monopolist sells multiple goods to an uninformed buyer. The buyer chooses to learn any one-dimensional linear signal of his values for the goods, anticipating the seller's mechanism. The seller designs an optimal mechanism, anticipating the buyer's learning choice. In a generalized Gaussian environment, we show that every equilibrium has \textbf{\textit{vertical learning}} where the buyer's posterior means are comonotonic, and every equilibrium is outcome-equivalent to \textit{\textbf{nested bundling}} where the seller offers a menu of nested bundles. In equilibrium, the buyer learns more about a higher-tier good, resulting in a higher posterior variance on the log scale.
\\

\end{abstract}

\newpage 
\section{Introduction}

When a new multiproduct firm enters a market, consumers are often uncertain about their willingness to pay for the firm’s various products.\footnote{For a concrete example, consider OpenAI, which launched new products such as ChatGPT (a language-generation model) and DALL$\cdot$E (an image-generation model) in 2022.} They therefore spend time learning about these products before buying. At the same time, the new firm often conducts experimentation to optimize prices and product offerings against the demand system, which is itself shaped by what consumers learn. What, then, should we expect in equilibrium about the endogenous demand system and product offerings resulting from consumer optimal learning and firm optimal pricing? 

In this paper, we answer this question with a model of equilibrium learning in multiproduct pricing. We consider a simultaneous-move game between a seller and a buyer. The seller has $K$ goods to sell.  She chooses a \textit{\textbf{selling mechanism}}, a menu of bundles and prices, allowing for lotteries. The buyer is initially uninformed about his vector of values for each good $\mathbf{v}=(v_k)_k$, which is drawn from an elliptical distribution supported on $V\subset\mathbb{R}^K_+$.\footnote{Elliptical distributions generalize Gaussian distributions; for foundations and economic applications of elliptical distributions, see e.g. \citet*{gupta2013elliptically}, \citet{frankel2019muddled}, \cite{he2023random}, and \cite{ball2025scoring}.} The buyer chooses a signal about his values $\mathbf{v}$ at no cost but faces a dimension restriction---he can only choose to observe a one-dimensional signal $\boldsymbol{\alpha}\cdot\mathbf{v}$, where $\boldsymbol{\alpha}\in\mathbb{R}^K$ is the vector of  \textit{\textbf{learning weights}}. In equilibrium, the buyer optimally chooses the learning weights to maximize his expected payoff when purchasing from the seller's menu; the seller optimally chooses a menu to maximize revenue given the endogenous demand system---the distribution of posterior means induced by the buyer's signal.  

By design, the buyer faces the choice of \textit{what} to learn: He can fully learn the value of any bundle, but must decide which one to learn. He can also learn about the differences between any two goods, or between two bundles, or more generally any linear combination of these signals.\footnote{Note that we do not allow for \textit{nonlinear} signals. If we were to allow the buyer to learn any real-valued signal, then the dimension restriction would have no bite, as every random vector can be embedded into a real-valued random variable by the Borel isomorphism theorem.} However, any two non-identical informative signals in our model are \textit{not} Blackwell ordered. Thus, the buyer's choice about what to learn depends on the equilibrium product offerings and their prices. Similarly, the seller's menu results in very different revenue depending on the endogenous distribution of posterior means, and if it fails to be revenue maximizing, the seller will re-optimize. Thus, our model captures a \textit{\textbf{two-way feedback}} between consumer learning and menu design: the buyer's learning shapes the demand system against which the seller optimizes, while the seller's menu shapes what the buyer finds valuable to learn.

Our main result (\Cref{thm:main}) shows that every equilibrium features a comonotonic posterior mean distribution (\emph{\textbf{vertical learning}}) and is outcome-equivalent to an equilibrium in which the seller offers a menu of nested bundles (\emph{\textbf{nested bundling}}). This result holds regardless of the correlation in the underlying value distribution. In particular, even if the values $\mathbf{v}$ are negatively correlated, the buyer's equilibrium types (posterior means) must be vertically ordered. Moreover, we show that every equilibrium outcome, which consists of a learning strategy and a nested menu, has a simple structure: In equilibrium, the buyer learns more about the higher-tier goods (the upgrades), resulting in a higher posterior variance on the log scale (\Cref{prop:ordering}). 

\begin{figure}[!t]
\begin{center}
\begin{tikzpicture}[scale=0.85, every node/.style={transform shape}]

  \draw[->, thick] (-0.1,0) -- (4.5,0) node[below] {$v_1$};
  \draw[->, thick] (0,-0.1) -- (0,4.5) node[left] {$v_2$};
 \filldraw[purple] (2,2) circle (2pt);
\fill[black!20, opacity=0.4] (2,2) circle (2);
\draw[thick, purple, dashed] (1.2,0.2) -- (3.83, 2.83) node[ right] {$s=1$};
\draw[thick, purple, dashed] (0.6,0.6) -- (3.42, 3.42) node[ right] {$s=0$};
\draw[thick, purple, dashed] (2.7,0.2) -- (3.9, 1.4) node[ right] {$s=2.5$};
\draw[thick] (0.1,2) -- (-0.1,2) node[left] {$2$};
   \draw[thick] (2,0.1) -- (2,-0.1) node[below] {$2$};
 \filldraw[purple] (2.5,1.5) circle (2pt);
 \filldraw[purple] (3.3,0.8) circle (2pt);
\node at (2, 4.5) {\small B learns $v_1-v_2$};

  \draw[->, thick] (-0.1+6,0) -- (4.5+6,0) node[below] {$v_1$};
  \draw[->, thick] (0+6,-0.1) -- (0+6,4.5) node[left] {$v_2$};
 \filldraw[black] (2+6,2) circle (1pt);
\fill[black!20, opacity=0.4] (2+6,2) circle (2);
\draw[thick, purple] (0.6+6,3.4) -- (3.4+6, 0.6);
\draw[thick] (0.1+6,2) -- (-0.1+6,2) ;
   \draw[thick] (2+6,0.1) -- (2+6,-0.1) ;
\node at (2+6, 4.5) {\small B learns $v_1-v_2$};
\draw[thick, blue, dashed] ( 0+6, 3.9) -- (3.9+6,0) ;
\draw[thick, blue] (4+6,0.1) -- (4+6,-0.1) node[below] {$4$};
\draw[thick, blue] (0.1+6,4) -- (-0.1+6,4) node[left] {$4$};
\draw[thick] (0.1+6,2) -- (-0.1+6,2) node[left] {$2$};
   \draw[thick] (2+6,0.1) -- (2+6,-0.1) node[below] {$2$};
   
\node at (2.8+6, 2.8) {${\color{blue} \{1,2\}}$};
\node at (1.2+6, 1.2) {${\color{blue} \varnothing}$};

  \draw[->, thick] (-0.1+12,0) -- (4.5+12,0) node[below] {$v_1$};
  \draw[->, thick] (0+12,-0.1) -- (0+12,4.5) node[left] {$v_2$};
 \filldraw[black] (2+12,2) circle (1pt);
\fill[black!20, opacity=0.4] (2+12,2) circle (2);
\draw[thick, purple] (0.6+12,0.6) -- (3.4+12, 3.4);
\draw[thick] (0.1+12,2) -- (-0.1+12,2) ;
   \draw[thick] (2+12,0.1) -- (2+12,-0.1) ;
   \draw[thick] (0.1+12,2) -- (-0.1+12,2) node[left] {$2$};
   \draw[thick] (2+12,0.1) -- (2+12,-0.1) node[below] {$2$};
\draw[thick, blue, dashed] ( 0+12, 3.1) -- (3.1+12,0) ;
\draw[thick, blue] (3.13+12,0.1) -- (3.13+12,-0.1) node[below] {$3.13$};
\draw[thick, blue] (0.1+12,3.13) -- (-0.1+12,3.13) node[left] {$3.13$};
\node at (1.4+12, 2.7) {${\color{blue} \{1,2\}}$};
\node at (0.5+12, 1.5) {${\color{blue} \varnothing}$};
\node at (2+12, 4.5) {\small B learns $v_1+v_2$};

\end{tikzpicture}
\end{center}
\caption{The shaded gray area is the set of possible values $V$. The left panel illustrates how the buyer updates his belief about goods' expected values upon learning $v_1-v_2=s$, for several realizations of $s$. The middle panel illustrates that the seller would offer only the bundle when the buyer chooses signal $v_1-v_2$, leading to a contradiction. The right panel illustrates an equilibrium with pure bundling. The full red line corresponds to the support of the buyer's type distribution. The dashed blue lines partition the type space based on optimal allocations under the seller's menu: types above the dashed line purchase $\{1,2\}$ while types below it purchase nothing $\varnothing$.  }\label{fig:ex_intro}
\end{figure}
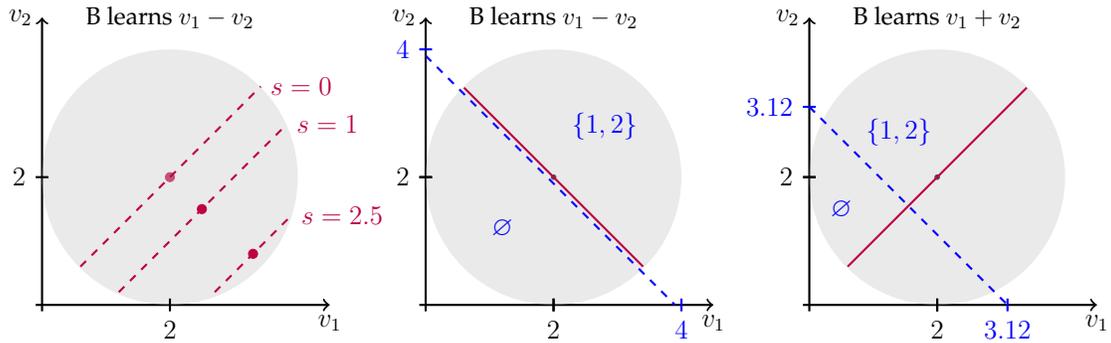

To illustrate the basic intuition behind our main result, consider the following example:

\vspace{-5mm}
\paragraph{Illustrative Example.}\hspace{-2mm}There are two goods whose values are drawn from a Gaussian distribution with $\mathbb{E}[v_1] = \mathbb{E}[v_2]=2$, $\text{Var}(v_1)=\text{Var}(v_2)=4$, and $\text{Corr}(v_1, v_2)=0$. The support of the distribution is truncated along an iso-density curve so as to lie in the positive quadrant as depicted in \Cref{fig:ex_intro}.\footnote{This truncation reduces the variance of each good to $0.92$.} 

Suppose for contradiction that we are in an equilibrium where the buyer chooses to learn about the difference between the two goods $v_1 - v_2$, which is a \textit{\textbf{horizontal learning}} strategy. If he learns $v_1-v_2=s$, then he knows that his values lie on the corresponding $45$-degree line segment (dashed red lines in the left panel of  \Cref{fig:ex_intro}). The goods are ex ante symmetric, so his posterior expected values for goods $1$ and $2$ are $2+0.5s$ and $2-0.5s$, respectively. Thus, the buyer's realized type always lies on the $-45$-degree line segment going through the prior mean (full red line in the middle panel).  
The seller correctly anticipates the buyer's chosen signal, but does not observe the signal realization. From the seller's point of view, the buyer's type then follows a (truncated) Gaussian distribution supported on the $-45$-degree line segment. Against this type distribution, to be revenue maximizing, the seller must offer the \textit{bundle} at price $4$, which extracts the full surplus since the sum of the posterior means for the two goods is always $(2+0.5s) + (2-0.5s) = 4$. However, the buyer then prefers to learn about the value of the bundle $v_1 + v_2$, which is a vertical learning strategy---hence, a contradiction. 

Now, suppose that the buyer indeed chooses the vertical learning strategy that reveals the bundle value $v_1+v_2$. Upon learning $v_1+v_2=s$, the buyer's posterior mean for each good is simply $0.5s$. The buyer's realized type now always lies on the $45$-degree line segment going through the prior mean (full red line in the right panel). Against this type distribution, it is an optimal strategy for the seller to offer a menu that consists only of the bundle $\{1,2\}$ at price $3.13$. Against this menu, it is indeed optimal for the buyer to learn his value for the bundle $v_1+v_2$. Thus, we have found an equilibrium. \qed

\vspace{10mm}

As the illustrative example shows, the key intuition behind our result builds on the long-standing insight from multiproduct pricing (\citealt{adams1976commodity}): Bundling is profitable when consumers have negatively correlated preferences, since it averages out the variation in their willingness to pay for different goods. We take this insight to its logical conclusion when consumers need to learn about their values: In equilibrium, consumers cannot spend too much effort learning about their relative values across different goods---if they did, the seller would re-optimize by offering a bundle for which the horizontal information is useless.

Importantly, however, this does not mean that horizontal learning is inherently suboptimal for the buyer---indeed, it can be optimal when the buyer moves first and commits to his learning strategy (see \Cref{subsec:timing}). Rather, our result is driven by the mutual-best-response requirement: The very menu that is optimal against horizontal learning induces the buyer to re-optimize what to learn. Consequently, in equilibrium, consumers must acquire information that updates their beliefs about different goods in the same direction---such as how much they value the firm's characteristics, the products' shared features, or general aspects of the underlying technology. Such vertical learning leads to positively correlated preferences. To screen such consumers, the seller then offers a menu of nested bundles, with larger bundles targeting consumers with higher posterior expected values for all the goods.

 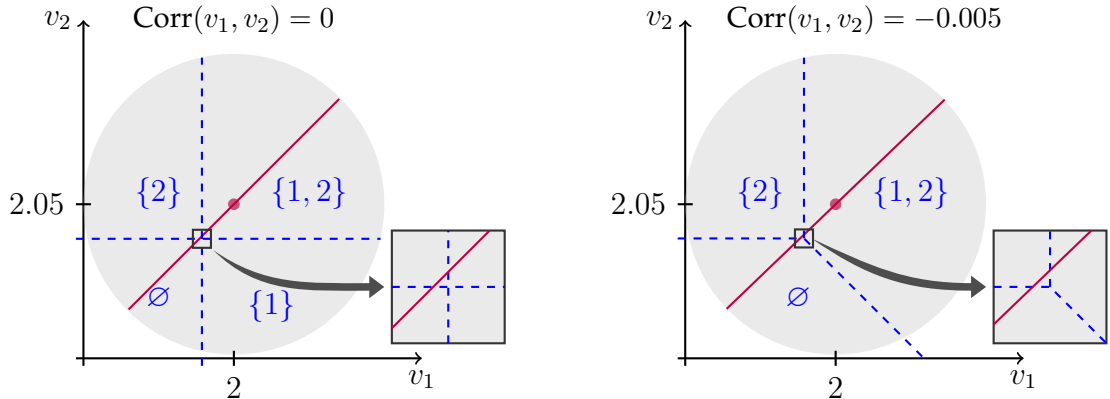
\begin{figure}[!t]
\begin{center}
\begin{tikzpicture}[scale=1, every node/.style={transform shape}]

  \draw[->, thick] (-0.1,0) -- (4.5,0) node[below] {$v_1$};
  \draw[->, thick] (0,-0.1) -- (0,4.5) node[left] {$v_2$};
 \filldraw[purple] (2,2.05) circle (2pt);
\fill[black!20, opacity=0.4] (2,2.05) circle (2);
\draw[thick] (0.1,2.05) -- (-0.1,2.05) node[left] {$2.05$};
   \draw[thick] (2,0.1) -- (2,-0.1) node[below] {$2$};
\draw[thick, purple] (0.6,0.6+0.05) -- (3.4, 3.4+0.05);
\draw[thick, blue, dashed] (-0.1,1.59) -- (3.95,1.59);
\draw[thick, blue, dashed] (3.1650-1.59,-0.1) -- (3.1650-1.59,4);
\draw[thick, black!80]  (3.1650-1.59+0.12,1.59-0.12) rectangle (3.1650-1.59-0.12, 1.59+0.12);
\fill[black!20, opacity=0.4] (4.1,0.2) rectangle (5.6, 1.7);
\draw[thick, black!80] (4.1,0.2) rectangle (5.6, 1.7);
\draw[thick, blue, dashed] (4.85,0.2) -- (4.85,1.7);
\draw[thick, blue, dashed] (4.1,0.95) -- (5.6, 0.95);
\draw[thick, purple] (4.1,0.4) -- (5.4, 1.7);
\fill[black!70]    (3.1650-1.59+0.15,1.59-0.15) to[out=-45,in=180] (3.8,0.9) to (3.8, 0.8) to (4, 0.95) to (3.8, 1.1) to (3.8, 1) to [out=180,in=-35] (3.1650-1.59+0.15,1.59-0.15);
\node at (2, 4.5) {\small $\text{Corr}(v_1,v_2)=0$};
\node at (3, 2.2) {${\color{blue}\{1, 2\}}$};
\node at (1, 2.2) {${\color{blue}\{2\}}$};
\node at (1, 0.8) {${\color{blue}\varnothing}$};
\node at (2.5, 0.7) {${\color{blue}\{1\}}$};

  \draw[->, thick] (-0.1+8,0) -- (4.5+8,0) node[below] {$v_1$};
  \draw[->, thick] (0+8,-0.1) -- (0+8,4.5) node[left] {$v_2$};
 \filldraw[purple] (2+8,2.05) circle (2pt);
\fill[black!20, opacity=0.4, rotate around={45: (2+8,2.05)}] (2+8,2.05) ellipse [x radius=2.0010, y radius= 1.9990 ];
\draw[thick] (0.1+8,2.05) -- (-0.1+8,2.05) node[left] {$2.05$};
   \draw[thick] (2+8,0.1) -- (2+8,-0.1) node[below] {$2$};
\draw[thick, purple] (0.55+8,{2.05-0.9569*(2-0.55)}) -- (3.45+8, {2.05+0.9569*(3.45-2)});
\draw[thick, blue, dashed] (-0.1+8,1.595) -- (3.174-1.595+8,1.595) -- (3.174+8,0);
\draw[thick, blue, dashed] (3.174-1.595+8,1.595) -- (3.174-1.595+8,4);
\draw[thick, black!80]  (3.174-1.595+8+0.12,1.595-0.12) rectangle (3.174-1.595+8-0.12, 1.595+0.12);
\fill[black!20, opacity=0.4] (4.1+8,0.2) rectangle (5.6+8, 1.7);
\draw[thick, black!80] (4.1+8,0.2) rectangle (5.6+8, 1.7);
\draw[thick, blue, dashed] (4.85+8,0.95) -- (4.85+8,1.7);
\draw[thick, blue, dashed] (4.1+8,0.95) -- (4.85+8, 0.95) --  (5.6+8, 0.2);
\draw[thick, purple] (4.1+8,0.45) -- (5.4+8, 1.7);
\fill[black!70]    (3.174-1.595+8+0.1, 1.595) to[out=-35,in=180] (3.8+8,0.9) to (3.8+8, 0.8) to (4+8, 0.95) to (3.8+8, 1.1) to (3.8+8, 1) to [out=180,in=-25]  (3.174-1.595+8+0.1, 1.595);
\node at (3+8, 2.2) {${\color{blue}\{1, 2\}}$};
\node at (1+8, 2.2) {${\color{blue}\{2\}}$};
\node at (1.5+8, 0.8) {${\color{blue}\varnothing}$};

\node at (2+8, 4.5) {\small $\qquad \text{Corr}(v_1,v_2)=-0.001$};
\end{tikzpicture}
\end{center}
\caption{Two perturbations of the above illustrative example. In the left panel, $\mathbb{E}[v_2]=2.05$ and $\text{Corr}(v_1,v_2)=0$. The figure depicts an equilibrium with separate sales. There no longer exists an equilibrium with pure bundling. In the right panel, $\mathbb{E}[v_2]=2.05$ and $\text{Corr}(v_1,v_2)=-0.001$. The figure depicts an equilibrium with nested bundling. There no longer exists an equilibrium with separate sales.}\label{fig:ex_intro2}
\end{figure}

The intuition behind why nested bundling is profitable against positively correlated preferences differs from the classic intuition that bundling averages out different values for different goods. Indeed, in the illustrative example, in the vertical-learning pure-bundling equilibrium, the bundle is not used to average out the willingness to pay but rather to screen the vertical information. In general, the seller offers more than the grand bundle, creating different tiers, and these tiers are ordered in such a way as to facilitate screening, as characterized in \Cref{prop:ordering}. For instance, in the previous illustrative example, if $\mathbb{E}[v_2]$ is perturbed to $\mathbb{E}[v_2]=2.05$, then pure bundling ceases to be an equilibrium. There exists, however, an equilibrium in which the buyer learns $1.05 v_1+v_2$ and the seller offers a nested menu $\big\{\{2\}, \{1, 2\}\big\}$ where the base bundle $\{2\}$ is priced at $1.59$, and the full bundle $\{1, 2\}$ is priced at $3.16$. The slightly higher mean and lower learning weight of good 2 lead the buyer's posterior means to be more concentrated on the log scale, making it strictly profitable for the seller to offer $\{2\}$ in addition to $\{1,2\}$.\footnote{See \Cref{subsec:example} for various other illustrative examples. The equilibria depicted in \Cref{fig:ex_intro2} and, later on, in \Cref{fig:ex_nb}, were computed numerically and verified as mutual best responses.}  The logic behind the buyer having a higher log-scale posterior variance about an upgrade good is due to the optimization by the \textit{seller}. If the buyer were to learn less about the higher-tier goods, then his posterior means would be more concentrated, but then the seller would benefit from switching the ordering of the goods to better screen the buyer. Intuitively, goods with lower log-scale posterior variance have more elastic demand curves and thus can be offered in the lower tiers to target the lower end of the consumers.  

Besides the screening property, nested bundling and vertical learning have a self-stabilizing aspect. To illustrate, note that in the above perturbed  example, another optimal mechanism  against vertical learning is to sell goods $1$ and $2$ separately, at price $1.57$ and $1.59$ respectively. When $\text{Corr}(v_1,v_2) = 0$, this still constitutes an equilibrium, which is depicted in \Cref{fig:ex_intro2} (left panel). This equilibrium is outcome equivalent to the nested bundling equilibrium: the buyer's learning is unchanged, and the added option in the seller's menu is never purchased. However, if we perturb the example once again and set $\text{Corr}(v_1,v_2) = -0.001$, separate sales cease to be an equilibrium, but nested bundling still is (\Cref{fig:ex_intro2}, right panel).  This is not specific to this example. Indeed, as we show, even though a separate sales mechanism can also be optimal against a vertical type distribution, it can never form an equilibrium when $\mathbf{v}$ has negative correlation because then the buyer strictly benefits from switching to a horizontal learning strategy (\Cref{prop:no-separate-sales}). Meanwhile, a nested bundling equilibrium always exists with symmetric distributions (\Cref{prop:existence}), even when $\mathbf{v}$ has negative correlation. By decreasing the instrumental value of horizontal comparisons, bundling favors vertical learning, which is necessary for equilibrium.

\subsection{Related Literature}

We build on a large literature on multiproduct pricing and optimal bundling (starting with \citealt*{stigler1963united, adams1976commodity, McAfee1989MultiproductValues, armstrong1996multiproduct, Rochet1998}). Following most of this literature, we assume that the buyer has additive values (\citealt*{mcafee1988multidimensional, manelli2006bundling, pavlov2011optimal, daskalakis2017strong}; \citealt{bergemann2022optimality}).\footnote{For models with non-additive values, see e.g. \citet{haghpanah2021pure}, \citet{ghili2021characterization}, and \citet{yang2023nested}.}  
There are two general insights from this literature: \textit{(i)} Bundling is often more profitable in settings with negatively correlated values (\citealt{stigler1963united}; \citealt{adams1976commodity}); \textit{(ii)} some form of bundling is generically profitable, but characterizing optimal mechanisms is analytically intractable (\citealt*{McAfee1989MultiproductValues,Rochet2003}).\footnote{Finding the optimal mechanism is also computationally intractable (\citealt*{Daskalakis2014}). The optimal mechanism often requires an infinite menu size (\citealt{hart2019selling}), and a small perturbation of virtually any incentive-compatible mechanism can make it optimal for some type distribution (\citealt{Manelli2007}; \citealt{lahr2024extreme}).}

Given the difficulty in multidimensional screening, our main conceptual contribution is to take a step back and model the buyer's learning process, which disciplines what type distributions are likely to arise endogenously in markets with new firms or new products. Our results take the classic insight from the bundling literature to its logical conclusion: When consumer preferences arise endogenously from optimal learning, they are likely to be positively correlated across goods; any negative correlation invites enough bundling responses to always disincentivize horizontal learning. As we explained, our model also brings out new insights about the self-stabilizing nature of nested bundling and vertical learning. As a consequence, we provide a \textit{microfoundation} for the exogenous type spaces studied in the bundling literature---in particular, \citet{yang2023nested}, which assumes a comonotonic type distribution to characterize demand conditions under which nested bundling is optimal (allowing for non-additive values).\footnote{For other sufficient conditions under which nested bundling is optimal for a monopolist, see also \citet{bergemann2022optimality} and \citet{yang2022costly}. Nested bundling can also arise as entry deterrence  via ``tying'' by a firm with monopoly power in one market to foreclose sales in, and thereby monopolize, another market (\citealt{whinstontying}).} 

In proving our main result, we also make a technical contribution to this literature by fully characterizing all optimal mechanisms for type distributions supported on any line segment in any dimension---in particular, characterizing optimal mechanisms when the buyer's types are not vertically ordered. The key difficulty is that the worst-off types are endogenous to a given mechanism. Our analysis uncovers and leverages a hidden connection between the recent works by \citet*{frick2024multidimensional} and \citet{loertscher2024optimal}---in particular, we provide a procedure that uses the \textit{dual multiplier} of the auxiliary program in \citet*{frick2024multidimensional} to explicitly construct a saddle point between mechanisms and worst-off types. Combined with the saddle-point formulation of \citet{loertscher2024optimal}, this construction yields our characterization of all optimal mechanisms.

Several recent papers adopt a robustness approach to study multidimensional screening (\citealt*{Carroll2017, brooks2024structure, debroesler2024, che2025robustly}). In these papers, the seller evaluates the performance of a mechanism against the worst-case distribution of buyer types within some set of admissible distributions. Like in our paper, the relevant distribution is then endogenous to the mechanism. However, it is not a result of the buyer's learning incentives. Most relevant for our analysis is \citet{debroesler2024}. They consider a setting where the seller and the buyer share a common prior about the buyer's values, but the seller is agnostic as to which additional information the buyer might have. Thus, the set of admissible type distributions is the set of all distributions that can be induced by \textit{some} signal, given the prior. Assuming an exchangeable prior, under which all goods have the same prior mean and variance, they show that randomized pure bundling is worst-case optimal where the worst-case signal reveals noisy information about the grand bundle in a way that generates a truncated Pareto distribution. This implies that when the buyer moves first, such that his chosen signal is observed by the seller, it leads to pure bundling with efficient trade.\footnote{This generalizes the single-good result of \citet{roesler2017buyer}.} We complement their analysis by considering a simultaneous-move game where the buyer cannot flexibly design a signal and cannot commit to the signal---he optimally chooses what to learn, given the seller's menu, from a set of Blackwell undominated signals. As a consequence, trade is inefficient in our model, and outcomes generally involve nested bundling with the buyer learning more about higher-tier goods.\footnote{See \Cref{subsec:timing} for further discussion of what happens if the buyer moves first in our model.}

Lastly, we contribute to the literature on mechanism design with information acquisition (\citealt*{bergemann2002information, shi2012optimal, mensch2022screening, mensch2025monopoly}). This literature has studied how the agent's learning incentives affect the principal's optimal mechanism in various settings, ranging from the design of efficient mechanisms (\citealt{bergemann2002information}) to monopoly pricing (\citealt{mensch2025monopoly}).\footnote{A smaller part of the literature studies post-purchase learning and product returns (e.g. \citealt{che1996customer}; \citealt{matthews2007information}); we abstract away from these concerns. For a discussion of bundling and product returns, see \citet*{haberman2025multidimensional}.} We depart from that literature in two main ways. First, to the best of our knowledge, this paper is the first to focus on learning incentives in multiproduct monopoly pricing. The complexity compared to the single-product case arises from the fact that the buyer learns about a multidimensional state.\footnote{A few papers also study multidimensional learning but in other contexts (\citealt*{gleyze2023informationally, bobkova2024information, bobkova2024optimality, pernoud2025competition}).} Second, we consider a simultaneous-move game between the buyer and seller, while most existing work lets the principal move first.\footnote{See \Cref{subsec:timing} for a discussion of what happens when the seller moves first; as we explain there, the timing of the moves is important for sustaining vertical learning.} A notable exception is \citet*{ravid2022learning} who also consider a simultaneous-move game, but in single-good monopoly pricing. 

\vspace{-3mm}
\paragraph{Overview.}\hspace{-2mm}The remainder of the paper proceeds as follows. \Cref{sec:model} presents our model. \Cref{sec:main} presents our main result and further illustrates the intuition. \Cref{sec:proof} sketches the proof of the main result. \Cref{sec:discuss} discusses assumptions and generalizations. \Cref{sec:conclude} concludes. All proofs are in \Cref{app:proof}. 

\section{Model}\label{sec:model}

We consider a simultaneous-move game between a seller and a buyer. The seller (she) has $K$ indivisible goods to sell to the buyer (he). The buyer's utility is additive across goods and quasilinear in money.\footnote{We relax this assumption and discuss robustness to non-additive values in  \Cref{subsec:nonadditive}.} His payoff from purchasing bundle $B\subseteq \{1,\dots, K\}$ at price $p$ is then
\[\sum_{k\in B}v_k -p\,,\]
where $v_k$ denotes his value for good $k$. We assume that the buyer's values are above the seller's costs, and normalize the cost for each good to be zero.\footnote{We relax this assumption and discuss the case of high production costs in \Cref{subsec:productioncosts}.}

The buyer's \textit{\textbf{values}} $\mathbf{v}=(v_k)_k$ follow a unimodal elliptical distribution with continuous density supported on a compact set $V\subset \mathbb{R}_+^K$.\footnote{Formally, a random vector $\mathbf{v} \in \mathbb{R}^{K}$ has a \textit{\textbf{unimodal elliptical distribution}} if the density of $\mathbf{v}$ only depends on the quadratic form: $f_{\mathbf{v}}(\mathbf{z}) = g\big((\mathbf{z}-\boldsymbol\mu)'\Sigma^{-1}(\mathbf{z}-\boldsymbol\mu)\big)$, where $g$ is an admissible density generator and is non-increasing. Iso-density curves are then ellipsoids centered at $\boldsymbol\mu$.} Let $\boldsymbol\mu = (\mu_k)_k$ and $\Sigma$ denote the mean vector and covariance matrix of $\mathbf{v}$. Goods can differ in their prior mean and variance, but we assume that they share the same correlation $\text{Corr}(v_i, v_j)=\rho\in(-1,1)$ for all pairs of goods $i$ and $j$. The values are \textbf{\textit{positively correlated}} if $\rho > 0$, \textit{\textbf{negatively correlated}} if $\rho < 0$, and \textbf{\textit{uncorrelated}} otherwise.

The buyer does not observe $\mathbf{v}$ but has access to a dimension-restricted \textit{\textbf{learning technology}}: he can choose any one-dimensional linear signal of the vector of values $\mathbf{v}$. That is, a learning strategy for the buyer consists of choosing \textit{\textbf{learning weights}} $\boldsymbol\alpha\in\mathbb{R}^K$, and the buyer gets to observe the realization of $\boldsymbol\alpha\cdot\mathbf{v}$.\footnote{Allowing instead any one-dimensional signal $s$ jointly elliptically distributed with $\mathbf{v}$ would not change our analysis. Moreover, the learning weights can be normalized so that $||\boldsymbol \alpha|| = 1$; we keep the un-normalized version for convenience.}

Without observing the buyer's choice of signal, the seller chooses a \textit{\textbf{selling mechanism}} $\mathcal{M} = (M, x, p)$, which consists of
\[\text{message space $M$, allocation rule $x:M\rightarrow \Delta(2^K)$, payment rule $p:M\rightarrow \mathbb{R}$}\,.\]
Equivalently, a mechanism can be represented as a \textbf{\textit{menu}} $\{(x, p)\}$ of lotteries over bundles and associated prices.

\vspace{-3mm}
\paragraph{Buyer's strategy and induced type distribution.}\hspace{-2mm}We now explain in more detail how the buyer's learning strategy maps into a type distribution.
The buyer is risk-neutral, so his purchasing decision depends only on his posterior expected value for each good. Given weights $\boldsymbol\alpha$ and signal realization $s$, the buyer's \textit{\textbf{type}} $\boldsymbol\theta = (\theta_k)_k$ consists of the expected value for each good $k$:  
\begin{align*}
    \theta_k(s;\boldsymbol\alpha) :=  \mathbb{E}[v_k\mid s]= \mu_k + \frac{\text{Cov}(v_k,\boldsymbol\alpha\cdot\mathbf{v})}{\text{Var}(\boldsymbol\alpha\cdot\mathbf{v})}[s-\boldsymbol\alpha\cdot\boldsymbol\mu]\,.
\end{align*}
Let $G_{\boldsymbol\alpha}$ denote the distribution of types induced by weights $\boldsymbol\alpha$. Note that conditional expectations, or types, are linear in the signal realization $s$.\footnote{The family of elliptical distributions is the most general class of distributions with this property (see \citealt*{gupta2013elliptically}).} This is an important property that we leverage in our analysis. It implies a \textit{\textbf{linear type distribution}}---i.e., the support of $G_{\boldsymbol\alpha}$ is a line segment in $\mathbb{R}^K$. It also implies that the buyer's type has a symmetric unimodal density, since unimodal ellipticity is preserved under affine transformations. The uninformative signal $\boldsymbol{\alpha} = \mathbf{0}$ is dominated by any other signal and never chosen in equilibrium. Without loss of generality, we assume that $\text{Cov}(v_k,\boldsymbol\alpha\cdot\mathbf{v}) > 0$ for at least one good $k$.\footnote{Indeed, if $\text{Cov}(v_k,\boldsymbol\alpha\cdot\mathbf{v}) \leq 0$ for all goods $k$, then the alternative signal $\boldsymbol\alpha'=-\boldsymbol\alpha$ induces the same type distribution, but satisfies $\text{Cov}(v_k,\boldsymbol\alpha' \cdot\mathbf{v}) \geq 0$ for all goods $k$.} 

A buyer with type $\boldsymbol{\theta}(s; \boldsymbol\alpha)$ who faces a mechanism $\mathcal{M}$ solves
\[\sup_{m\in M} \; \sum_k \theta_k(s;\boldsymbol\alpha)x_k(m) -p(m)\,. 
\]
Without loss of generality, we assume that there exists some $m_o \in M$ such that $x(m_o) = \varnothing$ and $p(m_o) = 0$ (i.e., the buyer can always walk away to obtain his outside option, which is normalized to have value $0$).
As is standard, we assume that when indifferent between two messages, the buyer breaks the indifference in favor of the seller. This guarantees the existence of an optimal mechanism for the seller given a type distribution. 

\paragraph{Solution concept.}\hspace{-2mm}Our solution concept is pure-strategy Nash equilibrium.\footnote{Our results hold even with a weaker solution concept as shown in \Cref{subsec:weak}.} A strategy profile $(\boldsymbol\alpha, \mathcal{M})$ forms an \textit{\textbf{equilibrium}} if the buyer's learning strategy $\boldsymbol\alpha$ is optimal against mechanism $\mathcal{M}$, and mechanism $\mathcal{M}$ is profit-maximizing against the type distribution  induced by learning strategy $\boldsymbol{\alpha}$. Two equilibria are \textit{\textbf{outcome-equivalent}} if they induce the same type distribution, allocation, and transfer. 

\paragraph{Interpretation of the learning technology.}\hspace{-2mm}Our main goal is to study the direction of learning (\emph{what} the buyer learns about) and not the extent of learning (\emph{how much} he learns). We thus model learning as being free but constrained in its dimensionality: The buyer can only learn along one direction, but can learn as much as possible along that direction. We could augment the model to also allow the buyer to control the precision of the signal at some cost. That is, after having chosen a direction, the buyer also chooses a level of noise, trading off higher precision for higher costs. As long as the cost is the same in all directions, all of our results go through. 

We view our framework as a device to model the direction of learning.  Intuitively, a signal that puts weights of the same sign on goods $i$ and $j$ can be interpreted as learning about some \emph{shared} feature of goods $i$ and $j$, while a signal that puts weights of different signs on $i$ and $j$ can be interpreted as learning about their \emph{difference} in some aspect. Formally, our one-dimensional linear signals can be characterized as the least informative signals under which the buyer can make one perfect \textit{\textbf{binary comparison}} between two stochastic bundles: 
\begin{observation}
Fix stochastic bundles $x, x'\in[0,1]^K$. A signal lets the buyer make the ex post correct choice between $(x,p)$ and $(x',p')$ for every pair of prices $(p,p')$ if and only if it is Blackwell more informative than the one-dimensional linear signal $(x-x')\cdot\mathbf{v}$. In particular, $(x-x')\cdot\mathbf{v}$ is the least informative signal that does so.
\end{observation}
Therefore, an interpretation of our model is that the buyer can only make one perfect binary comparison and chooses which pair of bundles (including the outside option) to compare.  One-dimensional linear signals not only constrain the buyer's learning but also provide tractability. We discuss the role played by this assumption and the possibility of multidimensional signals in  \Cref{subsec:multi}. 

The buyer's learning strategy, combined with the underlying distribution of values $\mathbf{v}$, pins down the \textit{observable demand system}---in particular, whether it features vertically differentiated types---which is the key focus of our analysis.

\section{Main Result}\label{sec:main}

To state our main result, we start by formally classifying the demand systems induced by consumer learning. 
\vspace{-3mm}
\paragraph{Classifying Induced Demand Systems.}\hspace{-2mm}We say that a learning strategy $\boldsymbol{\alpha}$ \textit{\textbf{induces a vertical demand system}}, or simply, is \textbf{\textit{vertical}} if the posterior mean distribution is comonotonic, i.e., its support can be totally ordered in $\R^K$: 
\[\text{for all distinct $\boldsymbol{\theta}, \boldsymbol{\theta}' \in  \supp  G_{\boldsymbol{\alpha}}$, we have either $\boldsymbol{\theta} \leq \boldsymbol{\theta}'$ or $\boldsymbol{\theta}' \leq \boldsymbol{\theta}$}. \tag{\textbf{Vertical}}\]
By contrast, we say that a learning strategy $\boldsymbol{\alpha}$ is \textbf{\textit{horizontal}} if no pair of types in the support can be ordered in $\R^K$: 
\[\text{\,\,\,\,for all distinct $\boldsymbol{\theta}, \boldsymbol{\theta}' \in  \supp  G_{\boldsymbol{\alpha}}$, we have both $\boldsymbol{\theta} \not \leq \boldsymbol{\theta}'$ and $\boldsymbol{\theta}' \not \leq \boldsymbol{\theta}$}. \tag{\textbf{Horizontal}}\]
In words, learning is vertical if it induces a demand system where consumers can be ordered so that higher types have higher values for \textit{every} good, whereas learning is horizontal if it induces a demand system where a higher value for one good must be associated with a lower value for \textit{some} other good. 

In our one-dimensional linear setting, because any learning strategy $\boldsymbol{\alpha}$ leads to a linear type distribution, the above classification is actually a \textit{\textbf{dichotomy}}---a learning strategy $\boldsymbol{\alpha}$ is either vertical or horizontal.\footnote{Indeed, $\boldsymbol{\alpha}$ is vertical if $\text{Cov}\big(\theta_i(\boldsymbol{\alpha}), \theta_j(\boldsymbol{\alpha})\big) \geq 0$ for every pair $(i, j)$, and horizontal otherwise.} 
However, with richer or multidimensional signals, the above classification is no longer a dichotomy---learning could have both vertical and horizontal components at once, inducing a mixed demand system.\footnote{We discuss the robustness of our insights to richer or multidimensional signals in \Cref{subsec:multi}.}
\vspace{-3mm}
\paragraph{Main Result.}\hspace{-2mm}Say that an equilibrium has \textbf{\textit{vertical learning}} if the buyer uses a vertical learning strategy, and an equilibrium has \textbf{\textit{nested bundling}} if the seller offers a menu of deterministic bundles that can be totally ordered by set-inclusion. 

\begin{theorem}\label{thm:main}
Every equilibrium has vertical learning, and is outcome-equivalent to a nested bundling equilibrium.
\end{theorem}

The proofs  of \Cref{thm:main} and all subsequent results are in the appendix. We provide the intuition in \Cref{subsec:example} and a sketch of the proof in \Cref{sec:proof}.  The first part of \Cref{thm:main} asserts that only vertical learning can be sustained in equilibrium, and hence the type distribution endogenously features vertically differentiated types.  This is true regardless of whether the underlying values $\mathbf{v}$ are positively or negatively correlated. Moreover, note that this is true even though the space of vertical learning strategies is vanishingly small as $K$ increases---indeed, with uncorrelated values, vertical learning requires $\boldsymbol{\alpha}\in\R^K$ to have weights being all positive or all negative, which are only two possibilities out of $2^K$ possible sign combinations.  

The second part of \Cref{thm:main} asserts that for every equilibrium, either the seller is using a nested bundling strategy, or there exists a nested bundling equilibrium in which both the seller and the buyer get exactly the same outcomes. More precisely, the buyer uses the same learning strategy in both equilibria, and each realized type of the buyer purchases the same bundle. 

\vspace{-3mm}
\paragraph{Learning and Tiering.}\hspace{-2mm}Our next result further characterizes the buyer's learning strategy in equilibrium. For a given nested menu $\{(B_1,p_1), \dots, (B_m, p_m)\}$,  where $B_1 \subseteq \cdots \subseteq B_m$, we define the \textit{\textbf{tier}} of an item $i$ as the index of the smallest bundle that includes item $i$. 

\begin{proposition}\label{prop:ordering}
Consider any nested bundling equilibrium. For any items $i, j$ where $\emph{tier}(i) \leq \emph{tier}(j)$, we have 
\[ \emph{Cov}(v_i/\mu_i,\boldsymbol{\alpha}\cdot\mathbf{v}) \leq \emph{Cov}(v_j/\mu_j,\boldsymbol{\alpha}\cdot\mathbf{v})  \,\,\text{ and } \,\,\emph{Var}(\log(\theta_i)) \leq \emph{Var}(\log(\theta_j))\,.\]
If the values are uncorrelated, then the buyer's adjusted learning weights are ordered: 
\[0 \leq \frac{\sigma^2_i}{\mu_i}\alpha_i  \leq \frac{\sigma^2_j}{\mu_j} \alpha_j \,,\]
where $\sigma^2_i := \emph{Var}(v_i)$ and $\sigma^2_j := \emph{Var}(v_j)$.
\end{proposition}

\Cref{prop:ordering} says that, in any nested bundling equilibrium, the buyer's signal covaries more with the higher-tier good when normalized by the mean, resulting in a higher posterior variance for the higher-tier good on the log scale. Under uncorrelated values, \Cref{prop:ordering} shows that this is only possible if the adjusted learning weights are  ordered, where the adjustment takes into account that the same learning weight may resolve more uncertainty for one good than the other due to differences in the prior distribution. We provide the intuition and illustrative examples in \Cref{subsec:example}. Note that if two goods belong to the same tier, then the inequalities must hold with equality.

Since \Cref{thm:main} shows that all equilibrium outcomes are characterized by vertical learning and nested bundling, \Cref{prop:ordering} and \Cref{thm:main} together then give a qualitative prediction of all equilibrium outcomes. 

\vspace{-3mm}
\paragraph{Equilibrium Existence.}\hspace{-2mm}Since we focus on pure-strategy equilibria, an equilibrium may not always exist, but the following result gives simple sufficient conditions for the existence of an equilibrium: 
\begin{proposition}\label{prop:existence}
An equilibrium exists if $\mathbf{v}$ is exchangeable. 
Moreover, holding everything else fixed, there exists $\underline{\rho} < 1$ such that for all $\rho \geq \underline{\rho}$, an equilibrium exists.  
\end{proposition}
When $\mathbf{v}$ is exchangeable, by the argument given in the introduction, a pure bundling equilibrium exists. Otherwise, when $\rho$ is sufficiently high, the buyer's problem becomes quasi-concave (while the seller's problem is linear), and hence existence of equilibrium is guaranteed by standard fixed-point arguments. We also consider a weaker solution concept in \Cref{subsec:weak} that only requires the buyer to choose a Blackwell-undominated signal given the bundles offered by the seller. As we show in \Cref{subsec:weak}, our main results continue to hold and equilibrium existence is guaranteed.  

\vspace{-3mm}
\paragraph{Nested Bundling versus Separate Sales.}\hspace{-2mm}The second part of \Cref{thm:main} only holds up to outcome equivalence. This comes from the fact that the seller's best response is not unique: Given some learning strategy $\boldsymbol{\alpha}$, there exist many mechanisms that are optimal for the seller, and several can be part of an equilibrium. For instance, against any vertical $\boldsymbol{\alpha}$, there always exists a separate-sales mechanism that is optimal. However, separate sales are less stable than nested bundling---the next result shows that it is impossible to sustain separate sales in equilibrium if the values are negatively correlated:

\begin{proposition} \label{prop:no-separate-sales}
When $\rho<0$, there exists \emph{no} separate sales equilibrium. 
\end{proposition}

Against a separate-sales mechanism, the buyer's purchasing decision is separable across goods---he can separately decide whether to buy each good $i$ at price $p_i$. Any correlation in the buyer's posterior values for goods $(\theta_i, \theta_j)$ is then irrelevant; only the marginals of the type distribution matter. When values are uncorrelated $(\rho=0)$, any vertical learning strategy can be matched to a ``flipped'' horizontal learning strategy that induces the same marginal type distributions, even though the joint distribution differs. The buyer is then indifferent between vertical and horizontal learning. Any negative correlation $(\rho<0)$ breaks the indifference in favor of horizontal learning. However, horizontal learning is precluded in equilibrium by \Cref{thm:main}---hence, separate sales, unlike nested bundling, cannot be sustained in equilibrium with negatively correlated values. 

\subsection{Illustrative Examples}\label{subsec:example}

We now go through several two-good examples to further illustrate and provide intuition for \Cref{thm:main} and \Cref{prop:ordering} step by step.

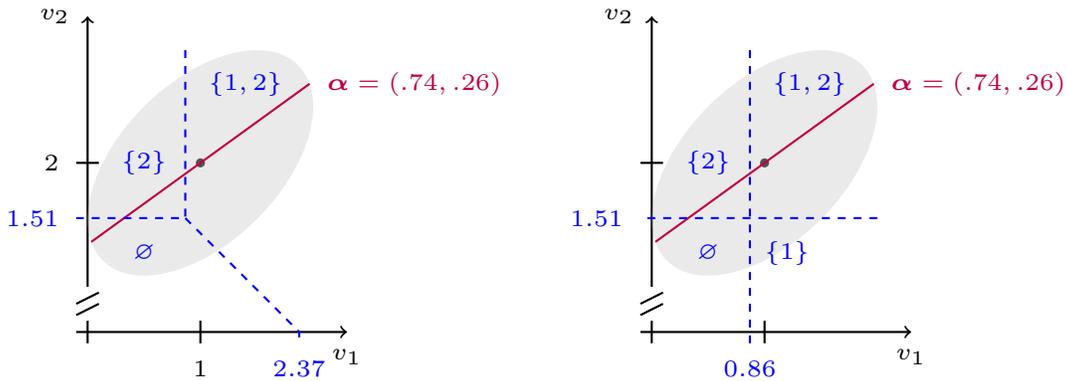
\begin{figure}[!t]
\begin{center}
\begin{tikzpicture}[scale=1.5, every node/.style={transform shape}]

  \draw[->, thick] (-0.1,0.5) -- (2.3,0.5) node[below] {\tiny$v_1$};
  \draw[->, thick] (0,0.9) -- (0,3.3) node[left] {\tiny$v_2$};
  \draw[thick] (0,0.4) -- (0,0.6);
   \draw[thick] (-0.1,0.65) -- (0.1,0.75);
   \draw[thick] (-0.1,0.75) -- (0.1,0.85);
 \filldraw[black] (1,2) circle (1pt);
\fill[black!20, opacity=0.4, rotate around={45: (1,2)}] (1,2) ellipse [x radius= 1.2247, y radius=0.7071];
\draw[thick, purple] (0.0321,1.2988) -- (1.9679, 2.7013) node[ right] {\tiny$\boldsymbol{\alpha}=(.74,.26)$};
\draw[thick] (0.1,2) -- (-0.1,2) node[left] {\tiny$2$};
   \draw[thick] (1,0.6) -- (1,0.4) node[below] {\tiny$1$};
 \draw[thick, blue, dashed] (2.376-1.52,3) -- (2.376-1.52, 1.52) -- (-0.1,1.52) node[left] {\tiny$ 1.52$};
 \draw[thick, blue, dashed] (2.376-1.52, 1.52) -- (2.376-0.5,0.5) -- (2.376-0.5,0.4)node[below] {\tiny$ 2.38$};
\node at (1.4, 2.7) {\tiny${\color{blue} \{1,2\}}$};
\node at (0.5, 1.2) {\tiny${\color{blue} \varnothing}$};
\node at (0.5, 2) {\tiny${\color{blue} \{2\}}$};

  \draw[->, thick] (-0.1+5,0.5) -- (2.3+5,0.5) node[below] {\tiny$v_1$};
  \draw[->, thick] (0+5,0.9) -- (0+5,3.3) node[left] {\tiny$v_2$};
  \draw[thick] (0+5,0.4) -- (0+5,0.6);
   \draw[thick] (-0.1+5,0.65) -- (0.1+5,0.75);
   \draw[thick] (-0.1+5,0.75) -- (0.1+5,0.85);
\filldraw[black] (1+5,2) circle (1pt);
\fill[black!20, opacity=0.4, rotate around={45: (1+5,2)}] (1+5,2) ellipse [x radius= 1.2247, y radius=0.7071];
 \draw[thick, purple] (0.0321+5,1.2988) -- (1.9679+5, 2.7013) node[ right] {\tiny$\boldsymbol{\alpha}=(.74,.26)$};
\draw[thick] (0.1+5,2) -- (-0.1+5,2) ;
   \draw[thick] (1+5,0.6) -- (1+5,0.4) ;
 \draw[thick, blue, dashed] ( 2+5, 1.52) -- (-0.1+5,1.52) node[left] {\tiny$ 1.52$};
 \draw[thick, blue, dashed]  (0.86+5,3) -- (0.86+5,0.4)node[below] {\tiny$  0.86$};
\node at (1.4+5, 2.7) {\tiny${\color{blue} \{1,2\}}$};
\node at (0.5+5, 1.2) {\tiny${\color{blue} \varnothing}$};
\node at (0.5+5, 2) {\tiny${\color{blue} \{2\}}$};
\node at (1.2+5, 1.2) {\tiny${\color{blue} \{1\}}$};

\end{tikzpicture}
\end{center}
\vspace{-10pt}
\caption{The shaded gray area is the set of possible values $V$. The left panel depicts an equilibrium with nested bundling. The buyer chooses signal $\boldsymbol{\alpha}=(.74,.26)$, leading to a type distribution supported on the full red line. The seller offers $\{2\}$ at price $1.52$ and $\{1,2\}$ at $2.38$. The right panel depicts a separate-sales equilibrium, which is outcome-equivalent to the nested-bundling equilibrium.}\label{fig:ex_nb}
\end{figure}

\vspace{-3mm}
\paragraph{Endogenous Ordering of Posterior Variance.}\hspace{-2mm}First, let the buyer's values be drawn from a Gaussian distribution with mean $\boldsymbol{\mu}=(1,2)$, unit variances, and correlation $\rho = 0.5$, whose support is truncated along an iso-density curve to lie in the positive quadrant, as depicted in \Cref{fig:ex_nb}. The induced distribution of $\mathbf{v}$ has $\sigma_1=\sigma_2 = 0.48$. There exists an equilibrium in which the buyer chooses signal $\boldsymbol{\alpha} = (.74, .26)$, and the seller offers a menu consisting of good $\{2\}$ at price $1.52$ and the bundle $\{1,2\}$ at $2.38$. This equilibrium is illustrated in the left panel of \Cref{fig:ex_nb}, where the red segment is the support of the type distribution induced by $\boldsymbol{\alpha}$. The red segment is increasing, which means that the buyer is using a vertical learning strategy as required by \Cref{thm:main}. The seller uses a nested bundling mechanism, where the low tier consists of good 2 and the high tier bundles good 1 with good 2. \Cref{prop:ordering} states that the buyer's posterior value for good 1 must be more dispersed on the log scale than his posterior value for good 2. This is indeed the case here since $\text{Var}(\log(\theta_1)) = 0.39$ while $\text{Var}(\log(\theta_2)) = 0.03$. To understand why the variation is measured on the log scale, note that the logic behind this comparison actually comes from the \textit{\textbf{seller's optimization}}. Indeed, if the log-scale dispersion of $\theta_1$ is strictly lower than that of $\theta_2$, then the endogenous demand curve of good  $1$ must be more elastic than that of good $2$---the seller then has an incentive to sell good $1$ to more consumers than good $2$, effectively swapping the base good and the upgrade good, to increase her revenue.\footnote{Indeed, as shown in \citet{yang2023nested}, items with more elastic demand curves are more profitably placed in lower tiers to target the lower end of the consumer base; for a broader analysis of how log-scale dispersion and demand elasticity affect optimal menu design, see \citet{yang2026screening}.}

\vspace{-3mm}
\paragraph{Outcome Equivalence to Nested Bundling.}\hspace{-2mm}The equilibrium described above features both vertical learning and nested bundling, but this is not the only equilibrium. Against the same type distribution, another optimal mechanism is a separate sales mechanism that offers good 2 at price $1.52$, good 1 at price $0.86$, and the grand bundle at the sum of the two prices (right panel). In this example, it happens that signal $\boldsymbol{\alpha} = (.74, .26)$ remains optimal under separate sales, and so this also constitutes an equilibrium. This equilibrium does \emph{not} feature nested bundling since the equilibrium mechanism offers both $\{1\}$ and $\{2\}$, but it is outcome-equivalent to a nested bundling equilibrium. Indeed, even though the buyer has the opportunity to buy each good by itself, he never does so in equilibrium, and makes the same purchasing decisions as in the nested bundling equilibrium. The players' payoffs and the equilibrium outcomes remain unchanged. This illustrates the second part of \Cref{thm:main}.

\vspace{-3mm}
\paragraph{Magnitude versus Direction of Information.}\hspace{-2mm}All the equilibria considered so far feature vertical learning, and \Cref{thm:main} states that only such equilibria can exist. This is true even when vertical learning seems to resolve little uncertainty for the buyer. For instance, take the extreme case of very negatively correlated values, as depicted in \Cref{fig:ex_negrho}. There is much more dispersion in the distribution of $\mathbf{v}$ along decreasing lines than along increasing lines. Thus, horizontal learning strategies seem to resolve much more uncertainty than vertical ones, and lead to type distributions with higher dispersion. 
 \begin{figure}[!t]
\begin{center}
\begin{tikzpicture}[scale=0.9, every node/.style={transform shape}]

  \draw[->, thick] (-0.1,0) -- (4.5,0) node[below right] {$v_1$};
  \draw[->, thick] (0,-0.1) -- (0,4.5) node[left] {$v_2$};
  \filldraw[black] (2,2) circle (1pt);
     \fill[black!20, opacity=0.4, rotate around={-45: (2,2)}] (2,2) ellipse [x radius=2.7568, y radius=0.6325 ];
  \draw[thick] (2,0.1) -- (2,-0.1) ;
  \draw[thick] (0.1, 2) -- (-0.1, 2) ;
  \draw[thick, purple]  (0.0,3.8) -- (4, 0.2) node[above right] {\small$\boldsymbol{\alpha}=(1,0)$};
\draw[thick, blue, dashed] (1.98,4) -- (1.98,1.97) -- (3.98,0);
\draw[thick, blue, dashed] (0,3.75) -- (1.98,1.97);
  \draw[thick] (0.1,2) -- (-0.1,2) node[left] {$2$};
  \draw[thick] (2,0.1) -- (2,-0.1) node[below] {$2$};
   \draw[thick] (3.97,0.1) -- (3.97,-0.1) node[below] {${\color{blue} 3.97}$};
   \draw[thick] (0.1, 3.8) -- (-0.1, 3.8) node[left] {${\color{blue} 3.80}$};
\node at (3, 2.5) {\small ${\color{blue} \{1,2\}}$};
\node at (1.6, 1.6) {\small${\color{blue} \varnothing}$};
\node at (1.2, 3.5) {\small${\color{blue} x^*}$};
\node at (2, 4.5) {\small$\rho = -0.9$};

  \draw[->, thick] (-0.1+8,0) -- (4.5+8,0) node[below right] {$v_1$};
  \draw[->, thick] (0+8,-0.1) -- (0+8,4.5) node[left] {$v_2$};
  \filldraw[black] (2+8,2) circle (1pt);
     \fill[black!20, opacity=0.4, rotate around={-45: (2+8,2)}] (2+8,2) ellipse [x radius=2.7568, y radius=0.6325 ];
  \draw[thick] (2+8,0.1) -- (2+8,-0.1) ;
  \draw[thick] (0.1+8, 2) -- (-0.1+8, 2) ;
  \draw[thick, purple] (1.55+8,1.55) -- (2.45+8,2.45) node[right] {\small$\boldsymbol{\alpha}=(1,1)$};
  \draw[thick] (0.1+8,2) -- (-0.1+8,2) node[left] {$2$};
  \draw[thick] (2+8,0.1) -- (2+8,-0.1) node[below] {$2$};
 \draw[thick, blue, dashed] ( 0+8, 3.23) -- (3.23+8,0) ;
\draw[thick] (3.23+8,0.1) -- (3.23+8,-0.1) node[below] {${\color{blue} 3.23}$};
\draw[thick] (0.1+8, 3.23) -- (-0.1+8, 3.23) node[left] {${\color{blue} 3.23}$};
\node at (2+8, 4.5) {$\rho = -0.9$};
\node at (3+8, 1.5) {\small ${\color{blue} \{1,2\}}$};
\node at (1+8, 1) {\small${\color{blue} \varnothing}$};

\end{tikzpicture}
\end{center}
\vspace{-10pt}
\caption{The left panel considers what happens when the buyer chooses horizontal learning strategy $\boldsymbol{\alpha}=(1,0)$. The type distribution is supported on the full red line. The associated optimal mechanism offers $\{1,2\}$ at price $3.97$ and the lottery $x^*=0.9\times \{1,2\}+0.1\times\{2\}$ at price $3.8$. This leaves little variation in the buyer's payoffs, which implies that the buyer's learning strategy must be quite uninformative about the optimal purchase decision given this menu. The right panel illustrates a pure bundling equilibrium. Even though the buyer's equilibrium signal $\boldsymbol{\alpha}=(1,1)$ leads to a less dispersed type distribution than $\boldsymbol{\alpha}=(1,0)$, it resolves all payoff-relevant uncertainty given the menu.}\label{fig:ex_negrho}
\end{figure}
However, \Cref{thm:main} asserts that horizontal learning cannot be sustained in equilibrium. To understand this, note that even though the \textit{\textbf{``magnitude of information''}} seems large for horizontal learning, any two non-identical signals in our model are \textit{not} Blackwell ordered. It turns out that the seller's optimal mechanism against any horizontal learning always leads to a decision problem for which the original strategy resolves uncertainty along the wrong \textit{\textbf{``direction of information''}} and is therefore suboptimal. Indeed, as shown in the left panel of \Cref{fig:ex_negrho}, the posterior-mean line (full red line) closely follows the buyer's indifference line (dashed blue line), along which each type is indifferent between participating and opting out---consequently, little \textit{\textbf{payoff-relevant information}} is resolved.

\vspace{-3mm}
\paragraph{Horizontal Learning and Zero-Dispersion Bundle.}\hspace{-2mm}To further understand the intuition behind the seller's best response, recall that horizontal learning leads to negatively correlated preferences: consumers with a higher willingness to pay for some good $i$ have a lower willingness to pay for some good $j$. As a result, the seller wants to bundle some goods to average out consumer heterogeneity and extract more rents. To illustrate, consider again the example depicted in \Cref{fig:ex_negrho}, where the distribution of $\mathbf{v}$ is exchangeable with $\rho=-0.9$.\footnote{Values are drawn from a symmetric Gaussian distribution with mean 2 and variance 4, truncated as depicted.} Signal $\boldsymbol{\alpha} = (1,0)$ fully reveals the buyer's value for good 1, but also provides information about good 2 since the values are correlated. The type distribution is then supported on a decreasing line segment, with slope $\text{Cov}(v_2,v_1)/\text{Var}(v_1) = -0.9$. Note that all types of the buyer have the same value for a particular \textit{\textbf{zero-dispersion bundle}} $x^*$ that bundles good $2$ and $0.9$ probability of good $1$. The seller's optimal mechanism, also depicted in \Cref{fig:ex_negrho}, leverages this zero-dispersion bundle to limit the variation in the buyer's payoffs. In particular, this bundle $x^*$ is \textit{\textbf{tailored}} to the buyer's original learning strategy: it is constructed so as to make $\boldsymbol{\alpha}=(1,0)$ completely uninformative about its value. Thus, even though the signal $\boldsymbol{\alpha}=(1,0)$ resolves a substantial amount of uncertainty about $\mathbf{v}$, it is actually very \textit{uninformative} about the payoff-relevant uncertainty---what to purchase when facing the seller's menu. Against horizontal learning, the seller's best response by design limits the informational value of the original learning strategy.

\vspace{-3mm}
\paragraph{Asymmetry of the Role of Bundles.}\hspace{-2mm}By contrast, vertical learning leads to positively correlated preferences, for which there exists no zero-dispersion bundle and the seller cannot ``average out'' the payoff variation.  The seller's optimal mechanism then takes the form of a \textit{\textbf{screening}} mechanism with nested bundles. In this case, the buyer's learning strategy can lead to substantial variations in payoffs, and hence can be quite informative about what to purchase under the seller's menu. For instance, in the right panel of \Cref{fig:ex_negrho}, there exists a pure-bundling equilibrium in which the seller offers $\{1, 2\}$ at price $3.23$ and the buyer chooses $\boldsymbol{\alpha} = (1,1)$. Even though the buyer's equilibrium signal $\boldsymbol{\alpha} = (1,1)$ leads to a much less dispersed type distribution than $\boldsymbol{\alpha} = (1,0)$, it resolves exactly the relevant uncertainty given the menu.  

\vspace{-3mm}
\paragraph{Mutual Best Responses versus Commitment.}\hspace{-2mm}Given the broad intuition that horizontal learning leads to little information rent, it might be tempting to conjecture that our result continues to hold even if the buyer moves first and commits to his learning strategy. However, as we discuss in \Cref{subsec:timing}, this conjecture is \textit{false}. Importantly, as already illustrated in \Cref{fig:ex_negrho}, the buyer can earn a strictly positive rent under horizontal learning---to maximize revenue, the seller offers not only the zero-dispersion bundle $x^*$ but also the grand bundle, on which the buyer does earn an information rent. If moving first, the buyer can commit to a horizontal learning strategy that maximizes this rent \textit{ex ante}, possibly earning more than under any vertical strategy (see \Cref{fig:commitment_ex2}). However, such a commitment solution is \textit{\textbf{unstable}}: given the menu, it does not maximize the buyer's payoff \textit{ex post}. Indeed, as we have emphasized, the seller's best response is built to neutralize exactly the variation that the buyer's strategy resolves, so the buyer wants to \textit{\textbf{re-optimize}} his learning against the resulting menu---possibly even to another horizontal learning strategy (see \Cref{fig:commitment_ex2}).

\section{Proof Sketch of \Cref{thm:main}}\label{sec:proof}

In this section, we sketch the proof of \Cref{thm:main} (see \Cref{subsec:proofmain} for the details).  \Cref{thm:main} consists of two parts: \textit{(i)} every equilibrium has vertical learning, and \textit{(ii)} every equilibrium is outcome-equivalent to a nested bundling equilibrium. Once we show part \textit{(i)}, part \textit{(ii)} follows relatively straightforwardly since the equilibrium type space must be comonotonic, and we can leverage known results from the literature. Thus, the main difficulty is to prove vertical learning. The proof is by contradiction and proceeds as follows: 
\begin{itemize}
    \item[\textbf{Step 1}.]  We characterize optimal mechanisms against any horizontal learning strategy, leveraging the linearity of the induced type distribution. 
    \item[\textbf{Step 2}.]  We characterize properties of the buyer's optimal learning strategies against any candidate optimal mechanism resulting from \textbf{Step 1}.
     \item[\textbf{Step 3}.] We show \textbf{Step 1} and \textbf{Step 2} together must lead to a contradiction if the buyer uses a horizontal learning strategy.
\end{itemize}

\Cref{subsec:mechanism} sketches \textbf{Step 1};  \Cref{subsec:learning} sketches \textbf{Step 2};   \Cref{subsec:complete} sketches \textbf{Step 3} which completes the proof for vertical learning and then shows how the nested bundling claim follows from there.

\subsection{Optimal Mechanisms}\label{subsec:mechanism}

As we have discussed, a key property of our elliptical setup is that any learning weights $\boldsymbol{\alpha}$ must lead to a distribution of posterior means $\boldsymbol{\theta} \in \mathbb{R}^K_+$ supported on a line segment in $\mathbb{R}^K_+$. This line segment must pass through the prior $\boldsymbol{\mu}$ but can point in any direction. When the buyer uses a horizontal learning strategy, this line segment must be \emph{strictly decreasing} when projected on some subspace $\theta_i\times \theta_j$. 

In equilibrium, the seller must use a mechanism that is optimal against this distribution. Unlike in the standard mechanism design problem, types are not vertically ordered. The key issue is then that the types with a binding IR constraint (the \textit{\textbf{worst-off types}}) are endogenous to the mechanism. Moreover, unlike standard mechanism design, we need to characterize the properties that hold for \textit{\textbf{all}} optimal mechanisms since the seller's indifference may be instrumental for sustaining an equilibrium. 

Assuming the buyer adopts a horizontal learning strategy, we first describe the structure of optimal mechanisms---which will be crucial to derive a buyer deviation---and then sketch how we characterize the optimal mechanisms. A complete characterization can be found in the appendix.

\vspace{-3mm}
\paragraph{Intuitive Description of Optimal Mechanisms.}\hspace{-2mm}Every optimal mechanism can be described as follows. First, we classify each good $i \in \{1, \dots, K\}$ according to whether its value $\theta_i$ is correlated positively (a \textit{\textbf{positive}} good) or non-positively (a \textit{\textbf{negative}} good) with the \textit{\textbf{grand bundle value}} $\sum_i \theta_i$. Intuitively, the negative goods are the ones for which the seller will be able to fully extract surplus without conceding any information rent---in particular, these negative goods are allocated to every type with full probability. 

Second, to achieve this surplus extraction, some of the positive goods (\textit{\textbf{positive balancing}} goods) will be (stochastically) bundled with the negative goods to create a \textit{\textbf{zero-dispersion bundle}} $x^*$. This bundle yields the \textit{same} value to all types, so that the surplus can be extracted while preserving incentive constraints. Importantly, the seller assigns the zero-dispersion bundle to the worst-off types, and assigns the positive balancing goods with full probability to all other types. 

Third, the rest of the positive goods (\textit{\textbf{positive non-balancing}} goods) are allocated in a standard fashion to types whose values exceed a good-specific threshold, as in a typical monopoly screening problem.  

These three sets of goods are mutually exclusive and collectively exhaustive. Naturally, this classification depends on the learning strategy chosen by the buyer. When learning is horizontal, there must be both negative and positive balancing goods---which is the key to obtaining a contradiction. In contrast, when learning is vertical, there are essentially only positive non-balancing goods---these are allocated in a standard fashion, leading to nested bundling.  

\vspace{-3mm}
\paragraph{Characterization of Optimal Mechanisms.}\hspace{-2mm}We now explain how we formally characterize the optimal mechanisms, building on the above intuitive description. Fix any horizontal learning strategy $\boldsymbol\alpha$. Since the associated distribution over posterior mean $\boldsymbol{\theta}\in \R^K_+$ is supported on a line segment, we can parameterize the types on the line by $t \in [0, 1]$, and write $\theta_i(t):= a_i t + b_i$. Let $F$ be the cumulative distribution over $t$ and $f$ its density.  Given our setup, we know that the distribution of $\boldsymbol\alpha\cdot\mathbf{v}$ is unimodal elliptical, and thus so is $F$. Note that $b_i\geq \max\{0,-a_i\}$ for each $i$ since values lie in the positive quadrant. 

Now, importantly, we make the following \textit{\textbf{sign convention}}:
\[\sum_i a_i \geq 0\,. \tag{Sign Convention}\]
Note that this is without loss of generality because if it fails, then we can simply redefine types $\tilde{t} = 1 - t \in [0, 1]$, and write 
\[a_i t + b_i = \underbrace{-a_i}_{\tilde{a}_i} \tilde{t} + \underbrace{a_i + b_i}_{\tilde{b}_i}\,,\]
which flips the sign for each good $i$. Importantly, this sign convention normalizes the direction of types so that a higher type $t$ has a higher grand bundle value $\sum_i \theta_i(t)$. Under this sign convention, we define 
\[I^+ := \Big\{i: a_i > 0\Big\}\quad \text{ and }\quad I^- :=\Big\{i: a_i \leq 0\Big\}\,, \]
and call the goods in $I^+$ the \textit{\textbf{(strictly) positive goods}}, and the goods in $I^-$ the \textit{\textbf{negative goods}}. Because of our sign normalization, whenever $\sum_i a_i > 0$, the positive goods are exactly the goods whose values are positively correlated with the grand bundle value, and the negative goods are exactly the goods whose values are (weakly) negatively correlated with the grand bundle value. 

Now, to characterize the positive balancing versus non-balancing goods, consider the following \textit{\textbf{auxiliary program}}: 
\begin{align*}
    \max_{\mathbf{x}\in [0,1]^K}&\sum_i b_ix_i \tag{Auxiliary Program} \\ 
    \text{ subject to } &\sum_i a_ix_i=0\,.
\end{align*}
Let $X^*$ be the set of all solutions to this problem. Intuitively, the auxiliary program finds the best \textit{\textbf{zero-dispersion bundle}} $x^*$ for surplus extraction: indeed, the constraint $\sum_i a_i x_i = 0$ guarantees that all types have the same value $\sum_i b_i x_i$ for the stochastic bundle. As we show, under an optimal mechanism, any type with $0$ payoff receives some $x^* \in X^*$. 

The auxiliary program can be viewed as a fractional knapsack problem and hence admits a greedy solution. In fact, every solution to the auxiliary program can be characterized as follows. As shown in \Cref{lem:auxiliary} in the appendix, there exists $\kappa \geq 0$ such that for any $x^*\in X^*$, we have:  
\begin{itemize}
    \item  For any negative good $i$, $x^*_i = 1$\,;
    \item For any strictly positive good $i$, $x^*_i = 1$ if $b_i/a_i > \kappa$ and $x^*_i = 0$ if $b_i/a_i < \kappa$\,.
\end{itemize}
For any strictly positive good $i$ where $b_i/a_i = \kappa$, an optimal solution has the freedom to ration it as long as the feasibility constraint is satisfied. Let $\lambda = - \kappa$ (and choose the largest $\kappa$ if it is not unique). It is not hard to see that $\lambda$ is exactly an optimal \textit{\textbf{dual multiplier}} on the equality constraint in \textbf{(Auxiliary Program)}. Now, define $I^* \subseteq I^+$ as follows: 
\[I^*:= \bigcup_{x^* \in X^*}\Big\{i \in I^+: x^*_i > 0\Big\}\,.\]
We call $I^*$ the \textit{\textbf{positive balancing goods}}, and $I^+ \backslash I^*$ the \textit{\textbf{positive non-balancing goods}}. Intuitively, positive balancing goods are bundled with negative goods so that $\sum_i a_i x^*_i = 0$ to construct the zero-dispersion bundle $x^*$. Indeed, under horizontal learning, there must exist a strictly negative and a positive balancing good. 

Now, for every optimal mechanism, we make the following claims: 
\begin{itemize}
    \item[\textbf{(Claim 1)}] \textit{The negative goods are allocated to every type with full probability.} 
    \item[\textbf{(Claim 2)}] \textit{The positive balancing goods are allocated to every type with full probability except the worst-off types for which the goods may be rationed.} 
    \item[\textbf{(Claim 3)}] \textit{The positive non-balancing goods are allocated in a standard fashion to types with their value above a threshold. The highest type gets all the goods.} 
\end{itemize}

We sketch the proofs of \textbf{(Claim 1)} to \textbf{(Claim 3)}. Our proof uncovers and leverages a hidden connection between \citet*{frick2024multidimensional} and \citet{loertscher2024optimal}. In particular, the auxiliary program above is a special case of the program in \citet*{frick2024multidimensional}, which finds the best lottery to give to types with binding IR in mechanism design problems with one-dimensional linear types.\footnote{Their program applies more generally as they allow for non-additive values.} Proposition 3 in \citet*{frick2024multidimensional} uses an extreme-point argument to show that it is without loss of optimality to consider ``almost deterministic'' mechanisms, which use deterministic allocations plus at most one lottery from \textbf{(Auxiliary Program)}. The same is true in our setting, but it does not identify the endogenous worst-off types or the allocations assigned to other types. Our key methodological contribution is a procedure (\Cref{lem:saddlepoint}) that uses the \textit{dual multiplier} from the \textbf{(Auxiliary Program)} to explicitly construct a \textit{saddle point} between mechanisms and worst-off types. Combined with the saddle-point formulation of \citet{loertscher2024optimal}, this explicit construction then immediately yields our characterization of all optimal mechanisms.

Indeed, any optimal direct mechanism solves the following problem:
    \begin{align*}
        &\max_{x, p}\quad \mathbb{E}\big[p(t)\big]\quad\text{s.t.}\\
        &U(t):=t\sum_ix_i(t)a_i+\sum_ib_ix_i(t) - p(t)\geq 0\quad \forall t\qquad &&[\text{IR}]\\
        &U(t)\geq t\sum_ix_i(t')a_i+\sum_ib_ix_i(t') - p(t')\quad \forall t, t'\qquad &&[\text{IC}]\,.
    \end{align*}
For the mechanism to be revenue-maximizing, the [IR] constraint must be binding for (at least) one type. Moreover, for any type $t_0$ such that $U(t_0)=0$, the standard characterization of [IC] applies in this setting: [IC] holds if and only if     \begin{align*}
        \sum_ia_ix_i(t) \text{ is nondecreasing in $t$ and } U(t) = \int_{t_0}^t \sum_ia_ix_i(z)dz  \text{ for all $t$}.
     \end{align*}
Given any IC mechanism, we can then use standard arguments to write the associated expected revenue as a function of the allocation rule $x(\,\cdot\,)$ and $t_0$ only: 
\[\mathrm{Revenue}\,(x; t_0) = \mathbb{E}\Bigg[\sum_i \Big(a_i x_i(t) \Phi(t; t_0) + b_i x_i(t)\Big)\Bigg],\]
where 
\[\Phi(t; t_0) :=\left(t+\frac{F(t)}{f(t)}\right)\mathbbm{1}\{t <  t_0\} + \left(t-\frac{1-F(t)}{f(t)}\right)\mathbbm{1}\{t \geq t_0\}\,\]
 corresponds to the virtual value for types $t\geq t_0$ and to the virtual cost for $t<t_0$. 
     
Again, a key complication is that the worst-off types, and in particular $t_0$, are endogenous to the mechanism. As in \citet{loertscher2019optimal} (Section 3.1) and \citet{loertscher2024optimal} (Lemma 2), given any IC mechanism, $t_0$ is a worst-off type if and only if 
 \[ t_0 \in \argmin_{\hat{t}}\; \mathrm{Revenue}(x; \hat{t})\,.\]
Indeed, applying the envelope formula with $\hat{t}$ as the reference type gives \[\mathrm{Revenue}(x;\hat{t}) = \mathbb{E}[p(t)] + U(\hat{t})\,;\]
hence, $\mathrm{Revenue}(x;\hat{t})$ and $U(\hat{t})$ share the same minimizers, i.e., the worst-off types. Thus, every optimal mechanism must have an allocation rule $x(\,\cdot\,)$ that solves
    \[\max_{x \in \text{MON}} \min_{t_0 \in[0,1]} \,\mathrm{Revenue}(x; t_0)\,,\]
     where 
     \[\text{MON}:= \Bigg\{x:[0,1] \rightarrow [0, 1]^K \text{ such that }  \sum_ia_ix_i(t) \text{ is nondecreasing in $t$ }\Bigg \}\,.\]
To solve the above problem, it suffices to construct a \textit{\textbf{saddle point}} $(x^*, t_0^*)$ of $\mathrm{Revenue}(\,\cdot\,; \,\cdot\,)$, i.e., $x^*$ maximizes $\mathrm{Revenue}(\,\cdot\,; t^*_0)$ and $t^*_0$ minimizes $\mathrm{Revenue}(x^*; \,\cdot\,)$. Indeed, we have the following \textit{\textbf{rectangularity property}} of the saddle points: if $(x^*, t_0^*)$ is a saddle point of $\mathrm{Revenue}(\,\cdot\,; \,\cdot\,)$, then an allocation rule $x$ is optimal if and only if $(x, t_0^*)$ is also a saddle point. Thus, such a construction  immediately yields a characterization of \textit{all} optimal mechanisms.

Our key methodological contribution is an explicit construction of a saddle point. Specifically, we show that a dual multiplier from the \textbf{(Auxiliary Program)} can be used to identify the worst-off types and construct the corresponding seller-optimal allocation rule.\footnote{\citet{loertscher2024optimal} use the minimax theorem to establish existence and then derive the saddle point by searching over candidate worst-off types in a two-good unit-demand setting. We instead construct the candidate directly for any number of goods with additive values.}  To state the construction, for any fixed $t_0$, let $\overline{\Phi}(t;t_0)$ denote the ironed version of $\Phi(t;t_0)$, as in \citet{Myerson1981}.

\begin{lemma}[Saddle-Point Construction]\label{lem:saddlepoint}
     Let $\lambda \leq 0$ be the optimal dual multiplier of \textbf{(Auxiliary Program)} selected above. Then, there exists a saddle point $(x^*,t_0^*)$ such that 
     \[\overline{\Phi}(t_0^*;t_0^*) = \lambda \tag{\textbf{$\lambda$-construction}}\,;\]
     and the ironing interval $\mathcal{I}:= \{t:\overline{\Phi}(t;t_0^*)=\lambda\}$ contains all types below $t_0^*$: $[0, t_0^*] \subset \mathcal{I}$. 
\end{lemma}

As a consequence of \Cref{lem:saddlepoint}, every optimal allocation rule $x$ must form a saddle point with the constructed $t^*_0$. In particular, every optimal  $x$ solves 
\[\max_{x \in \text{MON}}\,\mathbb{E}\Big[\sum_i \Big(a_i x_i(t) \Phi(t; t_0^*) + b_i x_i(t)\Big)\Big].\] 
As in \citet{Myerson1981}, every solution to this problem must also solve the pointwise maximization problem after ironing (see \Cref{lem:ironing} for a formal argument): 
\begin{equation}
    \max_{x:[0, 1] \rightarrow[0,1]^K} \mathbb{E}\Big[\Big(\sum_i a_i x_i(t) \Big) \overline{\Phi}(t; t^*_0) + \sum_i b_i x_i(t) \Big]\,.\label{eq:pointwise-text}
\end{equation}
First, consider all types $t \in \mathcal{I}$. By construction, on this interval the pointwise maximization problem becomes
\begin{equation}
\max_{x\in[0,1]^K} \Big\{ \sum_i b_i x_i+\lambda\sum_i a_i x_i \Big\}\,, \label{eq:pointwise-lambda}
\end{equation}
which is the Lagrangian of the \textbf{(Auxiliary Program)}. Moreover, consistency with ironing requires $\sum_i a_i x_i(t)$ to be constant on $\mathcal{I}$. Since $t_0^*$ is a worst-off type in the interior of $\mathcal{I}$, this constant must be zero. Thus, $x(t)$ both maximizes the Lagrangian and satisfies the constraint of the auxiliary program. Since $\lambda$ is an optimal multiplier, it follows that $x(t) \in X^*$. Thus, all types in $\mathcal I$ receive zero payoff, and purchase all negative goods and some positive balancing goods.

For \textbf{(Claim 1)}, consider any type $t  \not \in \mathcal{I}$ and any negative good $i \in I^-$. Then, 
\[a_i \overline{\Phi}(t; t^*_0) + b_i\geq a_i \overline{\Phi}(1; t^*_0) + b_i =  a_i + b_i \geq 0\,,\]
where the last inequality is due to  $\theta_i(t) \geq 0$ for all $t$, and in particular $t = 1$. Moreover, note that, for $t < 1$ and $a_i < 0$, the first inequality is strict since $t = 1$ cannot be in an ironing interval here,\footnote{Indeed, the revenue curve $R(q):=qF^{-1}(1-q)+\big(t_0^*-F^{-1}(1-q)\big)_+$ has slope $1$ at $q=0$ since $\Phi(1;t_0^*)=1$ (\Cref{lem:regular}), and any affine part of its concave envelope has slope strictly below $1$.} while for $a_i = 0$ the expression equals $b_i = \mu_i > 0$. Thus, any solution to the pointwise optimization problem also allocates all negative goods to (almost) all types $t  \not \in \mathcal{I}$. 

Now, for \textbf{(Claim 2)}, consider any positive balancing good $i \in I^*$. Note that for every type $t \not \in \mathcal{I}$, by construction of $\kappa$, we have 
\[a_i \overline{\Phi}(t; t^*_0) + b_i > a_i \overline{\Phi}(t^*_0; t^*_0) + b_i = a_i\lambda + b_i =  b_i - \kappa a_i \geq 0\,.\]
Thus, types $t \not \in \mathcal{I}$ also receive all positive balancing goods.

Finally, for \textbf{(Claim 3)}, take any positive non-balancing good $i \in I^+ \backslash I^*$. For all $t\not \in \mathcal{I}$, we have $\Phi(t;t_0^*) = t - \frac{1 - F(t)}{f(t)} =: \Psi(t)$. Then, as shown by \Cref{lem:regular} in the appendix, the function $t \mapsto a_i \Psi(t) + b_i$ is strictly single-crossing (from below) and strictly positive for all $t>0.5$; moreover, the same holds for $t \mapsto a_i \overline{\Phi}(t;t_0^*) + b_i$. Thus, there exists some interior threshold $t^*_i \not \in \mathcal{I}$ such that $x_i(t) = \mathbbm{1}\{t \geq t^*_i\}$ is the unique pointwise-optimal allocation for good $i$. 

\subsection{Optimal Learning}\label{subsec:learning}

Now, continue assuming that the buyer is using a horizontal learning strategy. From \textbf{Step 1}, we know  that \textbf{(Claim 1)} to \textbf{(Claim 3)} must hold for the seller's mechanism in this conjectured equilibrium. We now derive properties of the buyer's best response against such a mechanism, which will eventually lead to a contradiction. Conceptually, the buyer solves a constrained learning problem. Given the mechanism $\mathcal{M}$, there exists a menu of choices: 
\[O := \bigcup_{m \in M} \Big\{\big(x(m), p(m)\big)\Big\}\]
which is the set of potential outcomes the mechanism can induce. This menu induces an \textit{\textbf{indirect utility function}} for the buyer that depends on his type $\boldsymbol{\theta}$:
\[U(\boldsymbol{\theta}):=\max_{m\in M}\; \big\{ \boldsymbol{\theta}\cdot x(m) - p(m) \big\}\,. \]
Note that the function $U$ is convex. When choosing a learning strategy, the buyer then solves the following problem:
\[\max_{ \boldsymbol \alpha}\mathbb{E}\!\Bigl[\,U\Bigl(\theta_{1}(s;\boldsymbol\alpha),\dots,\theta_{K}(s;\boldsymbol\alpha)\Bigr)\Bigr]\,.\]

There are two key properties that we show must hold for every optimal $\boldsymbol{\alpha}^*$:\footnote{These properties must hold for any convex $U$ that is \textit{not} affine, so we must rule out the case where information is not strictly valuable in equilibrium. We prove this separately in \Cref{lem:infovaluable}.} 
\begin{itemize}
    \item[\textbf{(Claim 4)}] \textit{If $U(\boldsymbol{\theta})$ does not depend on $\theta_i$, then $\alpha^*_i = 0$.} 
    \item[\textbf{(Claim 5)}] \textit{If $U(\boldsymbol{\theta})$ depends on $\theta_i$ and $\theta_j$ symmetrically via $\theta_i + \theta_j$, then $\alpha^*_i = \alpha^*_j$.} 
\end{itemize}
\textbf{(Claim 4)} says that the buyer should not learn about a good whose value does not affect his choice from the menu, e.g., when every option in the menu contains the same quantity of good $i$, since then $U$ can be normalized so that it does not depend on $\theta_i$. \textbf{(Claim 5)} says that the buyer should learn symmetrically about two goods whose values affect his choice only via their sum.

These two properties are relatively easy to see in the uncorrelated case where $\rho = 0$, since then the learning weight put on good $i$ has no impact on how much is learned about good $j \neq i$. However, with correlated values, the learning weights may be chosen to balance learning across different goods. We show that even though the optimal strategy $\boldsymbol{\alpha}^*$ will in fact take into account the correlation structure, \textbf{(Claim 4)} and \textbf{(Claim 5)} must hold regardless of the correlation structure (see \Cref{lem:learnzero} and \Cref{lem:learnsame} in the appendix). The proofs exploit the \textit{\textbf{orthogonal decomposition}} property of elliptical distributions: for any $\boldsymbol{\alpha}$, and any $N < K$, we can write 
\[
\boldsymbol{\alpha}\cdot\mathbf{v}
=\sum_{i\le N}\alpha_{i}v_{i}+\sum_{j>N}\alpha_{j}v_{j}
=\sum_{i\le N}\tilde{\alpha}_{i}v_{i}+\varepsilon \,,
\]
for some $\boldsymbol{\tilde{\alpha}}$, where $\text{Cov}(v_{i},\varepsilon)=0$ for all $i\le N$, and $\varepsilon$ is a non-degenerate elliptical random variable if $(\alpha_j)_{j>N}\neq\mathbf{0}$. Now, to see \textbf{(Claim 4)}, suppose for contradiction that $\boldsymbol{\alpha}$ is optimal and yet $\alpha_i \neq 0$. Construct an alternative signal $\boldsymbol\alpha^{*}$ as follows. Apply the orthogonal decomposition to write $\boldsymbol{\alpha} \cdot \mathbf{v}$ as 
\[\boldsymbol{\alpha} \cdot \mathbf{v} = \sum_{j \neq i} \tilde{\alpha}_j v_j + \varepsilon\]
and set $\alpha^*_j := \tilde{\alpha}_j$ for all $j \neq i$ and $\alpha^*_i := 0$. Note that in this new coordinate, the original signal is $(\boldsymbol{\tilde{\alpha}}, 1)$ while the new signal is $(\boldsymbol{\tilde{\alpha}}, 0)$ and replaces the weight $1$ on the uncorrelated term $\varepsilon$ with $0$. One can then verify that such a change must lead to a new posterior mean distribution whose support, once projected on the relevant $\boldsymbol\Theta_{-i}$-subspace, is exactly the same as the support of the original distribution generated by $\boldsymbol{\alpha}$. Moreover, it generates a strictly higher posterior variance for each good $j \neq i$---the distribution of $\boldsymbol{\theta}^*_{-i}$ is strictly higher than the distribution of $\boldsymbol{\theta}_{-i}$ in the convex order. Since $U$ is convex and does not depend on good $i$, this must be a strict improvement---a contradiction.  \textbf{(Claim 5)} follows by a similar dominance argument after applying the orthogonal decomposition.

\subsection{Completion of the Proof of \Cref{thm:main}}\label{subsec:complete}

\paragraph{Vertical Learning.}\hspace{-2mm}Suppose for contradiction that there exists an equilibrium $(\boldsymbol{\alpha},\mathcal{M})$ with horizontal learning. Let $O:=\cup_{m \in M} \{(x(m), p(m))\}$ be the set of all possible outcomes that can be achieved under mechanism $\mathcal{M}$. Now let 
\[O^* := \bigcup_{t \in [0, 1]} \Big\{\big(x(t), p(t)\big)\Big\}\]
be the set of outcomes that are \emph{actually chosen} by some type $t$ under the seller's mechanism in equilibrium. Note that the buyer's equilibrium payoff is exactly the same, by construction, when facing $O^*$ or when facing $O \supseteq O^*$. This implies that there cannot be a profitable deviation by the buyer against menu $O^*$ since that would imply a profitable deviation against the original menu $O$. Moreover, there exist both strictly negative goods and positive balancing goods.
 
First, by \textbf{(Claim 1)} and \textbf{(Claim 2)} together, every option in $O^*$ that yields a strictly positive payoff for some equilibrium type $t$ must include all goods in $I^* \cup I^-$ with full probability. However, by \textbf{(Claim 5)}, this implies that the buyer's learning strategy $\boldsymbol{\alpha}$ must put \textit{equal weights} on all $i \in I^* \cup I^-$.

Second, by \textbf{(Claim 1)}, every option in $O^*$ must include all goods in $I^-$ with full probability. However, this implies that the buyer's decision problem, when facing menu $O^*$, does not depend on the values of the negative goods $\big(v_i\big)_{i \in I^-}$. Therefore, by \textbf{(Claim 4)}, the buyer must put \textit{zero weights} on all the goods in $I^-$. 

Together, these two observations imply that the buyer must put zero weight on \textit{every good} $i \in I^* \cup I^-$. But since the correlation $\rho$ is the same across all pairs of goods, this implies that $(a_{i})_{i\in I^* \cup I^-}$ must be either all weakly positive or all weakly negative. Indeed, for any $i\in I^* \cup I^-$, 
\[\text{sign}(a_i) =\text{sign}(\text{Cov}(\boldsymbol{\alpha}\cdot\mathbf{v},v_i)) =\text{sign}\bigg(\rho\sum_{j\in I^+\setminus I^* }\alpha_j\sigma_j\bigg),  \]
which does not depend on $i$.  However, by construction, $a_i > 0$ for all $i \in I^*$ and $a_i < 0$ for some $i \in I^{-}$---a contradiction.  
\vspace{-3mm}
\paragraph{Nested Bundling.}\hspace{-2mm}As we have shown, every equilibrium must have vertical learning, and hence a comonotonic type distribution.  Thus, the posterior mean distribution can be written as: for each $i$, $\theta_i = a_i t + b_i$ where $a_i \geq 0$, $b_i \geq 0$, and $t \in [0, 1]$. We claim that, against such a posterior mean distribution, there exists a unique optimal direct-revelation mechanism (up to measure zero) that is deterministic and can be represented by nested menus. This can be seen from the mechanism characterization in \Cref{subsec:mechanism} (which also holds under vertical learning). Now, there exists no good with $a_i < 0$, and hence every optimal solution $x$ to \textbf{(Auxiliary Program)} must have $x_i = 0$ for all the strictly positive goods $i$. Thus, there are no positive balancing goods, i.e., $I^* = \varnothing$. By \textbf{(Claim 1)}, all the goods with $a_i = 0$ must be allocated to all types with full probability. By \textbf{(Claim 3)}, all the goods with $a_i > 0$ must then be allocated in a monotone, deterministic fashion according to the threshold rules $\mathbbm{1}\{t \geq t^*_i\}$. 

Now, consider the outcomes chosen in equilibrium $O^* = \bigcup_{t \in [0, 1]} \big\{\big(x(t), p(t)\big)\big\}$. By the above argument, it must be that the allocations offered in $O^*$ are deterministic and totally ordered by set inclusion. Moreover, as argued before, there cannot be a profitable deviation by the buyer against menu $O^*$ since that would imply a profitable deviation against the original menu $O \supseteq O^*$. Then, the profile in which the seller offers the nested menu $O^*$ and the buyer uses the same strategy $\boldsymbol{\alpha}$ constitutes an outcome-equivalent equilibrium, completing the proof.

\section{Discussions}\label{sec:discuss}

\subsection{Commitment and Alternative Timing}\label{subsec:timing}

\paragraph{Seller Moves First.}\hspace{-2mm}Our main model considers a simultaneous-move game between the seller and the buyer. This best captures markets where the seller frequently readjusts prices and may not be able to commit not to do so. We now discuss what happens if instead the seller is a Stackelberg leader. That is, the seller first commits to a menu, which is observed by the buyer, who then learns and makes a purchasing decision. By choosing an appropriate menu, the seller can shape the buyer's learning incentives, and the optimal menu needs to account for this effect. A full analysis of such a  model is beyond the scope of this paper, but we go through an example to illustrate the role of commitment. 

\begin{figure}[!t]
\begin{center}
\begin{tikzpicture}[scale=0.8, every node/.style={transform shape}]

  \draw[->, thick] (-0.1,0) -- (4.5,0) node[below] {$v_1$};
  \draw[->, thick] (0,-0.1) -- (0,4.5) node[left] {$v_2$};
  \filldraw[black] (2,2) circle (1pt);
     \fill[black!20, opacity=0.4, rotate around={-45:(2,2)}] (2,2) ellipse [x radius=2.35, y radius=1.6];
  \draw[thick] (2,0.1) -- (2,-0.1) ;
  \draw[thick] (0.1, 2) -- (-0.1, 2) ;
  \draw[thick, purple] (0.87,0.87) -- (3.13, 3.13) node[above] {\small$\boldsymbol{\alpha}=(1,1)$};
   \draw[thick, blue, dashed] (0,3) -- (3,0);
  \draw[thick] (3,0.1) -- (3,-0.1) node[below] {${\color{blue} 3.01}$};
  \draw[thick] (0.1, 3) -- (-0.1, 3) node[left] {${\color{blue} 3.01}$};
\node at (3, 1.3) {${\color{blue} \{1,2\}}$};
\node at (1.55, 0.8) {${\color{blue} \varnothing}$};
\node at (2.25, -1.2) {Seller revenue $=$ 2.35};
\node at (2.25, -1.8) {Buyer surplus $=$ 1.10};

   \draw[->, thick] (5.4,0) -- (10,0) node[below] {$v_1$};
  \draw[->, thick] (5.5,-0.1) -- (5.5,4.5) node[left] {$v_2$};
  \filldraw[black] (7.5,2) circle (1pt);
     \fill[black!20, opacity=0.4, rotate around={-45: (7.5,2)}] (7.5,2) ellipse [x radius=2.35, y radius=1.6];
  \draw[thick] (7.5,0.1) -- (7.5,-0.1);
  \draw[thick] (5.6, 2) -- (5.4, 2) ;
  \draw[thick, purple] (5.5,2.82) -- (9.5,1.18) node[below] {\small$\boldsymbol{\alpha}=(1,0)$};
   \draw[thick, blue, dashed] (5.5,1.9) -- (5.5+3.66-1.9,1.9) --  (5.5+3.66-1.9,4) ;
   \draw[thick, blue, dashed]  (5.5+3.66-1.9,1.9) --(5.5+3.66, 0);
  \draw[thick] (5.6, 1.9) -- (5.4, 1.9) node[left] {${\color{blue} 1.9}$};
  \draw[thick] (5.5+3.66,0.1) -- (5.5+3.66,-0.1) node[below] {${\color{blue} 3.66}$};
\node at (5.5+1.55, 1) {${\color{blue} \varnothing}$};
\node at (5.5+2.6, 2.5) {${\color{blue} \{1,2\}}$};
\node at (5.5+1, 3) {${\color{blue} \{2\}}$};
\node at (2.25+5.5, -1.2) {Seller revenue $=$ 2.92};
\node at (2.25+5.5, -1.8) {Buyer surplus $=$  0.65};

   \draw[->, thick] (10.9,0) -- (15.5,0) node[below] {$v_1$};
  \draw[->, thick] (11,-0.1) -- (11,4.5) node[left] {$v_2$};
  \filldraw[black] (13,2) circle (1pt);
     \fill[black!20, opacity=0.4, rotate around={-45: (13,2)}] (13,2) ellipse [x radius=2.35, y radius=1.6];
  \draw[thick, purple] (11,2.82) -- (15,1.18) node[below] {\small$\boldsymbol{\alpha}=(1,0)$};
   \draw[thick, blue, dashed] (11,2.3) -- (11+3.5-2.3,2.3) --  (11+3.5-2.3,4) ;
   \draw[thick, blue, dashed]  (11+3.5-2.3,2.3) --(11+3.5, 0);
  \draw[thick] (11.1, 2.3) -- (10.9, 2.3) node[left] {${\color{blue} 2.32}$};
  \draw[thick] (11+3.5,0.1) -- (11+3.5,-0.1) node[below] {${\color{blue} 3.52}$};
       \draw[thick] (11+2,0.1) -- (11+2,-0.1) ;
  \draw[thick] (11+0.1, 2) -- (11-0.1, 2) ;
  
\node at (11+1.55, 1) {${\color{blue} \varnothing}$};
\node at (11+2.6, 2.5) {${\color{blue} \{1,2\}}$};
\node at (11+0.7, 3.1) {${\color{blue} \{2\}}$};
\node at (2.25+11, -1.2) {Seller revenue $=$ 3.22};
\node at (2.25+11, -1.8) {Buyer surplus $=$  0.60};

\end{tikzpicture}
\end{center}
\vspace{-10pt}
\caption{The shaded gray area is the set of possible values $V$. The left figure illustrates an equilibrium of the simultaneous-move game. The buyer chooses signal $\boldsymbol{\alpha}=(1,1)$ and the solid red line is the support of the associated type distribution. The dashed blue line represents optimal allocations given the seller's menu: types below the line buy nothing while types above the line buy the bundle. The middle figure illustrates an optimal deterministic menu when the seller has commitment and the buyer's associated learning strategy. The right figure illustrates an optimal deterministic menu against learning strategy $\boldsymbol{\alpha}=(1,0)$.}\label{fig:commitment_ex}
\end{figure}
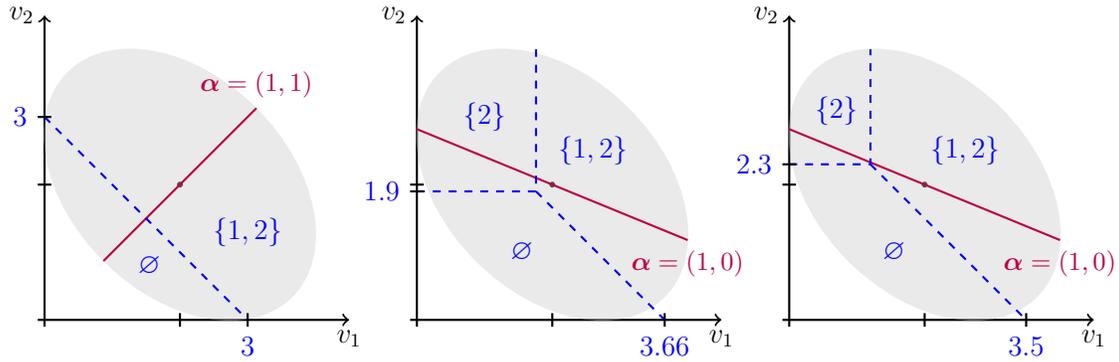

Suppose that there are $K=2$ goods. The buyer's values are drawn from a symmetric Gaussian distribution with mean $\mu_1=\mu_2=2$, standard deviation equal to $4$, and correlation $\rho = -0.4$, whose support is truncated to lie in the positive quadrant. The distribution of values is exchangeable, so the simultaneous-move game admits an equilibrium in which the seller only offers the grand bundle $\{1,2\}$ and the buyer only learns about his value for the grand bundle $\boldsymbol{\alpha}=(1,1)$. In this equilibrium, the grand bundle is priced at 3.01 and the expected revenue equals 2.35 (\Cref{fig:commitment_ex}, left panel).

Now, suppose that the seller moves first. We solve for an optimal deterministic menu numerically. An optimal menu is to offer good 2 at price 1.9 and the grand bundle at price  3.66. Against this menu, the buyer finds it optimal to only learn about his value for good 1. Because values are negatively correlated, this induces horizontal differentiation across types, as depicted in  \Cref{fig:commitment_ex} (middle panel). The expected revenue equals 2.92. 

Two features of the example are worth noting. First, commitment is valuable---the seller achieves a strictly greater expected revenue when she moves first. Second, horizontal learning can be sustained in equilibrium when the seller has commitment. This is not driven by the restriction to deterministic menus: Against the above horizontal learning strategy, the optimal deterministic mechanism sells good 2 at price 2.32 and the bundle at price 3.52 (\Cref{fig:commitment_ex}, right panel). Thus, absent commitment, the seller would have an incentive to increase the price of the base good and slightly decrease the price of the bundle. However, if she were to do that, then the buyer would want to \textit{deviate} to learning about the bundle, and this would not form an equilibrium. By committing to a lower price for the base good (good 2), the seller can incentivize the buyer to only learn about the upgrade good (good 1), leading to a higher expected revenue. 

Finally, commitment also affects the total surplus, which is higher when the seller moves first than when both move simultaneously. In our setting, the efficient outcome is to always allocate all goods to the buyer. Under seller commitment, the buyer always purchases good 2, whereas he purchases good 1 less often than under simultaneous move. The first effect dominates here, yielding higher welfare. However, this is not general; there are examples where total surplus goes down when the seller moves first.

\paragraph{Buyer Moves First.}\hspace{-2mm}We now discuss what happens when the buyer moves first. This can be viewed as a benchmark to understand which information structure benefits the buyer by shaping the seller's mechanism. This is also the timing considered by \citet{debroesler2024}. 

A simplistic intuition behind \Cref{thm:main} is that mechanisms that are optimal against horizontal learning leave little information rent to the buyer. However, this intuition is incomplete. Indeed, as we show next, in the model where the buyer moves first, there can exist horizontal learning strategies that secure a \textit{higher} information rent for the buyer than any vertical learning strategy.

\begin{figure}[!t]
\begin{center}
\begin{tikzpicture}[scale=1.1, every node/.style={transform shape}]

  \draw[->, thick] (-0.1,0) -- (4.5,0) node[below] {\footnotesize $v_1$};
  \draw[->, thick] (0,-0.1) -- (0,2.5) node[left] {\footnotesize $v_2$};
  \filldraw[black] (2,1) circle (1pt);
     \fill[black!20, opacity=0.4, rotate around={+71.6+90:(2,1)}] (2,1) ellipse [x radius=7.5*0.2814, y radius=7.5*0.0283];
   \draw[thick] (2,0.1) -- (2,-0.1) node[below] {\footnotesize $0.2$};
   \draw[thick] (0.1, 1) -- (-0.1, 1) node[left] {\footnotesize $0.1$};
   \draw[thick, purple] (2-1.15,{1+1.15*0.1710}) -- (2+1.15, {1-1.15*0.1710});
\node at (2+1.15+0.5, {1-1.15*0.1710+0.3}) {\footnotesize${\color{purple}\boldsymbol{\alpha}=(1,2.82)}$};
  \draw[thick, blue, dashed] (0,1.342) -- ({(2.566-1.342)/(1-0.1710)},{2.566-(2.566-1.342)/(1-0.1710)}) -- (2.559,0);
  \draw[thick, blue, dashed] ({(2.566-1.342)/(1-0.1710)},{2.566-(2.566-1.342)/(1-0.1710)}) -- ({(2.566-1.342)/(1-0.1710)},1.8);
\node at (3, 0.5) {\footnotesize ${\color{blue} \{1,2\}}$};
\node at (1, 0.5) {\footnotesize ${\color{blue} \varnothing}$};
\node at (1, 1.5) {\footnotesize ${\color{blue} x^*}$};
\node at (2, 2.5) {\footnotesize $\rho = -0.98$};
\node at (2.25, -1.3) {\footnotesize Seller revenue $=$ 0.243};
\node at (2.25, -0.9) {\footnotesize Buyer surplus $=$ 0.044};

  \draw[->, thick] (-0.1+6.5,0) -- (4.5+6.5,0) node[below] {\footnotesize $v_1$};
  \draw[->, thick] (0+6.5,-0.1) -- (0+6.5,2.5) node[left] {\footnotesize $v_2$};
  \filldraw[black] (2+6.5,1) circle (1pt);
     \fill[black!20, opacity=0.4, rotate around={+71.6+90:(2+6.5,1)}] (2+6.5,1) ellipse [x radius=7.5*0.2814, y radius=7.5*0.0283];
   \draw[thick] (2+6.5,0.1) -- (2+6.5,-0.1) node[below] {\footnotesize $0.2$};
   \draw[thick] (0.1+6.5, 1) -- (-0.1+6.5, 1) node[left] {\footnotesize $0.1$};
   \draw[thick, purple] ({6.5+2-2},{1+2*0.3267}) -- ({6.5+2+2},{1-2*0.3267});
\node[ purple] at  (2+1.95+6.5+0.4, 1-0.6242+0.3){{\footnotesize$\boldsymbol{\alpha}'=(1,0)$}};
    \draw[thick, blue, dashed] (0+6.5,1.342) -- ({(2.566-1.342)/(1-0.1710)+6.5},{2.566-(2.566-1.342)/(1-0.1710)}) -- (2.559+6.5,0);
  \draw[thick, blue, dashed] ({(2.566-1.342)/(1-0.1710)+6.5},{2.566-(2.566-1.342)/(1-0.1710)}) -- ({(2.566-1.342)/(1-0.1710)+6.5},1.8);
\node at (3+6.5, 0.3) {\footnotesize ${\color{blue} \{1,2\}}$};
\node at (1+6.5, 0.5) {\footnotesize ${\color{blue} \varnothing}$};
\node at (1+6.5, 1.7) {\footnotesize ${\color{blue} x^*}$};
\node at (2+6.5, 2.5) {\footnotesize $\rho = -0.98$};
\node at (6.5+2.25, -1.3) {{\footnotesize Seller revenue $=0.216$ }};
\node at (6.5+2.25, -0.9) {\footnotesize Buyer surplus $=0.061$ };
\end{tikzpicture}
\vspace{-10pt}
\end{center}
\caption{The left figure illustrates the optimal learning strategy  $\boldsymbol{\alpha}$ when the buyer has commitment, and the seller's associated mechanism. Notably, the optimal learning strategy is horizontal. The right figure shows that $\boldsymbol{\alpha}$ cannot be sustained in equilibrium---against the seller's mechanism, the buyer strictly prefers another horizontal learning strategy $\boldsymbol{\alpha}'$ to $\boldsymbol{\alpha}$.}\label{fig:commitment_ex2}
\end{figure}
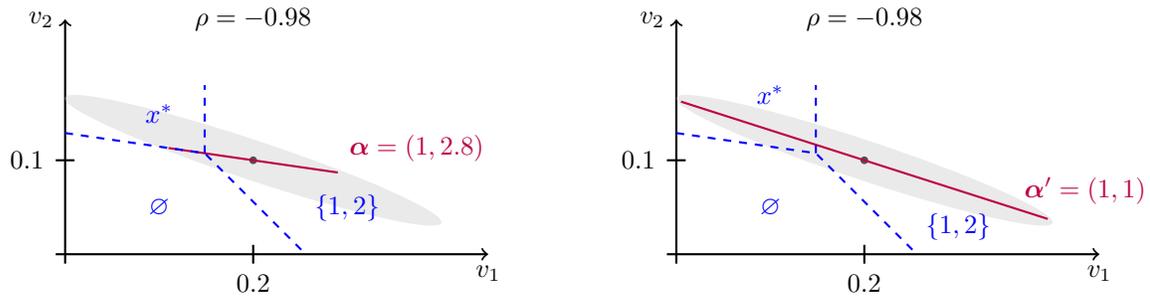

Consider again $K=2$ goods. Suppose that the buyer's values are drawn from a Gaussian distribution with mean $\boldsymbol{\mu}=(0.2, 0.1)$, variance-covariance matrix $[9,\;3\rho; \;3\rho,\; 1]$ with correlation $\rho = -0.98$, whose support is truncated to lie in the positive quadrant. For any learning strategy $\boldsymbol{\alpha}$, as part of our main analysis, we have characterized the seller's best response and the induced payoff to the buyer (see \Cref{lem:optx}). We find the learning strategy that maximizes the buyer's expected payoff numerically, which is $\boldsymbol{\alpha}=(1,2.82)$. As depicted in \Cref{fig:commitment_ex2}, this is a horizontal learning strategy. The seller's best response against $\boldsymbol{\alpha}=(1,2.82)$ is to offer a menu composed of the grand bundle $\{1,2\}$ at price $0.26$ and a rationing option $x^*  = (0.17, 1)$ at price $0.13$ (a lottery of getting good $1$ with probability $0.17$ and good $2$ for sure). As discussed in \Cref{subsec:example}, the rationing option is the zero-dispersion bundle that leaves no rent to any type who purchases it. However, there is sufficient rent for the types who purchase the grand bundle to make the strategy $\boldsymbol{\alpha}$ optimal for the buyer with commitment. 

We know from \Cref{thm:main} that $\boldsymbol{\alpha}$ cannot be part of an equilibrium. Indeed, given the menu $\big\{(x^*, 0.13), (\{1,2\}, 0.26)\big\}$, the buyer strictly benefits from deviating to only learning about good 1 (\Cref{fig:commitment_ex2}, right panel).\footnote{Note that because values are negatively correlated, learning  about good 1 here leads to negatively correlated posterior means---i.e., it is a horizontal learning strategy.} The key intuition behind the existence of such a deviation, as discussed in \Cref{subsec:example}, is that the seller's optimal mechanism, against horizontal learning, is designed to limit the \textit{variation} in payoffs along the posterior mean line direction. This implies that the buyer's original learning strategy cannot resolve too much uncertainty relevant for his purchase decision, and hence there exists another learning direction (which may or may not be vertical learning) that resolves more relevant uncertainty given the seller's menu---hence, horizontal learning is unstable and cannot be sustained in equilibrium.

This example also provides an interesting contrast with \citet{debroesler2024} who show that the \textit{buyer-optimal signal} induces vertical types, efficient trade, and pure bundling when the buyer moves first. The difference can be understood as follows. \citet{debroesler2024} allow the buyer to commit to arbitrary signals and assume that $\mathbf{v}$ has an exchangeable distribution. In particular, the buyer can commit to learning a noisy signal about the grand bundle such that \textit{(i)} the seller best responds by offering only the grand bundle and \textit{(ii)} any other signal leads to a weakly higher revenue for the seller. The noisy signal is constructed to induce a truncated Pareto posterior mean distribution such that the seller finds it optimal not to exclude any buyer types---hence, the signal must be buyer-optimal. In the above example, \textit{(i)} the buyer can only commit to a ``direction of learning'' which excludes the noisy Pareto signals and leads to inefficient trade, and \textit{(ii)} the seller would not best respond with pure bundling even if the buyer committed to learning only about the grand bundle because of the asymmetry in the distribution of $\mathbf{v}$.

\subsection{Beyond Elliptical Distributions and 1D Linear Signals}\label{subsec:multi}

We have assumed throughout that the buyer can only acquire a one-dimensional linear signal about his values $\mathbf{v}$ which follow an elliptical distribution. We now discuss the role played by these assumptions and the robustness of our result. 

\vspace{-3mm}
\paragraph{Interpretability and Tractability.}\hspace{-2mm}Our learning technology, combined with elliptical distributions, offers a combination of interpretability and tractability. As discussed in \Cref{sec:model}, one-dimensional linear \textit{signals} $\boldsymbol{\alpha}$ are precisely the least informative signals for which the buyer can make one perfect binary comparison, and they provide an intuitive understanding of what such learning entails. As emphasized in \Cref{sec:proof}, one-dimensional linear \textit{types} $\boldsymbol{\theta}(t)$ are what allow us to characterize the seller's optimal mechanisms against horizontal learning strategies where the worst-off types are endogenous to the mechanism. Elliptical distributions are not only sufficient but in fact necessary to deliver linear types from linear signals (\citealt*{gupta2013elliptically}). Together, it is this joint modeling choice that delivers the combination of interpretability and tractability, which allows us to extract economic insights that are otherwise hard to obtain. 

Indeed, as we demonstrate below, the assumptions on the prior can be relaxed while maintaining tractability as long as the buyer's signals lead to posterior mean distributions supported on a line segment. Therefore, the unimodality assumption and the additional properties implied by elliptical distributions, such as the symmetric density, are not crucial for tractability. Nevertheless, the elliptical-distribution assumption offers a simple way to interpret the signals that lead to linear types. 

By contrast, if the buyer could learn more flexibly, then we would generally lose tractability as some signal might induce a multidimensional type distribution, against which characterizing optimal mechanisms is known to be analytically intractable. Consequently, it would be difficult to characterize the induced demand systems across \textit{all} equilibria. Nevertheless, as we demonstrate below, we might still be able to leverage the insights learned from our baseline model to construct particular vertical-learning equilibria as well as to show that particular horizontal-learning strategies cannot survive in equilibrium.

\vspace{-3mm}
\paragraph{Relaxing Unimodal Elliptical Distributions.}\hspace{-2mm}For the following proposition only, let us drop the assumption that $\mathbf{v}$ is drawn from a unimodal elliptical distribution. We impose no shape restrictions on the distribution of $\mathbf{v}$, and instead only impose structure on the set of signals $\mathcal{S}$ available to the buyer. In particular, we assume that any feasible signal $s\in \mathcal{S}$ leads to a \textit{\textbf{linear type distribution}} (with a positive continuous density).\footnote{Equivalently, a feasible $s$ can be thought of as a potentially \textit{nonlinear} real-valued signal such that $\E[\mathbf{v} \mid s] = \boldsymbol{\mu} + s \boldsymbol{u}$ for some direction $\boldsymbol{u} \in \mathbb{R}^K$.} Say that $v_1, v_2$ are \textit{\textbf{mutually mean independent}} if $\mathbb{E}[v_2\mid v_1]=\mu_2$ and $\mathbb{E}[v_1\mid v_2]=\mu_1$. Say that an equilibrium is a \textit{\textbf{deterministic-menu equilibrium}} if the seller's menu has deterministic allocations. 

\begin{proposition}\label{prop:lineartypes}
Consider two goods that are mutually mean independent. Suppose that the linear-type signal set $\mathcal{S}$ includes learning either good and the bundle value, i.e., $\{v_1, v_2, v_1+v_2\} \subseteq \mathcal{S}$. Then, every equilibrium has vertical learning. If, in addition, the values are exchangeable, then a deterministic-menu equilibrium exists, and every deterministic-menu equilibrium is weakly Pareto-dominated by a nested-bundling one.
\end{proposition}

Assuming a weak form of independence, \Cref{prop:lineartypes} shows that our key insights generalize beyond unimodal elliptical distributions---all equilibria continue to exhibit vertically differentiated demand systems, and therefore, under exchangeability and proper selection of equilibria, nested bundling continues to arise in equilibrium. The proof is similar to that of \Cref{thm:main}. Take any horizontal-learning equilibrium. By our characterization, any optimal mechanism must allocate one of the goods to all types. The buyer then has a strict incentive to deviate to learning nothing about that good and everything about the other. That signal induces a linear type distribution under mean independence and is feasible. Furthermore, it is a vertical signal---a contradiction.\footnote{This simpler proof makes use of the weak form of independence between $v_1$ and $v_2$, which is satisfied with elliptical distributions under $\rho = 0$. With correlated values, the proof of \Cref{thm:main} also relies on the orthogonal decomposition property of elliptical distributions.} Nested bundling requires further equilibrium selection because, without any shape assumption on the value distribution, an optimal mechanism may use lotteries even under vertical learning. 

\vspace{-3mm}
\paragraph{Relaxing 1D Linear Signals.}\hspace{-2mm}We now consider what happens when the buyer has access to a richer set of feasible signals $\mathcal{S}$. Note that if the buyer has the choice to fully learn his value $\mathbf{v}$, then there always exists an equilibrium where the buyer fully learns his value $\mathbf{v}$ and the seller uses a (generally complex) optimal mechanism given the value distribution. However, there may exist other equilibria as well. In particular, \Cref{prop:pure} gives a sufficient condition for vertical learning and pure bundling to form an equilibrium. 

\begin{proposition}\label{prop:pure}
Suppose that it is feasible to learn the grand bundle value, i.e., $\sum_i v_i \in \mathcal{S}$, and that 
\[\mu_i = \frac{\sigma_i^2+\rho\sigma_i\sum_{k\neq i}\sigma_k}{\sigma_j^2+\rho\sigma_j\sum_{k\neq j}\sigma_k}\mu_j\,\, \text{for all $i\neq j$}\,.\]
Then, vertical learning and pure bundling form an equilibrium. 
\end{proposition}

The above condition guarantees that, when the buyer chooses to only learn about the grand bundle, the seller's optimal menu is indeed to sell only the grand bundle. In such an equilibrium, the buyer only acquires one signal even though information is costless and he can have access to an arbitrary set of signals. The self-stabilizing feature of vertical learning and bundling persists here.

It is also possible to use the logic of \Cref{thm:main} to rule out certain horizontal-learning equilibria, even if information is costly. For simplicity, consider $K=2$. Say that a feasible signal $s \in \mathcal{S}$ is \textit{\textbf{purely horizontal}} if it induces a horizontal demand system (with a positive continuous density) satisfying $\text{Corr}(\theta_1(s), \theta_2(s)) = -1$. The posterior mean distribution induced by such a signal is supported on a strictly decreasing line segment. Suppose a feasible signal $s \in \mathcal{S}$ costs the buyer $\mathcal{C}[s]$. We provide conditions to rule out purely-horizontal-learning equilibria:

\begin{proposition}\label{prop:cost}
Let $K = 2$ and $\rho = 0$. Suppose that learning either good is feasible, i.e., $\{v_1, v_2\} \subseteq \mathcal{S}$, and that  $\mathcal{C}[v_i]\leq\mathcal{C}[s]$ for all $i \in \{1, 2\}$ and all purely-horizontal-learning signals $s \in \mathcal{S}$. Then, no equilibrium has purely horizontal learning.
\end{proposition}

\Cref{prop:cost} follows by the same logic as \Cref{thm:main}---as in \Cref{prop:lineartypes}, with $\rho = 0$, it suffices to consider the deviation to fully learning one of the goods, which combined with the cost assumption, delivers the result. Under a simple information cost, we can in fact describe all equilibrium outcomes:  

\begin{corollary}\label{cor:bound}
Let $K = 2$ and $\rho = 0$. Suppose that the feasible set $\mathcal{S}$ consists of up-to-two linear signals, with the first signal being free and the second costing a constant $c>0$.  Then, in every equilibrium: either the buyer acquires one signal, learning is vertical, and the equilibrium is outcome-equivalent to nested bundling; or he acquires two signals and fully learns~$\mathbf{v}$. Moreover, any nested-bundling equilibrium from our baseline model remains an equilibrium here if either it is a pure-bundling equilibrium or 
    \[c \geq \mathbb{E}\big[\max\{v_{\emph{base good}}-p_{\emph{base good}},0\}\big]- \max \{ \mu_\emph{base good}-p_\emph{base good}, 0\}\,.\]
\end{corollary}

If the buyer acquires only one linear signal, then \Cref{thm:main} applies. If the buyer acquires two linear signals, then he should fully learn his values $\mathbf{v}$ given the simple cost structure. This establishes the first part of \Cref{cor:bound}. Moreover, the value of an additional signal is bounded above by the value of fully learning about the base good. If the base good is inexpensive, then the buyer has little incentive to acquire an additional signal; even a relatively low cost $c$ is enough for the one-signal equilibria of our main model to persist.

\section{Conclusion}\label{sec:conclude}

We study an equilibrium model of multiproduct pricing with consumer learning. The buyer chooses to learn any one-dimensional linear signal of his values for the goods, anticipating the seller's mechanism. The seller designs an optimal mechanism, anticipating the buyer's learning choice. In a generalized Gaussian environment, we show that every equilibrium has vertical learning where the buyer's posterior means are comonotonic, and every equilibrium is outcome equivalent to nested bundling where the seller offers a menu of nested bundles to screen the buyer. Our analysis highlights how the seller's ability to bundle goods affects the buyer's learning incentives. This insight has implications for merger analysis---a merger not only impacts how products are sold but also how consumers learn about them---which we investigate in future work.

\renewcommand*{\bibfont}{\linespread{0.8}\scriptsize\selectfont}
\setlength\bibsep{0pt}
\bibliographystyle{ecta} 
\bibliography{references}

@article{frankel2019muddled,
  title={Muddled information},
  author={Frankel, Alex and Kartik, Navin},
  journal={Journal of Political Economy},
  volume={127},
  number={4},
  pages={1739--1776},
  year={2019},
  publisher={The University of Chicago Press Chicago, IL}
}

@article{ball2025scoring,
  title={Scoring strategic agents},
  author={Ball, Ian},
  journal={American Economic Journal: Microeconomics},
  volume={17},
  number={1},
  pages={97--129},
  year={2025},
  publisher={American Economic Association 2014 Broadway, Suite 305, Nashville, TN 37203-2425}
}

@article{bobkova2024information,
  title={Information choice in auctions},
  author={Bobkova, Nina},
  journal={American Economic Review},
  volume={114},
  number={7},
  pages={1883--1915},
  year={2024},
  publisher={American Economic Association 2014 Broadway, Suite 305, Nashville, TN 37203}
}

@article{bobkova2024optimality,
  title={The Optimality of Majority Rule: An Information-Choice Perspective},
  author={Bobkova, Nina},
  journal={Available at SSRN 5018717},
  year={2024}
}

@article{gleyze2023informationally,
  title={Informationally simple incentives},
  author={Gleyze, Simon and Pernoud, Agathe},
  journal={Journal of Political Economy},
  volume={131},
  number={3},
  pages={802--837},
  year={2023},
  publisher={The University of Chicago Press Chicago, IL}
}

@article{pernoud2025competition,
  title={How Competition Shapes Information in Auctions},
  author={Pernoud, Agathe and Gleyze, Simon},
  year={2025},
  journal={Available at https://agathepernoud.com/Pernoud\_Gleyze\_InfoCompetition.pdf}
}

@article{mensch2022screening,
  title={Screening inattentive buyers},
  author={Mensch, Jeffrey},
  journal={American Economic Review},
  volume={112},
  number={6},
  pages={1949--1984},
  year={2022},
  publisher={American Economic Association 2014 Broadway, Suite 305, Nashville, TN 37203}
}

@article{mensch2025monopoly,
  title={Monopoly, product quality, and flexible learning},
  author={Mensch, Jeffrey and Ravid, Doron},
  journal={arXiv preprint arXiv:2202.09985},
  year={2025}
}

@article{shi2012optimal,
  title={Optimal auctions with information acquisition},
  author={Shi, Xianwen},
  journal={Games and Economic Behavior},
  volume={74},
  number={2},
  pages={666--686},
  year={2012},
  publisher={Elsevier}
}

@techreport{he2023random,
  title={Random Choice and Differentiation},
  author={He, Junnan and Natenzon, Paulo},
  year={2023},
  institution={Working Paper}
}

@article{roesler2017buyer,
  title={Buyer-optimal learning and monopoly pricing},
  author={Roesler, Anne-Katrin and Szentes, Bal{\'a}zs},
  journal={American Economic Review},
  volume={107},
  number={7},
  pages={2072--2080},
  year={2017},
  publisher={American Economic Association 2014 Broadway, Suite 305, Nashville, TN 37203}
}

@book{rockafellar1998variational,
  title={Variational analysis},
  author={Rockafellar, R Tyrrell and Wets, Roger JB},
  year={1998},
  publisher={Springer}
}

@article{brooks2024structure,
  title={On the structure of informationally robust optimal mechanisms},
  author={Brooks, Benjamin and Du, Songzi},
  journal={Econometrica},
  volume={92},
  number={5},
  pages={1391--1438},
  year={2024},
  publisher={Wiley Online Library}
}

@article{bergemann2002information,
  title={Information acquisition and efficient mechanism design},
  author={Bergemann, Dirk and V{\"a}lim{\"a}ki, Juuso},
  journal={Econometrica},
  volume={70},
  number={3},
  pages={1007--1033},
  year={2002},
  publisher={Wiley Online Library}
}

@article{ravid2022learning,
  title={Learning before trading: on the inefficiency of ignoring free information},
  author={Ravid, Doron and Roesler, Anne-Katrin and Szentes, Bal{\'a}zs},
  journal={Journal of Political Economy},
  volume={130},
  number={2},
  pages={346--387},
  year={2022},
  publisher={The University of Chicago Press Chicago, IL}
}

@article{che2025robustly,
    author = {Che, Yeon-Koo and Zhong, Weijie},
    title = {Robustly Optimal Mechanisms for Selling Multiple Goods},
    journal = {Review of Economic Studies},
    volume = {92},
    number = {5},
    pages = {2923-2951},
    year = {2024},
    month = {11},
    issn = {0034-6527},
    doi = {10.1093/restud/rdae107},
    url = {https://doi.org/10.1093/restud/rdae107},
    eprint = {https://academic.oup.com/restud/article-pdf/92/5/2923/60753318/rdae107.pdf},
}

@article{whinstontying,
 ISSN = {00028282},
 URL = {http://www.jstor.org/stable/2006711},
 author = {Michael D. Whinston},
 journal = {American Economic Review},
 number = {4},
 pages = {837--859},
 publisher = {American Economic Association},
 title = {Tying, Foreclosure, and Exclusion},
 urldate = {2025-10-07},
 volume = {80(4)},
 year = {1990}
}

@article{lahr2024extreme,
  title={Extreme Points in Multi-Dimensional Screening},
  author={Lahr, Patrick and Niemeyer, Axel},
  journal={arXiv preprint arXiv:2412.00649},
  year={2024}
}

@article{matthews2007information,
  title={Information acquisition and refunds for returns},
  author={Matthews, Steven A and Persico, Nicola},
  year={2007},
  publisher={PIER working paper}
}

@inproceedings{haberman2025multidimensional,
  title={Multidimensional Screening with Returns},
  author={Haberman, Alexander and Jagadeesan, Ravi and Yang, Frank},
  booktitle={Proceedings of the 26th ACM Conference on Economics and Computation},
  pages={36--36},
  year={2025}
}

@article{yang2026screening,
  title={Screening Frontiers},
  author={Yang, Frank},
  journal={arXiv preprint arXiv:2602.20087},
  year={2026}
}

@article{che1996customer,
  title={Customer return policies for experience goods},
  author={Che, Yeon-Koo},
  journal={Journal of Industrial Economics},
  pages={17--24},
  year={1996},
  publisher={JSTOR}
}

@article{debroesler2024,
journal={Review of Economic Studies},
author={Rahul Deb and Anne-Katrin Roesler},
title={Multi-Dimensional Screening: Buyer-Optimal Learning and Informational Robustness},
year={2024},
month={None},
pages={2744-2770},
volume={91},
number={5},
}

@article{adams1976commodity,
  title={Commodity Bundling and the Burden of Monopoly},
  author={Adams, William James and Yellen, Janet L},
  journal={Quarterly Journal of Economics},
  pages={475--498},
  year={1976},
  volume = {90(3)},
  number = {3}
}

@book{gupta2013elliptically,
  title={Elliptically contoured models in statistics and portfolio theory},
  author={Gupta, Arjun K and Varga, Tamas and Bodnar, Taras},
  volume={2},
  year={2013},
  publisher={Springer}
}

@article{hart2019selling,
  title={Selling Multiple Correlated Goods: Revenue Maximization and Menu-Size Complexity},
  author={Hart, Sergiu and Nisan, Noam},
  journal={Journal of Economic Theory},
  volume={183},
  pages={991--1029},
  year={2019},
  publisher={Elsevier}
}

@article{manelli2006bundling,
  title={Bundling as an Optimal Selling Mechanism for a Multiple-Good Monopolist},
  author={Manelli, Alejandro M and Vincent, Daniel R},
  journal={Journal of Economic Theory},
  volume={127(1)},
  number={1},
  pages={1--35},
  year={2006},
  publisher={Elsevier}
}

@article{daskalakis2017strong,
  title={Strong Duality for a Multiple-Good Monopolist},
  author={Daskalakis, Constantinos and Deckelbaum, Alan and Tzamos, Christos},
  journal={Econometrica},
  volume={85(3)},
  number={3},
  pages={735--767},
  year={2017},
  publisher={Wiley Online Library}
}

@article{armstrong1996multiproduct,
  title={Multiproduct Nonlinear Pricing},
  author={Armstrong, Mark},
  journal={Econometrica},
  pages={51--75},
  volume={64(1)},
  number={1},
  year={1996},
  publisher={JSTOR}
}

@book{muller2002comparison,
  title={Comparison Methods for Stochastic Models and Risks},
  author={M{\"u}ller, Alfred and Stoyan, Dietrich},
  year={2002},
  publisher={New York: Wiley}
}

@article{ghili2021characterization,
  title={A characterization for optimal bundling of products with nonadditive values},
  author={Ghili, Soheil},
  journal={American Economic Review: Insights},
  volume={5},
  number={3},
  pages={311--326},
  year={2023},
  publisher={American Economic Association 2014 Broadway, Suite 305, Nashville, TN 37203}
}

@techreport{yang2023comparison,
  title={Comparison of Screening Devices},
  author={Yang, Frank and Dworczak, Piotr and Akbarpour, Mohammad},
  year={2023},
  institution={GRAPE Group for Research in Applied Economics}
}

@article{Rochet1998,
    title = {Ironing, Sweeping, and Multidimensional Screening},
    year = {1998},
    journal = {Econometrica},
    author = {Rochet, Jean-Charles and Chone, Philippe},
    volume = {66(4)},
    pages  = {783--826},
    number = {4}
}

@article{pavlov2011optimal,
  title={Optimal Mechanism for Selling Two Goods},
  author={Pavlov, Gregory},
  journal={B.E. Journal of Theoretical Economics},
  volume={11(1)},
  number={1},
  year={2011},
  publisher={De Gruyter}
}

@article{Manelli2007,
    title = {{Multidimensional Mechanism Design: Revenue Maximization and the Multiple-good Monopoly}},
    year = {2007},
    journal = {Journal of Economic Theory},
    author = {Manelli, Alejandro M and Vincent, Daniel R},
    number = {1},
    month = {11},
    pages = {153--185},
    volume = {137(1)},
    publisher = {Academic Press},
    doi = {10.1016/j.jet.2006.12.007},
    issn = {00220531}
}

@article{stigler1963united,
  title={United States v. Loew's Inc.: A Note on Block-booking},
  author={Stigler, George J},
  journal={Supreme Court Review},
  volume={1963},
  pages={152--157},
  year={1963},
  publisher={The University of Chicago Press}
}

@article{Myerson1981,
    title = {{Optimal Auction Design}},
    year = {1981},
    journal = {Mathematics of Operations Research},
    author = {Myerson, Roger B.},
    volume = {6(1)},
    number = {1},
    pages = {58--73}
}

@article{Carroll2017,
    title = {{Robustness and Separation in Multidimensional Screening}},
    year = {2017},
    journal = {Econometrica},
    author = {Carroll, Gabriel},
    number = {2},
    pages = {453--488},
    volume = {85(2)},
    doi = {10.3982/ecta14165},
    issn = {0012-9682}
}

@inproceedings{Daskalakis2014,
    title = {{The Complexity of Optimal Mechanism Design}},
    year = {2014},
    booktitle = {Twenty-Fifth Annual ACM-SIAM Symposium on Discrete Algorithms},
    author = {Daskalakis, Constantinos and Deckelbaum, Alan and Tzamos, Christos},
    arxivId = {arXiv:1211.1703v2}
}

@incollection{Rochet2003,
    title = {{The Economics of Multidimensional Screening}},
    year = {2003},
    booktitle = {Advances in Economics and Econometrics: Theory and Applications, Eighth World Congress, Volume 1},
    author = {Rochet, Jean-Charles and Stole, Lars A.},
    publisher= {Cambridge: Cambridge University Press} }

@article{bergemann2022optimality,
  title={The Optimality of Upgrade Pricing},
  author={Bergemann, Dirk and Bonatti, Alessandro and Haupt, Andreas and Smolin, Alex},
  journal={\emph{ arXiv:2107.10323 [econ.TH]}},
  year={2022}
}

@article{mcafee1988multidimensional,
  title={Multidimensional Incentive Compatibility and Mechanism Design},
  author={McAfee, R Preston and McMillan, John},
  journal={Journal of Economic Theory},
  volume={46(2)},
  number={2},
  pages={335--354},
  year={1988},
  publisher={Elsevier}
}

@article{haghpanah2021pure,
  title={When is Pure Bundling Optimal?},
  author={Haghpanah, Nima and Hartline, Jason},
  journal={Review of Economic Studies},
  volume={88(3)},
  number={3},
  pages={1127--1156},
  year={2021},
  publisher={Oxford University Press}
}

@article{McAfee1989MultiproductValues,
    title = {{Multiproduct Monopoly, Commodity Bundling, and Correlation of Values}},
    year = {1989},
    journal = {Quarterly Journal of Economics},
    author = {McAfee, R. Preston and McMillan, John and Whinston, Michael D.},
    number = {2},
    volume = {104},
    doi = {10.2307/2937852},
    issn = {15314650}
}

@article{yang2022costly,
    author = {Yang, Frank},
    title = {Costly Multidimensional Screening},
    journal = {Review of Economic Studies},
    pages = {rdaf105},
    year = {2025},
    month = {12},
    issn = {0034-6527},
    doi = {10.1093/restud/rdaf105},
    url = {https://doi.org/10.1093/restud/rdaf105},
    eprint = {https://academic.oup.com/restud/advance-article-pdf/doi/10.1093/restud/rdaf105/66124356/rdaf105.pdf}
}

@article{yang2023nested,
  title={Nested bundling},
  author={Yang, Frank},
  journal={American Economic Review},
  volume={115(9)},
  number={9},
  pages={2970--3013},
  year={2025},
  publisher={American Economic Association 2014 Broadway, Suite 305, Nashville, TN 37203}
}

@article{frick2024multidimensional,
  title={Multidimensional screening with precise seller information},
  author={Frick, Mira and Iijima, Ryota and Ishii, Yuhta},
 journal={Econometrica},
  volume={94},
  number={1},
  pages={35--70},
  year={2026},
  publisher={Wiley Online Library}
}

@article{loertscher2024optimal,
  title={Optimal Hotelling Auctions},
  author={Loertscher, Simon and Muir, Ellen V},
  journal={Mimeo},
  year={2024}
}

@article{loertscher2019optimal,
  title={Optimal structure and dissolution of partnerships},
  author={Loertscher, Simon and Wasser, C{\'e}dric},
  journal={Theoretical Economics},
  volume={14},
  number={3},
  pages={1063--1114},
  year={2019},
  publisher={Wiley Online Library}
}

\newpage

\begin{centering}
 \qquad   {\Large \textbf{Online Appendix to ``Bundling against Learning''}}
\end{centering}

\[\text{\centering \Large Agathe Pernoud and Frank Yang}\]
\appendix

\section{Proofs}\label{app:proof}

\subsection{Proof of \Cref{thm:main}}\label{subsec:proofmain}

The proof is organized as follows. First, we prove that every equilibrium must have vertical learning. Second, we prove its outcome equivalence to a nested bundling equilibrium.

Most of the proof is contained in \Cref{sec:proof}. To avoid repeating the same arguments, we only present the steps that were omitted from the proof sketch. 

\subsubsection{Vertical Learning}\label{subsec:vertical}

\paragraph{Preliminaries.}\hspace{-2mm}We first explain in more detail our normalization of types. Recall that a buyer's type is linear in the signal realization, with
\[\theta_i(s;\boldsymbol\alpha)=\frac{\text{Cov}(v_i,\boldsymbol{\alpha}\cdot\mathbf{v})}{\text{Var}(\boldsymbol{\alpha}\cdot\mathbf{v})}s+\mu_i - \frac{\text{Cov}(v_i,\boldsymbol{\alpha}\cdot\mathbf{v})}{\text{Var}(\boldsymbol{\alpha}\cdot\mathbf{v})}\boldsymbol\alpha\cdot\boldsymbol\mu\,.\]
Let $[\underline{s}, \overline{s}]$ denote the set of possible signal realizations. That is, $\underline{s}:=\min_{\mathbf{v}\in V}\boldsymbol\alpha\cdot\mathbf{v}$ and $\overline{s}:=\max_{\mathbf{v}\in V}\boldsymbol\alpha\cdot\mathbf{v}$. We can re-index types by some parameter $t$ so that $t\in [0,1]$. Indeed, set $t=(s-\underline{s})/(\overline{s}-\underline{s})$. Rewriting the above expression yields
\[\theta_i(t;\boldsymbol\alpha)=(\overline{s}-\underline{s})\frac{\text{Cov}(v_i,\boldsymbol{\alpha}\cdot\mathbf{v})}{\text{Var}(\boldsymbol{\alpha}\cdot\mathbf{v})}t+\mu_i - \frac{\text{Cov}(v_i,\boldsymbol{\alpha}\cdot\mathbf{v})}{\text{Var}(\boldsymbol{\alpha}\cdot\mathbf{v})}(\boldsymbol\alpha\cdot\boldsymbol\mu-\underline{s})=: a_it + b_i\,.\]
Let $F$ be the cumulative distribution of $t$ induced by the strategy $\boldsymbol\alpha$, and let $f$ be its density. By \Cref{sec:model}, $f$ is symmetric about $0.5$ and unimodal. The following two lemmas establish properties of $F$ that will prove useful in the characterization of optimal mechanisms.

\begin{lemma}\label{lem:regular}
    Let $\Psi(t):=t-\frac{1-F(t)}{f(t)}$. The function $\Psi$ is strictly increasing for $t\leq 0.5$ and strictly positive for $t> 0.5$. Moreover, $\Psi(0):=\displaystyle \lim_{t\rightarrow 0}\Psi(t) = -\infty$ and $\Psi(1):=\displaystyle \lim_{t\rightarrow 1}\Psi(t)=1$. Thus, for any $a_i>0, b_i \geq 0$, we have $a_i \Psi(t) + b_i$ is strictly single-crossing with a crossing point $t^*_i \in (0, 0.5]$. 
\end{lemma}

\begin{proof}
We first prove that $\Psi$ is strictly increasing for $t\leq 0.5$. Recall that $f$ is symmetric about $0.5$ and unimodal, so that $f$ is nondecreasing on $[0, 0.5]$ and nonincreasing on $[0.5, 1]$.
Thus, for any $0 < t < t' \leq 0.5$, since $f(s) \geq f(t)$ for $s\in[t, t']$ and $f(t)\leq f(t')$, we have
\[1-F(t) = 1-F(t') + \int_t^{t'} f(s)\,ds \;\geq\; 1-F(t') + f(t)\,(t'-t)\,, \]
and hence, $\frac{1-F(t)}{f(t)} \;\geq\; \frac{1-F(t')}{f(t')} + (t'-t)$. Therefore, $\Psi(t) \leq \Psi(t') - 2(t'-t) < \Psi(t')$.

We now prove that $\Psi$ is strictly positive for $t> 0.5$. Using the fact that $f$ is nonincreasing on $[0.5,1]$, we get that for all $t>0.5$:
\[1-F(t) = \int_t^1f(s) ds \leq \int_t^1f(t) ds = f(t) (1-t). \]
Thus, 
\[\Psi(t) = t-\frac{1-F(t)}{f(t)}\geq t- (1-t) = 2t-1>0\,. \]
This directly implies $\Psi(1)=1$. Indeed, we have $2t-1\leq \Psi(t) \leq t$, where both the lower and upper bounds go to 1 as $t$ goes to 1. 

Finally, we show that $f(0)=0$, which implies $\displaystyle \lim_{t \rightarrow 0}\Psi(t) = -\infty$ since $f$ is continuous. Recall that the buyer has type $t=0$ if and only if he receives the lowest possible signal realization $\underline{s} = \min_{\mathbf{v}\in V}\boldsymbol\alpha\cdot\mathbf{v}$. The overall set of values $V$ is an ellipsoid, so the subset of values at which the buyer receives $\underline{s}$ is a single point $|\{\mathbf{v}'\in V\mid \boldsymbol\alpha\cdot\mathbf{v}' = \underline{s}\}|=1$, which has $(K-1)$-dimensional Lebesgue measure $0$, while $\{\mathbf{v}'\in V\mid \boldsymbol\alpha\cdot\mathbf{v}' = s\}$ has strictly positive $(K-1)$-dimensional Lebesgue measure for all $s \in(\underline{s}, \overline{s})$. Thus, by Fubini's theorem, $f(0) = 0$. 

Now, fix any $i$ with $a_i >0$ and $b_i \geq 0$. Suppose that $a_i \Psi(t) +b_i \geq 0$. We claim that for all $\hat{t} > t$, $a_i \Psi(\hat{t}) +b_i > 0$. Indeed, if not, then
\[a_i \Psi(\hat{t}) +b_i \leq 0 \]
and hence 
\[ \Psi(\hat{t})  \leq - \frac{b_i}{a_i} \leq 0\,,\]
but then it implies that $\hat{t} \leq 0.5$, and hence it must be that  
\[ \Psi(t) < \Psi(\hat{t}) \,.\]
But then it follows that 
\[a_i \Psi(t) +b_i < 0\,,\]
a contradiction. Finally, we claim that the crossing point 
\[t^*_i := \inf \big\{t: a_i \Psi(t) + b_i > 0\big\} \in ( 0,0.5]\,.\]
Indeed, if $t^*_i=0$, then for any $t > 0$, we have 
\[a_i \Psi(t) + b_i > 0\]
but then 
\[\lim_{t \rightarrow 0}\Psi(t)  \geq -\frac{b_i}{a_i}\,,\]
a contradiction. Similarly, for any $t>0.5$, we have 
\[\Psi(t)>0 \implies a_i\Psi(t)+b_i>0,\]
so $t_i^*\leq 0.5$.
\end{proof}

\begin{lemma}\label{lem:continuous}
     For any $t_0$, let   
     \[\Phi(t; t_0) :=\left(t+\frac{F(t)}{f(t)}\right)\mathbbm{1}\{t< t_0\} + \left(t-\frac{1-F(t)}{f(t)}\right)\mathbbm{1}\{t\geq t_0\}\,.\]    
Let $\overline{\Phi}(t; t_0)$ be the ironed $\Phi(\,\cdot\,;t_0)$. Then $g(t_0) := \overline{\Phi}(t_0;t_0)$ is continuous in $t_0$. 
\end{lemma}

\begin{proof}
By definition, we can write 
\[H(q; t_0) := \int^{q}_0 \Phi(F^{-1}(u);t_0) d u    \,, \]
and 
\[\overline{\Phi}(t;t_0) = \partial_{+} \text{conv}[H(\,\cdot\,;t_0)](F(t))\,.\]
Note first that 
\[\sup_{q\in[0, 1]}\big|H(q; t'_0) - H(q; t_0)  \big|  \rightarrow 0 \]
as $t'_0 \rightarrow t_0$. Indeed, for $t_0 < t'_0$ we have $\Phi(t;t'_0)-\Phi(t;t_0) = 1/f(t)$ for $t \in [t_0, t'_0)$ and $0$ otherwise; hence, substituting $u=F(t)$, we have 
\[\big|H(q; t'_0) - H(q; t_0)\big| = \int_{F(t_0)}^{\min\{q,\,F(t'_0)\}} \frac{du}{f(F^{-1}(u))} = \big(\min\{F^{-1}(q),\,t'_0\}-t_0\big)_+ \leq |t'_0 - t_0|\]
for every $q\in[0,1]$ (and symmetrically for $t_0'<t_0$). Therefore, $H(\,\cdot\,;t_0)$ is continuous in $t_0$ in the sup norm. We claim that 
\[||\text{conv}[H_1] -\text{conv}[H_2]||_{\sup}\leq ||H_1 - H_2||_{\sup}\,.\]
Indeed, note that if $H_2(z) - \delta \leq H_1(z) \leq H_2(z) + \delta$, then we have
\[\text{conv}[H_2 - \delta] \leq \text{conv}[H_1] \leq \text{conv}[H_2 + \delta]\]
and hence 
\[\text{conv}[H_2] - \delta \leq \text{conv}[H_1] \leq \text{conv}[H_2] + \delta\,.\]
Now, we claim that for any interval $\mathcal{I}$ on which a sequence of convex $H_n$ converge to $H$ uniformly, and $t_n \rightarrow t \in \mathcal{I}$, where $H$ is differentiable at $t$, we have 
\[\partial_{+} H_n(t_n) \rightarrow \partial H(t)\,.\]
Indeed, the uniform convergence implies epi-convergence, which by Attouch's theorem, implies that the subdifferentials must converge in the sense of graphical convergence (\citealt{rockafellar1998variational}). Then, the claim follows immediately given that $H$ is differentiable at $F(t_0)$. 

Now let 
\[H_n(\,\cdot\,):= \text{conv}H(\,\cdot\,;t^n_0)\,.\]
Combining earlier observations, we have that $H_n(\,\cdot\,)$ converges uniformly to 
\[H^\star(\,\cdot\,) := \text{conv} H(\,\cdot\,;t_0)\] as $t^n_0 \rightarrow t_0$. By construction $H^\star(\,\cdot\,)$ is differentiable at $F(t_0)$. It follows by the previous claim that 
\[\partial_+ H_n(F(t^n_0)) \rightarrow \partial H^\star(F(t_0))\,,\]
as $t^n_0 \rightarrow t_0$. Therefore, we have 
\[g(t^n_0) = \overline{\Phi}(t^n_0;t^n_0) = \partial_+ H_n(F(t^n_0)) \rightarrow \partial H^\star(F(t_0)) = \overline{\Phi}(t_0;t_0) = g(t_0)\,\]
as $t^n_0\rightarrow t_0$, proving the result. 
\end{proof}

\paragraph{Outline.}\hspace{-2mm}As in \Cref{sec:proof}, the remainder of the proof proceeds by contradiction as follows: 
\begin{itemize}
    \item[\textbf{Step 1}.]  We characterize optimal mechanisms against any distribution supported on a line segment in $\mathbb{R}^K_+$, assuming a horizontal learning strategy.  
    \item[\textbf{Step 2}.]  We characterize properties of the buyer's optimal learning strategies against any candidate optimal mechanism using \textbf{Step 1}.
     \item[\textbf{Step 3}.] We show these together lead to a contradiction. 
\end{itemize}

Our mechanism characterization results also hold when the buyer uses a vertical learning strategy and, in fact, become substantially simpler then. However, to streamline the proof, in this subsection, we focus on deriving the contradiction assuming the buyer uses a horizontal learning strategy. 

\vspace{-3mm}
\paragraph{Optimal Mechanisms.}\hspace{-2mm}We solve for every optimal direct revelation mechanism $\mathcal{M}^{DR}$ under the type distribution induced by any horizontal-learning strategy $\boldsymbol\alpha$. For the seller's strategy $\mathcal{M}$ to be a best response, the mechanism must be outcome-equivalent to some optimal $\mathcal{M}^{DR}$, when restricting to the type space induced by $\boldsymbol\alpha$.\footnote{Since we focus on pure-strategy Nash equilibria, the seller correctly anticipates the buyer's learning strategy $\boldsymbol\alpha$ and does not benefit from eliciting it.}

Recall the sign convention: 
\[\sum_i a_i \geq 0\,, \tag{Sign Convention}\]
as well as the definitions of \textit{\textbf{strictly positive goods}} and \textit{\textbf{negative goods}}:
\[I^+ := \{i: a_i > 0\}\quad \text{ and }\quad I^- :=\{i: a_i \leq 0\}\,. \]
Note that under horizontal learning, there must exist two goods $i$, $j$ such that \[\text{sign}(\text{Cov}(v_i,\boldsymbol{\alpha}\cdot\mathbf{v}))\neq\text{sign}(\text{Cov}(v_j,\boldsymbol{\alpha}\cdot\mathbf{v}))\,,\]
and hence $\text{sign}(a_i)\neq\text{sign}(a_j)$---there must exist both strictly positive and strictly negative goods. Under vertical learning, by the sign convention, all the goods must be either strictly positive goods, or zero-sign goods (i.e., $a_i = 0$). 

While either $a_i$ or $b_i$ can be $0$, they cannot both be zero. Indeed, Bayesian plausibility requires that $\mathbb{E}[a_it+b_i] = \mu_i>0$ for any good $i$. Since $\mathbb{E}(t)=0.5$, this implies $(a_i, b_i) \neq (0, 0)$.  Combined with the observation that $b_i\geq \max\{0, -a_i\}$, this implies: 
\[a_i \leq 0 \implies b_i > 0\,.\]

Recall the \textbf{(Auxiliary Program)} introduced in \Cref{sec:proof}: 
\begin{equation}
\max_{\mathbf{x}\in [0,1]^K}\sum_i b_ix_i \quad \text{ subject to }\quad \sum_ia_ix_i=0\,.\label{eq:auxiliary}  
\end{equation}
Let $X^*$ be the set of all optimal solutions to the above program. We first characterize every solution to \textbf{(Auxiliary Program)}: 
\begin{lemma}\label{lem:auxiliary}
There exists $\kappa \geq 0$ such that for any optimal solution $x^*$ to \eqref{eq:auxiliary}, we have:  
\begin{itemize}
    \item[(i)]  For any negative good $i$, $x^*_i = 1$. 
    \item[(ii)] For any strictly positive good $i$, $x^*_i = 1$ if $b_i/a_i > \kappa$ and $x^*_i = 0$ if $b_i/a_i < \kappa$.
\end{itemize}
\end{lemma}
\begin{proof}
For part \textit{(i)}, we prove by contradiction. Suppose by contradiction that $x^*$ is an optimal solution such that $x^*_i < 1$ for some negative good $i$. By the previous observation, this implies that $b_i > 0$. Now, if $a_i = 0$, then increasing $x^*_i$ would be strictly improving the objective while satisfying the constraint. Thus, it must be the case that $a_i < 0$. However, then it must be that there exists some strictly positive good $j$ such that $x^*_j < 1$, because otherwise we cannot satisfy the equality constraint: 
\[\sum_{k} a_k x^*_k > a_i + \sum_{k \neq i} a_k x^*_k \geq a_i + \sum_{k \in I^{+}} a_k + \sum_{k \in I^{-}, k \neq i} a_k \geq 0\,,\]
where the last inequality is by the sign convention. Then, strictly increasing $x^*_i$ and $x^*_j$ can strictly increase the objective while keeping the equality constraint (given that $b_i > 0$ by the previous observation again). For part \textit{(ii)}, note that given all negative goods must have $x^*_i = 1$, the problem of assigning the positive goods reduces to a fractional knapsack problem. Thus, by an exchange argument, every optimal solution must satisfy the greedy property of assigning in the order of $b_i / a_i$ with possible randomization for the goods with the same $b_i / a_i$. The claim follows immediately. 
\end{proof}

Recall the definition of \textit{\textbf{positive balancing goods}}: 
\[I^*:= \bigcup_{x^* \in X^*}\Big\{i \in I^+: x^*_i > 0\Big\}\,.\]
Note that under any horizontal learning strategy, some positive balancing good always exists: Indeed, under horizontal learning, there must exist a strictly positive good and a strictly negative good. Thus, for every $x^* \in X^*$, by \Cref{lem:auxiliary}, if $x^*_i = 0$ for all $i \in I^+$, then 
\[\sum_i a_i x^*_i = \sum_{i \in I^-} a_i x^*_i = \sum_{i \in I^-} a_i < 0\,,\]
which would violate the equality constraint of $\eqref{eq:auxiliary}$. Therefore, for every $x^* \in X^*$, $\Big\{i \in I^+: x^*_i > 0\Big\} \neq \varnothing$, and hence $I^* \neq \varnothing$.

We now characterize the structure of every optimal mechanism against any linear projection of a unimodal elliptical distribution in $\mathbb{R}^{K}_+$ (\Cref{lem:optx}). The general argument is the same as in \Cref{sec:proof}. We first prove the two lemmas that we have used in the main text (\Cref{lem:ironing} and \Cref{lem:saddlepoint}), where the statement of \Cref{lem:saddlepoint} (\textbf{Saddle-Point Construction}) is given in the main text. 

Let $\Phi(t ; t_0)$ be defined as in \Cref{lem:continuous}, and let $\overline{\Phi}(\,\cdot\,;t_0)$
be the ironed version of $\Phi(\,\cdot\,;t_0)$ exactly as in \citet{Myerson1981}.

\begin{lemma}\label{lem:ironing}
Fix any $t_0$ and let $x \in \emph{MON}$. Then $x$ solves
\begin{align}
    \max_{x' \in \emph{MON}}\,&\mathbb{E}\Bigg[\sum_i \Big(a_i x'_i(t) \Phi(t; t_0) + b_i x'_i(t)\Big)\Bigg] \label{eq:constrained}
\end{align}
if and only if $x$ solves
\begin{align}
    &\max_{x':[0, 1] \rightarrow[0,1]^K} \mathbb{E}\Bigg[\Big(\sum_i a_i x'_i(t) \Big) \overline{\Phi}(t; t_0) + \sum_i b_i x'_i(t) \Bigg]  \label{eq:unconstrained}
\end{align}
and is consistent with ironing, in the sense that $\sum_i a_i x_i(t)$ is constant on the interior of every ironing interval where $\overline{\Phi}(t; t_0)$ is constant and $\overline{\Phi}(t; t_0)\neq \Phi(t; t_0)$.
\end{lemma}

\begin{proof}
By \citet{Myerson1981}, we have that for any $x' \in \text{MON}$, 
\[\mathbb{E}\Bigg[\sum_i \Big(a_i x'_i(t) \Phi(t; t_0) + b_i x'_i(t)\Big)\Bigg] \leq \mathbb{E}\Bigg[\Big(\sum_i a_i x'_i(t) \Big) \overline{\Phi}(t; t_0) + \sum_i b_i x'_i(t) \Bigg]\,,\]
where the inequality holds with equality if and only if $x'$ is consistent with ironing. Moreover, by an argument similar to that of \citet{Myerson1981}, we also know that there exists some $\tilde{x} \in \text{MON}$ that solves the ironed, unconstrained problem \eqref{eq:unconstrained} and is consistent with ironing. Therefore, the optimal value of \eqref{eq:unconstrained}  is given by 
\[\text{OPT}:=\mathbb{E}\Bigg[\Big(\sum_i a_i \tilde{x}_i(t) \Big) \overline{\Phi}(t; t_0) + \sum_i b_i \tilde{x}_i(t) \Bigg] = \mathbb{E}\Bigg[\Big(\sum_i a_i \tilde{x}_i(t) \Big) \Phi(t; t_0) + \sum_i b_i \tilde{x}_i(t) \Bigg]\,.\]
This implies that every $x$ solving the MON-constrained problem \eqref{eq:constrained} must also solve \eqref{eq:unconstrained} since by construction 
\begin{align*}
   \mathbb{E}\Bigg[\Big(\sum_i a_i \tilde{x}_i(t) \Big) \Phi(t; t_0) + \sum_i b_i \tilde{x}_i(t) \Bigg] &\leq \mathbb{E}\Bigg[\Big(\sum_i a_i x_i(t) \Big) \Phi(t; t_0) + \sum_i b_i x_i(t) \Bigg] \\
   & \leq \mathbb{E}\Bigg[\Big(\sum_i a_i x_i(t) \Big) \overline{\Phi}(t; t_0) + \sum_i b_i x_i(t) \Bigg] \,,
\end{align*}
but the left-hand side is the optimal value of \eqref{eq:unconstrained} and hence these inequalities must hold with equality. This also implies that any such $x$ must be consistent with ironing, since otherwise the second inequality above would be strict. 

Conversely, for any $x \in \text{MON}$ that solves the ironed, unconstrained problem \eqref{eq:unconstrained} and is consistent with ironing, we have 
\begin{align*}
   \mathbb{E}\Bigg[\Big(\sum_i a_i x_i(t) \Big) \Phi(t; t_0) + \sum_i b_i x_i(t) \Bigg] &= \mathbb{E}\Bigg[\Big(\sum_i a_i x_i(t) \Big) \overline{\Phi}(t; t_0) + \sum_i b_i x_i(t) \Bigg] \\
   &\geq \mathbb{E}\Bigg[\Big(\sum_i a_i x'_i(t) \Big) \overline{\Phi}(t; t_0) + \sum_i b_i x'_i(t) \Bigg] \\
    &\geq \mathbb{E}\Bigg[\Big(\sum_i a_i x'_i(t) \Big) \Phi(t; t_0) + \sum_i b_i x'_i(t) \Bigg]\,,
\end{align*}
for any $x' \in \text{MON}$. Thus, any such $x$ must be optimal for the MON-constrained problem \eqref{eq:constrained}, as desired.
\end{proof}

\begin{proof}[Proof of \Cref{lem:saddlepoint}]
We now prove \Cref{lem:saddlepoint}, which states that there exists a saddle point $(x^*, t_0^*)$ such that $\overline{\Phi}(t_0^*;t_0^*)=\lambda$ and $[0,t_0^*]\subset\{t:\overline{\Phi}(t;t_0^*)=\lambda\}$. 

Let 
 \[g(t_0) := \overline{\Phi}(t_0;t_0) \]
be the value of the ironed part including $t_0$. Note that $g(t_0)$ is continuous in $t_0$ by \Cref{lem:continuous}, and satisfies $g(1) > 0$. Moreover, by \Cref{lem:regular}, we have that\[g(0) =  \overline{\Phi}(0;0)   < -\max_{i \in I^+}\Big\{\frac{b_i}{a_i}\Big\}\,.\]
Indeed, \Cref{lem:regular} implies that $a_i \Phi(t; 0) + b_i$ is strictly single-crossing for all $i \in I^+$ with a  strictly positive crossing point, which implies that the ironed version  $a_i \overline{\Phi}(t; 0) + b_i$ must also be strictly single-crossing with the same crossing point, and hence $a_i \overline{\Phi}(0; 0) + b_i < 0$ for all $i \in I^+$.\footnote{To see this, note that the concave envelope of a strictly quasi-concave function must touch the original function at the original peak.}

Note that under horizontal learning, at least one $a_i < 0$, and hence by our previous observation, for any $x^* \in X^*$, there exists some $i \in I^+$ such that $x^*_i > 0$. Let $\kappa \geq 0$ be the largest constant with the properties in \Cref{lem:auxiliary}. Let
\[
\lambda := -\kappa
\]
and note that $\lambda$ must be an optimal dual multiplier for the problem \eqref{eq:auxiliary}. 
As a consequence, it must be that $\kappa \leq \max_{i \in I^+}\{\frac{b_i}{a_i}\}$ and hence 
\[0 \geq \lambda  \geq -\max_{i \in I^+}\Big\{\frac{b_i}{a_i}\Big\}\,.\]
Therefore, by the continuity of $g(\,\cdot\,)$ and the intermediate value theorem, there exists some $t^*_0$ such that 
\[\overline{\Phi}(t^*_0; t^*_0) = \lambda \,.\]
We claim that the ironing interval including $t^*_0$ must also include $0$. Indeed, if not, then we have both that $\overline{\Phi}(t^*_0; t^*_0) \leq 0$ and that $\overline{\Phi}(0^+; t^*_0) > 0$ (since that value would become the ironed virtual cost), contradicting the monotonicity of $\overline{\Phi}(\,\cdot\,;t^*_0)$. As a consequence, there must exist an ironing interval $\mathcal{I} \supset [0, t^*_0]$ (in particular, note that $t^*_0$ must be in the interior of the ironing interval $\mathcal{I}$ by construction).

Now, we claim that $t^*_0$ is part of a saddle point. Indeed, fix $t^*_0$ as a conjectured worst-off type. Consider the ironed, pointwise maximization problem \eqref{eq:unconstrained}.

First, consider the interval $t \in \mathcal{I}$, note that on that interval the pointwise maximization problem, by construction, is equivalent to 
\begin{equation}
    \max_{x \in [0, 1]^K} \sum_i a_i \overline{\Phi}(t; t^*_0) x_i + \sum_i  b_i x_i = \max_{x \in [0, 1]^K} \sum_i b_i x_i + \lambda \sum_i a_i x_i  \,, \label{eq:pointwise-lambda-ap}
\end{equation}
which is the Lagrangian of \eqref{eq:auxiliary}. By construction of $\lambda$, there must exist a solution $x^\dagger \in X^*$ to this pointwise maximization problem. Note that $\sum_i a_i x^\dagger_i = 0$, and $x^\dagger_i = 1$ for all $i \in I^-$. 

Now we consider any $t  \not \in \mathcal{I}$.  For any $i \in I^-$, we have that for any type $t \not \in \mathcal{I}$, 
\[a_i \overline{\Phi}(t; t^*_0) + b_i\geq a_i \overline{\Phi}(1; t^*_0) + b_i =  a_i + b_i \geq 0\,,\]
where the last inequality is due to  $\theta_i(t) \geq 0$ for all $t$, and in particular $t = 1$. Moreover, note that, for $t < 1$ and $a_i < 0$, the first inequality is strict since $t = 1$ cannot be in an ironing interval, while for $a_i = 0$ the expression equals $b_i > 0$. 

For any $i \in I^*$, we have that for any type $t \not \in \mathcal{I}$, 
\[a_i \overline{\Phi}(t; t^*_0) + b_i > a_i \overline{\Phi}(t^*_0; t^*_0) + b_i = a_i\lambda + b_i \geq 0\,.\]
For any $i \in I^+ \backslash I^*$, note that since 
\[a_i \overline{\Phi}(t; t^*_0) + b_i \]
is a monotone function that starts at a strictly negative value (on $\mathcal{I}$), there exists some threshold $t^*_i \not \in \mathcal{I}$ such that $x_i(t) = \mathbbm{1}\{t \geq t^*_i\}$ is pointwise optimal. Indeed, for $t  \not \in \mathcal{I}$, we have $\Phi(t;t_0^*)=\Psi(t)$ where $a_i \Psi(t) + b_i$ is strictly single-crossing by \Cref{lem:regular}. Now, note that outside of the ironing interval $\mathcal{I}$, $a_i \overline{\Phi}(t; t^*_0) + b_i$ coincides with the ironing of $a_i \Psi(t) + b_i$ when restricted to the remaining interval $[0, 1] \backslash \mathcal{I}$. Consequently, as argued before, $a_i \overline{\Phi}(t; t^*_0) + b_i$ then must preserve the strict single-crossing property of $a_i\Psi(t)+b_i$ for all $i \in I^+ \backslash I^*$. 

Now, simply define the allocation rule $x^*(\,\cdot\,)$ as: for all $t \in \mathcal{I}$, $x^*(t) = x^\dagger \in X^*$, and for all $t \not \in \mathcal{I}$, $x^*_i(t) = 1$ for all $i \in I^* \cup I^-$ and $x^*_i(t) = \mathbbm{1}\{t \geq t^*_i\}$ for all $i \in I^+ \backslash I^*$. By the above argument, $x^*(\,\cdot\,)$ solves \eqref{eq:unconstrained}. Note that $\sum_i a_i x^*_i(t)$ is nondecreasing since we keep adding strictly positive goods as we move from $t = 0$ to $t = 1$. Moreover, it is a \textit{\textbf{consistent}} solution with respect to ironing intervals. Together, by \Cref{lem:ironing}, these imply that the constructed solution $x^*$ solves 
\[\max_{x \in \text{MON}} \mathbb{E}\Bigg[\sum_i \Big(a_i x_i(t) \Phi(t; t^*_0) + b_i x_i(t)\Big)\Bigg]\,.\]
Now, we verify that $t^*_0$ must be a worst-off type given $x^*$, which then implies that it solves 
\[\min_{t_0 \in [0, 1]} \mathbb{E}\Bigg[\sum_i \Big(a_i x^*_i(t) \Phi(t; t_0) + b_i x^*_i(t)\Big)\Bigg]\,.\]
But that is clear by construction: $U(t) = 0$ for all $t \in \mathcal{I}$ by construction, and $U(t) = \int_{t^*_0}^t \sum_i a_i x^*_i(z) d z \geq 0$ for all $t \not\in \mathcal{I}$ since $\sum_i a_i x^*_i(\,\cdot\,) \geq 0$ by construction.  

Therefore, we have found a saddle point $(x^*, t^*_0)$. As argued in \Cref{sec:proof}, this implies $x^*$ is optimal and that any optimal mechanism $x$ also forms a saddle point with $t_0^*$.     
\end{proof}

\Cref{lem:saddlepoint} directly yields a full characterization of optimal mechanisms. For any given mechanism, let $U(t)$ denote the indirect utility of the type-$t$ buyer.

\begin{lemma}\label{lem:optx}
There exists a threshold $\bar{t}_0 > 0$, and unique thresholds $t^*_i \in ( \bar{t}_0,1)$ for all $i \in I^+\backslash I^*$ such that for any optimal mechanism $\mathcal{M}^{DR}$, up to a measure-zero set of types, we have
    \begin{itemize}
        \item[(i)] $x(t) \in X^*$ and $U(t) = 0$ for all $t \in [0, \bar{t}_0]$\,; 
        \item[(ii)] For all $i \in I^{*}$, $x_i(t) = 1$ for all $t \in (\bar{t}_0, 1]$\,; 
        \item[(iii)] For all $i \in I^{-}$, $x_i(t) = 1$ for all $t \in [0, 1]$\,;
        \item [(iv)] For all $i \in I^{+}\backslash I^*$, $x_i(t) = 1_{t \geq t^*_i}$ for all $t \in [0, 1]$\,. 
        \end{itemize}
\end{lemma}

\begin{proof}
Consider the saddle point $(x^*, t^*_0)$ constructed in \Cref{lem:saddlepoint}. Now, let 
\[\overline{t}_0 := \sup \,\mathcal{I}\,,\]
and let 
\[t^*_i := \inf\big\{t: a_i \overline{\Phi}(t;t^*_0) + b_i > 0\big\}\]
for all $i \in I^+ \backslash I^*$. By construction, for such $i$ 
\[a_i \overline{\Phi}(\overline{t}_0;t^*_0) + b_i = a_i \lambda + b_i < 0 \,,\]
and hence $t^*_i > \overline{t}_0$. Furthermore, by \Cref{lem:regular} and the fact that $t = 1$ cannot be in an ironing interval, we have 
\[a_i \overline{\Phi}(1;t^*_0) + b_i = a_i + b_i > 0\,,\]
and hence $t_i^*<1$. 

Since every optimal $x$ must form a saddle point with $t^*_0$, by \Cref{lem:ironing}, they must solve the ironed, unconstrained problem \eqref{eq:unconstrained} in a way such that $t^*_0$ is a worst-off type. In particular, this implies that for \textit{every} optimal $x$, we must have 
\[\sum_i a_i x_i(t) = 0 \]
for all $t$ in the interior of $\mathcal{I}$, i.e., $(0, \overline{t}_0)$. To see this, note that, by \Cref{lem:ironing}, $x$ must also be consistent with respect to the ironing interval $\mathcal{I}$ and hence  
\[\sum_i a_i x_i(t) = \sum_i a_i x_i(t^*_0) \]
for all $t$ in the interior of $\mathcal{I}$, which, combined with the fact that $t^*_0$ is a worst-off type, implies that 
\[\sum_i a_i x_i(t) = \sum_i a_i x_i(t^*_0) = 0\,,\]
for all $t$ in the interior of $\mathcal{I}$. As a consequence, for almost all $t \in (0, \overline{t}_0)$, every $x(t)$ must be maximizing \eqref{eq:pointwise-lambda-ap} in a way such that $\sum_i a_i x_i(t) = 0$, which happens, by construction, if and only if $x(t) = \hat{x}$ for some $\hat{x} \in X^*$ given that $\lambda$ is an optimal dual multiplier of \eqref{eq:auxiliary}. Therefore, any optimal $x$ must satisfy part \textit{(i)}. 

Now, for the types $t \not \in \mathcal{I}$, note that the pointwise maximization problem \eqref{eq:unconstrained} in fact has a unique solution almost everywhere by inspecting our previous inequalities in \Cref{lem:saddlepoint}. Thus, by the proof of  \Cref{lem:saddlepoint}, we have that any optimal mechanism must satisfy parts \textit{(ii)} to \textit{(iv)} (up to a measure-zero set of types). In particular, the thresholds $t^*_i$ must be unique.
\end{proof}

\paragraph{Optimal Learning.}\hspace{-2mm}We start with showing that, all else equal, the buyer strictly benefits from a more dispersed type distribution  whenever his indirect utility from the mechanism is not affine. 

\begin{lemma}\label{lem:strictJensen}
    Let $U$ be a piecewise-linear convex function, and let $t$ be a random variable whose support $T$ is an interval. Let $t'=t+\varepsilon$ be  a strict mean-preserving spread in the sense that $\mathbb{E}[\varepsilon\mid t]=0$ and, for any $t$ in the interior of $T$, the conditional distribution of $\varepsilon$ given $t$ has a density that is strictly positive on $(-r(t), r(t))$, with $r(\,\cdot\,)>0$ continuous on the interior of $T$. If $U$ is not affine on $T$, then $\mathbb{E}[U(t')]>\mathbb{E}[U(t)]$.
\end{lemma}

\begin{proof}
    By Jensen's inequality, $\mathbb{E}[U(t')\mid t] \geq U(t)$ for every $t$, so it suffices to find a set of $t$ of positive probability on which the inequality is strict. Since $U$ is not affine on the interval $T$, it has a kink $t_1$ in the interior of $T$. Pick a compact interval $[a,b]$ in the interior of $T$ with $a<t_1<b$, and let $\eta := \min_{t \in [a,b]} r(t) > 0$ and $\delta:=\min\{\eta/2,\, t_1-a\}$. For any $t\in (t_1 - \delta, \, t_1)$, we have $t\in[a,b]$ and $0<t_1-t<r(t)$, so the event $\{\varepsilon > t_1-t\}$ has positive probability given $t$. On this event, $t'$ lies strictly beyond the kink, where $U$ lies strictly above the supporting line of $U$ at $t$; hence $\mathbb{E}[U(t') \mid t] > U(t)$. Since $T$ is the support of $t$ and $(t_1-\delta,\, t_1)\subset T$ is open, it has positive probability, and the claim follows.
\end{proof}

We now show two properties of optimal learning strategies. 

\begin{lemma}\label{lem:learnzero}
Let $N<K$. Consider the following optimization problem:
\[
\max_{\boldsymbol{\alpha}}\mathbb{E}\bigl[\,U\bigl(\theta_{1}(s;\boldsymbol\alpha),\dots,\theta_{N}(s;\boldsymbol\alpha)\bigr)\bigr]\,,
\]
where $U$ is a piecewise-linear convex function. Then, every optimal solution $\boldsymbol{\alpha^{*}}$ must put zero weight on 
$(v_{N+1},\dots,v_{K})$ unless $\boldsymbol{\alpha^{*}} = \boldsymbol{0}$ is optimal.     
\end{lemma}

\begin{proof}
Assume that $\boldsymbol{0}$ is suboptimal. Suppose for contradiction that $\boldsymbol\alpha$ is optimal and that it puts non-zero weights on $(v_{N+1},\dots, v_K)$. Since $\boldsymbol{0}$ is suboptimal, 
\[\mathbb{E}\bigl[\,U\bigl(\theta_{1}(s;\boldsymbol\alpha),\dots,\theta_{N}(s;\boldsymbol\alpha)\bigr)\bigr] > \mathbb{E}\bigl[\,U\bigl(\mu_1,\dots, \mu_N\bigr)\bigr]\,,\]
and thus along the posterior mean line generated by $\boldsymbol\alpha$, the function $U$ is not affine. 

Note that
\[
\boldsymbol{\alpha}\cdot\mathbf{v}
=\sum_{i\le N}\alpha_{i}v_{i}+\sum_{j>N}\alpha_{j}v_{j}
=\sum_{i\le N}\tilde{\alpha}_{i}v_{i}+\varepsilon\,,
\]
for some $\boldsymbol{\tilde{\alpha}}$, where $\text{Cov}(v_{i},\varepsilon)=0$ for all $i\le N$, and $\varepsilon$ is a non-degenerate elliptical random variable. Indeed, by the linear‐projection property, we can write for each $j>N$ 
\[
v_{j}
=\sum_{i\le N}\beta_{i}v_{i}+\varepsilon_{j}\,,
\]
with $\text{Cov}(v_{i},\varepsilon_{j})=0$ for all $i\le N$. Note that $\boldsymbol{\tilde{\alpha}} \neq \boldsymbol{0}$ by the previous observation. 

Consider the signal structure 
\[
\boldsymbol\alpha^{*}=(\tilde{\alpha}_{1},\dots,\tilde{\alpha}_{N},0,\dots,0)\,.
\]
Under the original $\boldsymbol\alpha$, each posterior mean is
\[
\theta_{i}(s;\boldsymbol\alpha)=a_{i}s+b_{i},
\quad
a_{i}
=\frac{\text{Cov}\bigl(v_{i},\sum_{k\le N}\tilde{\alpha}_{k}v_{k}\bigr)}
{\text{Var}(\boldsymbol{\alpha}\cdot\mathbf{v})}
=\frac{\sum_{k\le N}\tilde{\alpha}_{k}\,\text{Cov}(v_{i},v_{k})}
{\text{Var}(\boldsymbol{\alpha}\cdot\mathbf{v})}\,,
\]
whereas under $\boldsymbol\alpha^{*}$,
\[
\theta_{i}^{*}(s^*,\boldsymbol\alpha^*)=a_{i}^{*}s^{*}+b_{i}^{*},
\quad
a_{i}^{*}
=\frac{\text{Cov}\bigl(v_{i},\sum_{k\le N}\tilde{\alpha}_{k}v_{k}\bigr)}
{\text{Var}(\boldsymbol{\alpha^*}\cdot\mathbf{v})}
=\frac{\sum_{k\le N}\tilde{\alpha}_{k}\,\text{Cov}(v_{i},v_{k})}
{\text{Var}(\boldsymbol{\alpha^*}\cdot\mathbf{v})}
= a_{i}\,\frac{\text{Var}(\boldsymbol{\alpha}\cdot\mathbf{v})}{\text{Var}(\boldsymbol{\alpha^*}\cdot\mathbf{v})}\,.
\]

Clearly the posterior lines 
\[
\bigl(\theta_{1}(s;\boldsymbol\alpha),\dots,\theta_{N}(s;\boldsymbol\alpha)\bigr)
\quad\text{and}\quad
\bigl(\theta_{1}(s^*;\boldsymbol\alpha^*),\dots,\theta_{N}(s^*;\boldsymbol\alpha^*)\bigr)
\]
in $\mathbb{R}^{N}$ both pass through the posterior mean $(\mu_{1},\dots,\mu_{N})$, and their directional derivatives are
\[
\partial_{s}\,\theta(s)=(a_{1},\dots,a_{N}),
\qquad
\partial_{s^{*}}\,\theta^{*}(s^{*})
=\frac{\text{Var}(\boldsymbol{\alpha}\cdot\mathbf{v})}{\text{Var}(\boldsymbol{\alpha}^*\cdot\mathbf{v})}(a_{1},\dots,a_{N})\,,
\]
so the two lines are collinear.  

Leveraging this collinearity, we can write for each $i\leq N$: 
\[\theta_{i}^{*}(s^*,\boldsymbol\alpha^*)=a_{i}\,\frac{\text{Var}(\boldsymbol{\alpha}\cdot\mathbf{v})}{\text{Var}(\boldsymbol{\alpha^*}\cdot\mathbf{v})}s^*+b_i^* = a_{i}\hat{s}^*+b_i\,,\]
where 
\[\hat{s}^* := \frac{\text{Var}(\boldsymbol{\alpha}\cdot\mathbf{v})}{\text{Var}(\boldsymbol{\alpha}^*\cdot\mathbf{v})}s^* + \Big(\boldsymbol{\alpha} \cdot \boldsymbol{\mu} -  \frac{\text{Var}(\boldsymbol{\alpha}\cdot\mathbf{v})}{\text{Var}(\boldsymbol{\alpha}^*\cdot\mathbf{v})} \boldsymbol{\alpha}^* \cdot \boldsymbol{\mu}\Big) = \frac{\text{Var}(\boldsymbol{\alpha}\cdot\mathbf{v})}{\text{Var}(\boldsymbol{\alpha^*}\cdot\mathbf{v})}s^* +  \frac{b_i^*-b_i}{a_i}\,. \]
In the above, for any $i\leq N$ with $a_i\neq 0$ (such an $i$ exists, since otherwise $\theta_1,\dots,\theta_N$ would be constant and $\boldsymbol\alpha$ could not outperform $\boldsymbol{0}$), we have used that $b_i = \mu_i-a_i (\boldsymbol{\alpha}\cdot\boldsymbol{\mu})$, $b_i^* = \mu_i - a_i^*(\boldsymbol{\alpha}^*\cdot\boldsymbol{\mu})$, and $a_i^*/a_i=\text{Var}(\boldsymbol{\alpha}\cdot\mathbf{v})/\text{Var}(\boldsymbol{\alpha}^*\cdot\mathbf{v})$.

Then, 
\[\mathbb{E}\bigl[\,U\bigl(\theta_{1}(s^*;\boldsymbol\alpha^*),\dots,\theta_{N}(s^*;\boldsymbol\alpha^*)\bigr)\bigr]=\mathbb{E}\bigl[\,U\bigl(\theta_{1}(\hat{s}^*;\boldsymbol\alpha),\dots,\theta_{N}(\hat{s}^*;\boldsymbol\alpha)\bigr)\bigr]\,.\]
To reach a contradiction, we have left to show that 
\[\mathbb{E}\bigl[\,U\bigl(\theta_{1}(\hat{s}^*;\boldsymbol\alpha),\dots,\theta_{N}(\hat{s}^*;\boldsymbol\alpha)\bigr)\bigr]>\mathbb{E}\bigl[\,U\bigl(\theta_{1}(s;\boldsymbol\alpha),\dots,\theta_{N}(s;\boldsymbol\alpha)\bigr)\bigr]\,.\]
Recall that 
\[s = \boldsymbol{\alpha}\cdot v = \sum_{i\leq N}\tilde{\alpha}_i v_i + \varepsilon = s^* + \varepsilon\,.\]
Thus, 
\[\hat{s}^* = \frac{\text{Var}(\boldsymbol{\alpha}\cdot\mathbf{v})}{\text{Var}(\boldsymbol{\alpha^*}\cdot\mathbf{v})}(s-\varepsilon) + \underbrace{\Big(\boldsymbol{\alpha} \cdot \boldsymbol{\mu} -  \frac{\text{Var}(\boldsymbol{\alpha}\cdot\mathbf{v})}{\text{Var}(\boldsymbol{\alpha}^*\cdot\mathbf{v})} \boldsymbol{\alpha}^* \cdot \boldsymbol{\mu}\Big)}_{=:Z}\,.\]
Note that  
\begin{align*}
    \mathbb{E}[\hat{s}^*\mid s] &= \mathbb{E}\left[\frac{\text{Var}(\boldsymbol{\alpha}\cdot\mathbf{v})}{\text{Var}(\boldsymbol{\alpha^*}\cdot\mathbf{v})}s^* + \frac{b_i^*-b_i}{a_i}\mid s\right] = \frac{\text{Var}(\boldsymbol{\alpha}\cdot\mathbf{v})}{\text{Var}(\boldsymbol{\alpha^*}\cdot\mathbf{v})}\mathbb{E}[s^*\mid s] + Z\\
    &=\frac{\text{Var}(\boldsymbol{\alpha}\cdot\mathbf{v})}{\text{Var}(\boldsymbol{\alpha^*}\cdot\mathbf{v})}\left(\boldsymbol{\alpha^*}\cdot \boldsymbol{\mu}+\frac{\text{Cov}(\boldsymbol{\alpha^*}\cdot \mathbf{v}, \boldsymbol{\alpha}\cdot \mathbf{v})}{\text{Var}(\boldsymbol{\alpha}\cdot \mathbf{v})}(s-\boldsymbol{\alpha}\cdot \boldsymbol{\mu})\right) + Z\\
    &=\frac{\text{Cov}(\boldsymbol{\alpha^*}\cdot \mathbf{v}, \boldsymbol{\alpha}\cdot \mathbf{v})}{\text{Var}(\boldsymbol{\alpha^*}\cdot\mathbf{v})}s +\frac{\text{Var}(\boldsymbol{\alpha}\cdot\mathbf{v})}{\text{Var}(\boldsymbol{\alpha^*}\cdot\mathbf{v})}\boldsymbol{\alpha^*}\cdot \boldsymbol{\mu}-\frac{\text{Cov}(\boldsymbol{\alpha^*}\cdot \mathbf{v}, \boldsymbol{\alpha}\cdot \mathbf{v})}{\text{Var}(\boldsymbol{\alpha^*}\cdot\mathbf{v})}\boldsymbol{\alpha}\cdot \boldsymbol{\mu} + Z\\
     &=s +\frac{\text{Var}(\boldsymbol{\alpha}\cdot\mathbf{v})}{\text{Var}(\boldsymbol{\alpha}^*\cdot\mathbf{v})}\boldsymbol{\alpha^*}\cdot \boldsymbol{\mu}-\boldsymbol{\alpha}\cdot \boldsymbol{\mu} + Z\\
    &= s\,,
\end{align*}
since $\text{Cov}(\boldsymbol{\alpha^*}\cdot \mathbf{v}, \boldsymbol{\alpha}\cdot \mathbf{v}) = \text{Cov}(\boldsymbol{\alpha^*}\cdot \mathbf{v}, \boldsymbol{\alpha^*}\cdot \mathbf{v}+\varepsilon) = \text{Var}(\boldsymbol{\alpha^*}\cdot\mathbf{v})$. Moreover, 
\begin{align*}
    \text{Var}(\hat{s}^*\mid s) &= \left(\frac{\text{Var}(\boldsymbol{\alpha}\cdot\mathbf{v})}{\text{Var}(\boldsymbol{\alpha^*}\cdot\mathbf{v})}\right)^2\text{Var}(\varepsilon\mid s) = \left(\frac{\text{Var}(\boldsymbol{\alpha}\cdot\mathbf{v})}{\text{Var}(\boldsymbol{\alpha^*}\cdot\mathbf{v})}\right)^2\kappa(s)\left[\text{Var}(\varepsilon) - \frac{\text{Cov}(\varepsilon, \boldsymbol\alpha\cdot \mathbf{v})^2}{\text{Var}(\boldsymbol\alpha\cdot \mathbf{v}) }\right]\\
    &= \left(\frac{\text{Var}(\boldsymbol{\alpha}\cdot\mathbf{v})}{\text{Var}(\boldsymbol{\alpha^*}\cdot\mathbf{v})}\right)^2\kappa(s)\text{Var}(\varepsilon)\left(1-\frac{\text{Var}(\varepsilon)}{\text{Var}(\boldsymbol\alpha\cdot \mathbf{v})}\right)>0\,,
\end{align*}
where the scaling factor $\kappa(s)>0$ for all $s$ in the interior of the support and is continuous in $s$ (\citealt*{gupta2013elliptically}). 

Thus, $\hat{s}^*$ is a strict mean-preserving spread of $s$---it can be written as $\hat{s}^*=s+\hat{\varepsilon}$ where $\mathbb{E}[\hat{\varepsilon}\mid s]=0$ and $\hat{\varepsilon}$ satisfies the hypothesis of \Cref{lem:strictJensen}.  By \Cref{lem:strictJensen}, this implies
\[\mathbb{E}\bigl[\,U\bigl(\theta_{1}(\hat{s}^*;\boldsymbol\alpha),\dots,\theta_{N}(\hat{s}^*;\boldsymbol\alpha)\bigr)\bigr]>\mathbb{E}\bigl[\,U\bigl(\theta_{1}(s;\boldsymbol\alpha),\dots,\theta_{N}(s;\boldsymbol\alpha)\bigr)\bigr]\,,\]
contradicting the optimality of $\boldsymbol\alpha$. Hence, every optimal $\boldsymbol\alpha^{*}$ must place zero weight on $(v_{N+1},\dots,v_{K})$.
\end{proof}

\begin{lemma}\label{lem:learnsame}
Let $N<K$. Consider the following optimization problem
\[
\max_{ \boldsymbol \alpha}\mathbb{E}\!\Bigl[\,U\Bigl(\theta_{1}(s;\boldsymbol\alpha),\dots,\theta_{N-1}(s;\boldsymbol\alpha),\sum_{j\ge N}\theta_{j}(s;\boldsymbol\alpha)\Bigr)\Bigr]\,,
\]
where $U$ is a piecewise-linear convex function. Then, every optimal solution $\boldsymbol \alpha^*$ assigns equal weights to each of $v_N,\dots,v_K$ unless $\boldsymbol{\alpha^{*}} = \boldsymbol{0}$ is optimal.  
\end{lemma} 
\begin{proof}
Consider the elliptical random vector
\[
\bigl(v_1,\dots,v_{N-1},\,w_N,\,v_N,\dots,v_K\bigr),
\qquad
w_N \;=\;\sum_{j\ge N}v_j\,.
\]
Clearly any linear signal in the original space is equivalent to one in this $(K+1)$-dimensional space, and vice versa.  By \Cref{lem:learnzero} (applied with dimensions $N<K+1$) to this augmented vector, any optimal signal must be equivalent to some signal that puts zero weights on each of $v_N,\dots,v_K$, unless $\boldsymbol{0}$ is optimal (in particular, note that the constructed $\varepsilon$ in \Cref{lem:learnzero} is again non-degenerate here). Therefore, any optimal signal must be equivalent to one that puts non-zero weights only on $v_1,\dots,v_{N-1},w_N$, which implies that in the original coordinates it assigns the same weights to each of $v_N,\dots,v_K$, unless $\boldsymbol{0}$ is optimal.  This completes the proof.
\end{proof}

We next show that in every horizontal learning equilibrium, information must be strictly valuable:
\begin{lemma}\label{lem:infovaluable}
    For any horizontal learning strategy $\boldsymbol{\alpha}$ and any mechanism $\mathcal{M}$ that is optimal against the associated type distribution, information is strictly valuable under $\mathcal{M}$, i.e.,  $\tilde{{\boldsymbol{\alpha}}} = \mathbf{0}$ is strictly suboptimal against $\mathcal{M}$. 
\end{lemma}

\begin{proof}
Fix any such $(\boldsymbol{\alpha},\mathcal{M})$. We denote the posterior mean distribution under $\boldsymbol{\alpha}$ by $(a_i t + b_i)_{i} \in \mathbb{R}^{K}_+$ following the previous notation. Suppose for contradiction that $\tilde{{\boldsymbol{\alpha}}} = \mathbf{0}$ is optimal against $\mathcal{M}$.

We consider two cases. Case \textit{(i)}: the indirect utility function $U(t)$ of the types $t$ is not affine on $[0, 1]$. By \Cref{lem:optx}, we know that $U$ must be a convex (piecewise linear) function. But then given that type $t$ has an elliptical distribution with full-support, by the proof of \Cref{lem:strictJensen}, we have
\[\int U(t) d F(t) > U\Bigg(\int t d F(t)\Bigg) = U\Big(\frac{1}{2}\Big)\,.\]
Note that, by construction, type $t = \frac{1}{2}$ has posterior mean given by ${\boldsymbol{\theta}}(t) = \boldsymbol{\mu}$. Therefore, by definition, 
\[U\Big(\frac{1}{2}\Big)\]
is the buyer's (ex ante) payoff under strategy $\tilde{{\boldsymbol{\alpha}}} = \mathbf{0}$. Since $\int U(t) d F(t)$ is the buyer's (ex ante) payoff under strategy $\boldsymbol{\alpha}$, it follows immediately that $\tilde{{\boldsymbol{\alpha}}} = \mathbf{0}$ cannot be optimal. 

Now, consider case \textit{(ii)}: $U$ is affine on $[0, 1]$. Since, by \Cref{lem:optx}, $U(t) = 0$ for all $t \in [0, \bar{t}_0]$ with $\bar{t}_0 > 0$, it must be that for all types $t$, we have
\[U(t) = 0\,.\]
In particular, this holds for $t = \frac{1}{2}$ and hence the buyer's (ex ante) payoff under strategy $\tilde{{\boldsymbol{\alpha}}} = \mathbf{0}$ must be $0$. 
Note that the seller cannot offer $x(t) = 0$ to all types $t$ since that would imply $0$ revenue, while the seller can clearly secure a strictly positive revenue (by even selling one good).
Then, by \Cref{lem:optx}, we know that there must exist some type $t$ such that $x(t) = x^*$ for some non-zero $x^* \in X^*$ and $p(t) = \sum_i b_i x^*_i$. Since $\sum_i a_i x^*_i = 0$, we know that $p(t) = \sum_i b_i x^*_i = \sum_i \mu_i x^*_i$\,. Now consider the strategy $\hat{\boldsymbol{\alpha}} = x^*$. Note that the buyer's (ex ante) payoff from this strategy must be bounded from below by 
\[\int \max\Big\{0, s - p^* \Big\} d G(s)> 0\,,\]
where $G$ is the distribution of $\boldsymbol{v} \cdot x^*$ which is non-degenerate, and $p^* = \sum_i \mu_i x^*_i = \E[\boldsymbol{v} \cdot x^*]$. But then $\tilde{{\boldsymbol{\alpha}}} = \mathbf{0}$ cannot be optimal, since it gives $0$ payoff. 
\end{proof}
Now, we exploit the characterization of the optimal mechanism in \Cref{lem:optx} to further pin down the buyer's learning in equilibrium: 
\begin{lemma}\label{lem:learnfinal}
 Let $(\boldsymbol{\alpha},\mathcal{M})$ be an equilibrium that exhibits horizontal learning.  Then $\boldsymbol{\alpha}$ must put zero weights on all goods in $I^{*}\cup I^{-}$.
\end{lemma} 
\begin{proof}
Fix any horizontal learning equilibrium $(\boldsymbol{\alpha},\mathcal{M})$. We follow the same notation as before. As noted before, under any horizontal learning, we must have the set $I^* \neq \varnothing$. Let 
\[O:=\Big\{\big(x(t), p(t)\big)\Big\}_{t\in [0, 1]}\]
denote the minimal menu that implements the seller's optimal mechanism (which would give the same ex ante payoff to the buyer under strategy $\boldsymbol{\alpha}$). Note that as long as we can construct a profitable deviation for the buyer against this menu, then it must be a profitable deviation considering the other possible options offered by the seller. 

Now, note that by \Cref{lem:optx}, for any $(x, p) \in O$ such that $x_i < 1$ for some $i \in I^*$, we must have that 
\[x \in X^*\,,\]
and hence $(x, p) \in O$ yields zero payoff to every realized type $t$ under the equilibrium strategy. Thus, we may treat them as the outside option $0$ (in the deviation strategy we construct these options can only bring non-negative payoffs, and we bound them from below by $0$). Moreover, for any other $(x, p) \in O$, we must have 
\[x_i = 1 \text{ for all $i \in I^* \cup I^-$}\,.\]
By \Cref{lem:infovaluable}, information must be strictly valuable for the buyer against mechanism $\mathcal{M}$. Therefore, by \Cref{lem:learnsame}, it must be the case that $\boldsymbol{\alpha}$ puts equal weights on all goods in $I^{*}\cup I^{-}$, since otherwise the buyer has a profitable deviation of assigning equal weights on all goods in $I^{*}\cup I^{-}$. Indeed, by \Cref{lem:learnsame}, there exists one such strategy that results in a mean‐preserving spread on 
\[
\Bigl((\theta_{i})_{i\in I^{+} \backslash I^*},\sum_{j\in I^{*}\cup I^{-}}\theta_{j}\Bigr)
\]
that strictly improves the expected payoff. Thus, the buyer's strategy $\boldsymbol{\alpha}$ must put equal weights, say $\alpha_{c}$, on all goods in $I^{*}\cup I^{-}$. 

Note that if $\alpha_{c}=0$, then we are done. Otherwise, consider another signal
\[
\boldsymbol{\alpha}'=\Bigl((\tilde{\boldsymbol{\alpha}}_{i})_{i\in I^{+}},0,\dots,0\Bigr)
\]
where $\boldsymbol{\alpha}'$ modifies $\boldsymbol{\alpha}$ by changing the weights to $0$ for all the goods in $I^{-}$ and keeping the posterior‐mean line of $(\theta_{i})_{i\in I^{+}}$ in the space $\mathbb{R}^{|I^{+}|}$ collinear---such a signal exists by the proof of \Cref{lem:learnzero} and leads to a strict mean-preserving spread along the posterior‐mean line of $(\theta_{i})_{i\in I^{+}}$. Since every element $x^* \in X^*$ also has $x^*_i = 1$ for all $i \in I^-$, every option $(x, p)$ in $O$ satisfies that $x_i = 1$ for all $i \in I^-$. Therefore, for the buyer's decision problem from menu $O$, the negative goods are irrelevant. By the proof of \Cref{lem:learnzero}, this implies that $\boldsymbol{\alpha}'$ must be a strict improvement unless $\varepsilon=0$ in \Cref{lem:learnzero}, but that could only happen if
\[
\alpha_{c}\sum_{j\in I^{-}}\Bigg(v_{j}-\sum_{i\in I^{+}}\beta_{i}v_{i}\Bigg)=0\,.
\]
However, that is impossible since the random vector \[\text{$(\varepsilon_{j})_{j\in I^{-}}$,\, where $\varepsilon_{j}=v_{j}-\sum_{i\in I^{+}}\beta_{i}v_{i}$}\,,\]
is a full‐dimensional elliptical distribution by the assumption that $\rho\in(-1,1)$. This concludes the proof. 
\end{proof}

\paragraph{Completion of the Proof.}\hspace{-2mm}Suppose for contradiction that there exists a horizontal-learning equilibrium $(\boldsymbol{\alpha},\mathcal{M})$. As noted before, we must then have $I^{*}\neq\varnothing$ and $I^{-}\neq\varnothing$. By \Cref{lem:learnfinal}, it must be the case that for all $i \in I^* \cup I^-$, we have 
\[\alpha_i = 0\,.\]
Since the correlation $\rho$ is the same across all pairs of goods, this implies that $(a_{i})_{i\in I^* \cup I^-}$ must be either \textit{(i)} all weakly positive or \textit{(ii)} all weakly negative. Indeed, for any $i\in I^* \cup I^-$, 
\[\text{sign}(a_i) =\text{sign}(\text{Cov}(\boldsymbol{\alpha}\cdot\mathbf{v},v_i)) =\text{sign}\left(\rho\sum_{j\in I^+\setminus I^* }\alpha_j\sigma_j\right)\,,  \]
which does not depend on $i$.  However, by construction, $a_i > 0$ for all $i \in I^*$ and $a_i \leq 0$ for all $i \in I^{-}$. Moreover, since $(\boldsymbol{\alpha}, \mathcal{M})$ is a horizontal learning equilibrium, there exists some $i \in I^-$ such that $a_i < 0$. A contradiction.

\subsubsection{Nested Bundling}
By the previous results, we know that in every equilibrium, the buyer uses a vertical learning strategy. Thus, the posterior mean distribution can be written as: for each $i$,  
\[\theta_i = a_i t + b_i\]
where $a_i \geq 0$, $b_i \geq 0$, and $t \in [0, 1]$. Moreover, $(a_i, b_i) \neq (0, 0)$\footnote{If $\boldsymbol{\alpha} = 0$, then the seller would extract full surplus, but that cannot form an equilibrium.} and $I^*=\varnothing$. 

Note that in this case our characterization result continues to apply and becomes much simpler---by a similar argument as in \Cref{lem:optx}, the optimal direct-revelation mechanism is unique up to a measure-zero set of types and characterized by maximizing the standard virtual value function pointwise. Consequently, under the optimal mechanism, the type space can be partitioned into a finite number of intervals: $[0,1]=[t_0, t_1)\cup\dots\cup [t_{L-1}, t_L]$ such that all types $t\in [t_{l}, t_{l+1})$ get allocated a bundle $B_l\subseteq \{1,\dots, K\}$ at price $p_l$ with $B_l\subset B_{l+1}$ for all $l$. For the equilibrium menu $\mathcal{M}$ to be optimal, it must then include the bundles $B_0,\dots, B_{L-1}$ and only these bundles must be chosen with positive probability in equilibrium. 

Now, consider another strategy profile $(\boldsymbol\alpha, \mathcal{M}^{NB})$ where $\mathcal{M}^{NB}$ is a nested bundling mechanism with message space $M^{NB}=\{0,\dots, L-1\}$, allocation rule $x^{NB}_i(l) = \mathbbm{1}\{i\in B_l\}$ and payment rule $p^{NB}(l) = p_l$, where $p_l$ is the price of $B_l$ in menu $\mathcal{M}$. Fixing $\boldsymbol\alpha$, the mechanism $\mathcal{M}^{NB}$ induces the exact same revenue as $\mathcal{M}$, and so is optimal. It remains to argue that learning strategy $\boldsymbol\alpha$ is a best response to $\mathcal{M}^{NB}$. What matters for the buyer is not the label of the message sent to the seller, but the induced allocation and payment. Let $O^{NB}=\bigcup_{m\in M^{NB}}(x^{NB}(m), p^{NB}(m))$ and $O = \bigcup_{m\in M}(x(m), p(m))$ be the set of outcomes that can be induced under mechanisms $\mathcal{M}^{NB}$ and $\mathcal{M}$, respectively. By contradiction, suppose that $\boldsymbol\alpha$ is not a best response. This means that there exists $\boldsymbol\alpha'$ such that 
\begin{align*}
    \mathbb{E}_{\boldsymbol\alpha'}\left[\max_{(x,p)\in O^{NB}}\sum_i \theta_i(s';\boldsymbol\alpha' )x_i - p\right]> \mathbb{E}_{\boldsymbol\alpha}\left[\max_{(x,p)\in O^{NB}}\sum_i \theta_i(s;\boldsymbol\alpha )x_i - p\right]\,.
\end{align*}
However, \[\mathbb{E}_{\boldsymbol\alpha}\left[\max_{(x,p)\in O^{NB}}\sum_i \theta_i(s;\boldsymbol\alpha )x_i - p\right]=\mathbb{E}_{\boldsymbol\alpha}\left[\max_{(x,p)\in O}\sum_i \theta_i(s;\boldsymbol\alpha )x_i - p\right]\]
by construction. Furthermore, 
\begin{align*}
     \mathbb{E}_{\boldsymbol\alpha'}\left[\max_{(x,p)\in O}\sum_i \theta_i(s';\boldsymbol\alpha' )x_i - p\right]\geq  \mathbb{E}_{\boldsymbol\alpha'}\left[\max_{(x,p)\in O^{NB}}\sum_i \theta_i(s';\boldsymbol\alpha' )x_i - p\right]\,,
\end{align*}
since $O^{NB}\subseteq O$. Therefore, we have 
\[ \mathbb{E}_{\boldsymbol\alpha'}\left[\max_{(x,p)\in O}\sum_i \theta_i(s';\boldsymbol\alpha' )x_i - p\right]> \mathbb{E}_{\boldsymbol\alpha}\left[\max_{(x,p)\in O}\sum_i \theta_i(s;\boldsymbol\alpha )x_i - p\right]\,,\]
and hence $\boldsymbol\alpha'$ is also a profitable deviation under $\mathcal{M}$, which contradicts the initial assumption that $(\boldsymbol\alpha,\mathcal{M})$ is an equilibrium.

Thus, $(\boldsymbol\alpha, \mathcal{M}^{NB})$ is also an equilibrium. Therefore, any equilibrium is outcome-equivalent to a nested bundling equilibrium.

\subsection{Proof of \Cref{prop:ordering}}

Let $(\boldsymbol\alpha, \mathcal{M})$ be a nested-bundling equilibrium. Suppose for contradiction that there exist some goods $i, j$ where $\text{tier}(i) \leq \text{tier}(j)$ and 
\[\text{Var}(\log(\theta_i)) > \text{Var}(\log(\theta_j))\,.\]
By \Cref{thm:main}, the equilibrium learning strategy is vertical. As shown before, there exists a type parameterization $t \in \mathbb{R}$ such that for all goods $k$,
\[\theta_k  = a_k t + b_k\,,\]
with $a_k \geq 0$. Clearly, we have a contradiction if $a_i = 0$. Moreover, if $a_j = 0$, then it implies that $a_i = 0$ by the proof of \Cref{thm:main}. Hence, assume $a_i, a_j > 0$. Now, write 
\[\text{Var}(\log(\theta_k)) = \text{Var}\Big[\log\Big(t + \frac{b_k}{a_k}\Big) + \log a_k \Big] = \text{Var}\Big[\log\Big(t + \frac{b_k}{a_k}\Big) \Big]\,.\]
By the proof of \Cref{thm:main}, since item $j$ has a weakly higher tier than item $i$, there exist some $t_1 \leq t_2$ and some nondecreasing function $\Phi^*$ such that 
\[a_i\Phi^*(t_2) + b_i \geq 0\,,\text{ and } a_j\Phi^*(t_2) + b_j \geq  0\,;\]
\[a_i\Phi^*(t_1) + b_i \geq 0\,,\text{ and } a_j\Phi^*(t_1) + b_j \leq 0\,,\]
which implies that 
\[\frac{b_i}{a_i} \geq -\Phi^*(t_1) \geq \frac{b_j}{a_j}\,.\]
Therefore, we can write the random variable 
\[t + \frac{b_i}{a_i} = t + \frac{b_j}{a_j} + \underbrace{(\frac{b_i}{a_i} - \frac{b_j}{a_j})}_{\geq 0}\,,\]
and hence\footnote{For this claim, see e.g. Lemma 2 of \citet*{yang2023comparison}.} 
\[\log\Big(t + \frac{b_i}{a_i}\Big)  \preceq_{\text{disp}} \log\Big(t + \frac{b_j}{a_j}\Big)\]
where $\preceq_{\text{disp}}$ is the \textit{\textbf{dispersive order}}, i.e., $X \preceq_{\text{disp}} Y$ if $F^{-1}_X(q') - F^{-1}_X(q) \leq F^{-1}_Y(q') - F^{-1}_Y(q)$ for all $q < q'$, where $F_X, F_Y$ are the CDFs (\citealt{muller2002comparison}). Since the variance operator respects the dispersive order, we immediately have that 
\[\text{Var}\Big[\log\Big(t + \frac{b_i}{a_i}\Big)\Big]  \leq \text{Var}\Big[\log\Big(t + \frac{b_j}{a_j}\Big)\Big]\,,\]
which is a contradiction. 

Now, note that for any good $k$, we have the following 
\[\theta_k(s;\boldsymbol{\alpha}) = \mu_k + \frac{\text{Cov}(v_k,\boldsymbol{\alpha}\cdot\mathbf{v})}{\text{Var}(\boldsymbol{\alpha}\cdot\mathbf{v})}[s-\boldsymbol\alpha\cdot\boldsymbol\mu]\,.\]
Therefore, 
\[\theta_k(s;\boldsymbol{\alpha}) = \frac{\text{Cov}(v_k,\boldsymbol{\alpha}\cdot\mathbf{v})}{\text{Var}(\boldsymbol{\alpha}\cdot\mathbf{v})} s + \Bigg(\mu_k - \frac{\text{Cov}(v_k,\boldsymbol{\alpha}\cdot\mathbf{v})}{\text{Var}(\boldsymbol{\alpha}\cdot\mathbf{v})} \boldsymbol{\alpha} \cdot \boldsymbol{\mu}\Bigg)  \,.\]
Write 
\[\theta_k(s;\boldsymbol{\alpha}) = a_k s + b_k\]
where 
\[a_k = \frac{\text{Cov}(v_k,\boldsymbol{\alpha}\cdot\mathbf{v})}{\text{Var}(\boldsymbol{\alpha}\cdot\mathbf{v})} , \qquad b_k = \mu_k - \frac{\text{Cov}(v_k,\boldsymbol{\alpha}\cdot\mathbf{v})}{\text{Var}(\boldsymbol{\alpha}\cdot\mathbf{v})} \boldsymbol{\alpha} \cdot \boldsymbol{\mu}  \,.\]
As argued before, we can assume $a_i, a_j > 0$. Thus, for $k = i, j$, we can write 
\[\frac{b_k}{a_k} = \frac{\text{Var}(\boldsymbol{\alpha}\cdot\mathbf{v})}{\text{Cov}(v_k,\boldsymbol{\alpha}\cdot\mathbf{v})} \mu_k - \boldsymbol{\alpha} \cdot \boldsymbol{\mu}\,.\]
Thus, by the previous argument, it follows that 
\[\text{Var}(\log(\theta_i)) \leq  \text{Var}(\log(\theta_j)) \iff \frac{b_i}{a_i} \geq \frac{b_j}{a_j} \iff \text{Cov}(v_i/\mu_i,\boldsymbol{\alpha}\cdot\mathbf{v}) \leq \text{Cov}(v_j/\mu_j,\boldsymbol{\alpha}\cdot\mathbf{v}) \,.\]
Moreover, note that 
\[\text{Cov}(v_i,\boldsymbol{\alpha}\cdot\mathbf{v}) =  \alpha_i \sigma_i^2 + \alpha_j \rho \sigma_i \sigma_j + \text{Cov}\Big(v_i, \sum_{k\neq i, j} \alpha_k v_k\Big)\,.\]
\[\text{Cov}(v_j,\boldsymbol{\alpha}\cdot\mathbf{v}) =  \alpha_j \sigma_j^2 + \alpha_i \rho \sigma_i \sigma_j + \text{Cov}\Big(v_j, \sum_{k\neq i, j} \alpha_k v_k\Big)\,.\]
If $\rho = 0$, then these together imply that 
\[\big(\alpha_i \sigma^2_i/\mu_i -\alpha_j \sigma^2_j/\mu_j\big) \leq 0\,,\]
and hence 
\[ 0 \leq \alpha_i \sigma^2_i/\mu_i \leq \alpha_j \sigma^2_j/\mu_j\,,\]
where the first inequality is due to $a_i \geq 0$.

\subsection{Proof of \Cref{prop:existence}}

If $\mathbf{v}$ is exchangeable, then the condition in \Cref{prop:pure} holds and hence there exists a pure bundling equilibrium. 

Now, we prove equilibrium existence for $\rho$ high enough. We first state three lemmas, then prove the existence of a separate sales equilibrium given the lemmas, and finally prove the lemmas.\footnote{Focusing on separate sales simplifies the buyer's problem.} 

Throughout, we normalize learning weights to have unit length. Let $\boldsymbol{\alpha}(\boldsymbol{\varphi})$ denote the $\boldsymbol{\alpha} \in \R^K$ vector on the unit sphere $\mathbb{S}^{K-1}$ with spherical coordinates $\boldsymbol{\varphi}$. Let $\underline{\theta}_i:=\displaystyle\lim_{\rho\rightarrow 1}\min_{\mathbf{v}\in V}v_i$. 

The first lemma shows that it is without loss of optimality for the buyer to only consider learning weights with $\boldsymbol{\varphi} \in [0, \frac{\pi}{2}]^{K-1}$, whenever he faces a separate sales mechanism and $\rho\geq 0$. The second lemma shows that, once the correlation is sufficiently high, it is without loss of optimality for the seller to choose a separate sales mechanism with prices in $\boldsymbol{p}\in\prod_i[\underline{\theta}_i+\varepsilon, \mu_i]$ when the buyer's learning strategy is $\boldsymbol{\varphi} \in [0, \frac{\pi}{2}]^{K-1}$. Finally, the third lemma shows that the buyer's optimization problem becomes quasiconcave over $\boldsymbol{\varphi} \in [0, \frac{\pi}{2}]^{K-1}$ against any undominated separate sales mechanism, once the correlation is sufficiently high.

\begin{lemma}\label{lem:dominance}
If $\rho \geq 0$, then for any $\boldsymbol{\varphi}$, and any $\boldsymbol{p}$, there exists some $\boldsymbol{\varphi}' \in [0, \frac{\pi}{2}]^{K-1}$ such that $U(\boldsymbol{\alpha}(\boldsymbol{\varphi}')) \geq U(\boldsymbol{\alpha}(\boldsymbol{\varphi}))$, where 
\[ U(\boldsymbol{\alpha}(\boldsymbol{\varphi})) := \sum_i  \mathbb{E}\Big[ \max\big\{\theta_i(s;\boldsymbol{\alpha}(\boldsymbol{\varphi})\big) - p_i, 0\big\} \Big]\,.\]
\end{lemma}

\begin{lemma}\label{lem:pricerange}
There exist some $\varepsilon>0$ and $\underline{\rho} < 1$ such that for all $\rho \in [\underline{\rho}, 1)$, the following holds: for any separate sales prices $\boldsymbol{p}$, and any $\boldsymbol{\varphi}\in [0, \frac{\pi}{2}]^{K-1}$, there exists some $\boldsymbol{p}' \in \prod_i[\underline{\theta}_i+\varepsilon, \mu_i]$ such that $\boldsymbol{p}'$ leads to weakly higher expected revenue than $\boldsymbol{p}$.    
\end{lemma}

\begin{lemma}\label{lem:quasiconcave}
For any $\varepsilon'>0$, there exists some $\underline{\rho}' < 1$ such that for all $\rho \in [\underline{\rho}', 1)$, for all $\boldsymbol{p} \in \prod_i[\underline{\theta}_i+\varepsilon', \mu_i]$, we have that $U$ has a unique maximizer on $[0, \frac{\pi}{2}]^{K-1}$.
\end{lemma}

\paragraph{Proof of Equilibrium Existence.}\hspace{-2mm}Fix any $\rho \geq \max\{\underline{\rho}, \underline{\rho}', 0\}$, where $\underline{\rho}'$ is given by \Cref{lem:quasiconcave}, $\underline{\rho}$ is given by \Cref{lem:pricerange}. Let $\varepsilon$ also be that given by \Cref{lem:pricerange}.  Define the buyer's strategy space as $[0, \frac{\pi}{2}]^{K-1}$, with payoff function given by $U(\boldsymbol{\alpha}(\boldsymbol{\varphi}))$. Define the seller's strategy space as $\prod_{i} [\underline{\theta}_{i} + \varepsilon, \mu_i]$ with payoff function given by $\sum_i \E\big[p_i \mathbbm{1}\{\theta_i(s;\boldsymbol{\alpha}(\boldsymbol{\varphi})) \geq p_i\}\big]$. It is easy to verify that both the buyer's payoff and the seller's payoff are continuous in their joint action. By \Cref{lem:quasiconcave}, the buyer's best-reply correspondence $\text{BR}_B(\boldsymbol{p})$ is single-valued. By Berge's maximum theorem, $\text{BR}_B(\boldsymbol{p})$ is also continuous. By \Cref{lem:regular}, the seller has a unique best reply for any action of the buyer, and hence $\text{BR}_S(\boldsymbol{\varphi})$ is nonempty and single-valued. $\text{BR}_S(\boldsymbol{\varphi})$  is also continuous by Berge's theorem. Now, define the map: $\mathbf{B}(\boldsymbol{\varphi}, \boldsymbol{p}):= (\text{BR}_B(\boldsymbol{p}), \text{BR}_S(\boldsymbol{\varphi}))$. By Kakutani's fixed point theorem, $\mathbf{B}$ has a fixed point $(\boldsymbol{\varphi}^*, \boldsymbol{p}^*)$ such that $\boldsymbol{\varphi}^* \in \text{BR}_B(\boldsymbol{p}^*)$ and $\boldsymbol{p}^* \in \text{BR}_S(\boldsymbol{\varphi}^*)$. Thus, the defined game has an equilibrium. 

We now show that $(\boldsymbol{\varphi}^*, \boldsymbol{p}^*)$ forms a Nash equilibrium in the original game. Note that $\boldsymbol{\varphi}^*$ defines a vertical learning strategy, given that $\boldsymbol{\alpha}(\boldsymbol{\varphi}^*)\geq \mathbf{0}$ and $\rho \geq 0$. By the proof of \Cref{thm:main}, note that, by \Cref{lem:pricerange}, $\boldsymbol{p}^*$ as a separate sales mechanism is actually optimal against $\boldsymbol{\varphi}^*$ even if the seller can choose any mechanism.\footnote{In particular, by the proof of the nested bundling part of \Cref{thm:main}, under vertical learning, one can also implement the optimal direct-revelation mechanism as a separate sales mechanism.} Therefore, the seller has no profitable deviation  in the original game. Now, by \Cref{lem:dominance}, the buyer also has no profitable deviation since  $\boldsymbol{\varphi}^*$ must yield an optimal payoff for the buyer against $\boldsymbol{p}^*$ even if the buyer can choose any $\boldsymbol{\varphi}$. Thus, we have found an equilibrium. 

\subsubsection{Proof of \Cref{lem:dominance}}

\paragraph{Rewriting of the buyer's payoff.}\hspace{-2mm}We start by rewriting the buyer's expected payoff under any separate sales mechanism from any learning strategy $\boldsymbol{\alpha}$, in a way that highlights the geometry of the buyer's problem. 

Under separate sales, the buyer's payoff is separable across goods: 
\[U(\boldsymbol{\alpha}) = \mathbb{E}\left[\sum_i\left(\theta_i(s; \boldsymbol{\alpha})-p_i\right)_+\right]\]
with 
\[\boldsymbol{\theta}(s; \boldsymbol{\alpha}) = \boldsymbol{\mu} +\Sigma \boldsymbol{\alpha} \, (\boldsymbol{\alpha}^\top \Sigma\boldsymbol{\alpha})^{-1} \, \bigl(s - \boldsymbol{\alpha}^\top \boldsymbol{\mu}\bigr)\quad\text{and}\quad s = \boldsymbol{\alpha}^\top\mathbf{v}\,.\]
Instead of optimizing over $\boldsymbol{\alpha}$, it is equivalent to optimize over $\boldsymbol{a}:= \Sigma \boldsymbol{\alpha}$. Note that we can write 
\begin{align*}
    \boldsymbol{\mu} +\Sigma \boldsymbol{\alpha} \, (\boldsymbol{\alpha}^\top \Sigma\boldsymbol{\alpha})^{-1} \, \bigl(s - \boldsymbol{\alpha}^\top \boldsymbol{\mu}\bigr) &= \boldsymbol{\mu} +\boldsymbol{a} \, (\boldsymbol{a}^\top \Sigma^{-1} \boldsymbol{a})^{-1} \, \bigl(\boldsymbol{a}^\top \Sigma^{-1} \mathbf{v} - \boldsymbol{a}^\top \Sigma^{-1}\boldsymbol{\mu}\bigr) \\
   &= \boldsymbol{\mu} +\boldsymbol{a} \, (\boldsymbol{a}^\top \Sigma^{-1} \boldsymbol{a})^{-1/2} \overline{s}\,,
\end{align*}
where 
\[\overline{s}:= \frac{\boldsymbol{a}^\top \Sigma^{-1} \mathbf{v} - \boldsymbol{a}^\top \Sigma^{-1}\boldsymbol{\mu}}{\sqrt{\boldsymbol{a}^\top \Sigma^{-1} \boldsymbol{a}}}\]
is a standardized signal. Note that  $\overline{s}$ is elliptical, scalar-valued, has mean $0$ and variance $1$.  Importantly, the distribution of $\overline{s}$ is then independent of $\boldsymbol{\alpha}$. 

Therefore, for any choice $\boldsymbol{a}$, the buyer's expected payoff is: 
\[\overline{U}(\boldsymbol{a}):=\sum_i\mathbb{E}\left[\left(\theta_i(s; \Sigma^{-1}\boldsymbol{a})-p_i\right)_+\right]= \sum_i\mathbb{E}\left[\left(\mu_i-p_i+\frac{a_i}{\sqrt{\boldsymbol{a}^\top \Sigma^{-1}\boldsymbol{a}}}\overline{s}\right)_+\right]\,.\]
The following facts will prove important. First, consider the function
\[g_i(l):=\mathbb{E}\left[\left(\mu_i-p_i+l\overline{s}\right)_+\right]\,.\]
Note that 
\[g_i'(l) = \mathbb{E}\big[\overline{s}\mathbbm{1}\{\mu_i-p_i+l\overline{s}\geq 0\}\big]=\mathbb{E}\big[\overline{s}\mathbbm{1}\{\overline{s}\geq -(\mu_i-p_i)/l\}\big]\geq \mathbb{E}\big[\overline{s}\big]=0\,,\]
where the last inequality comes from $l>0$. Furthermore, the distribution of $\overline{s}$ is symmetric around zero, and so the function $g_i$ is even: $g_i(l) = g_i(-l)$ for all $l$, $i$. 

Second, in the above expression for $\overline{U}$, the correlation $\rho$ only enters through $\boldsymbol{a}^\top \Sigma^{-1}\boldsymbol{a}$. We can simplify this term further to make the dependence on $\rho$ explicit. Because $\text{Corr}(v_i, v_j)$ is constant and equal to $\rho$ for all $i$, $j$, we can decompose the variance-covariance matrix into
\[\Sigma = \text{diag}(\sigma_i)\, R\, \text{diag}(\sigma_i)\quad\text{where}\quad R = (1-\rho)I+\rho\mathbf{1}\mathbf{1}^\top\,. \]
Letting $\boldsymbol{\omega}:=(a_i/\sigma_i)_i$, we get
\begin{align*}
    \boldsymbol{a}^\top \Sigma^{-1}\boldsymbol{a}= \boldsymbol{a}^\top  \text{diag}(1/\sigma_i)\, R^{-1}\, \text{diag}(1/\sigma_i)\boldsymbol{a} &= \boldsymbol{\omega}^\top \, R^{-1}\, \boldsymbol{\omega}\\
    &=\boldsymbol{\omega}^\top \, \frac{1}{1-\rho}\left(I-\frac{\rho}{1+\rho (K-1)}\mathbf{1}\mathbf{1}^\top\right)\, \boldsymbol{\omega}    \\
    &=\frac{1}{1-\rho} \, \left(||\boldsymbol{\omega}||^2-\frac{\rho}{1+\rho (K-1)}(\mathbf{1}^\top\boldsymbol{\omega})^2\right)\,.
\end{align*}
Note that by construction we also have $1+\rho (K-1) > 0$ (for $\Sigma$ to be positive definite). 

\paragraph{Completion of the proof.}\hspace{-2mm}We first show that there exists an optimal solution $\boldsymbol{a} \in \R^K_{+}$, and then show that in fact, we can focus on 
\[\boldsymbol{a} \in \mathcal{A}:=\left\{\frac{\boldsymbol{a}'}{||\boldsymbol{a}'||}:\boldsymbol{a}' = \sum_i c_i\frac{\Sigma e_i}{||\Sigma e_i||}\text{ for some } \boldsymbol{c} \geq 0\right\}\,.
 \]
Fix any $\boldsymbol{a}$. Let $\hat{\boldsymbol{a}}:= (|a_i|)_i$.  
Using the above notation, let $\boldsymbol{\omega}=(a_j/\sigma_j)_j$ and $\hat{\boldsymbol{\omega}}=(\hat{a}_j/\sigma_j)_j$. By construction:
\[|\hat{a}_j| = |a_j|\quad \forall j\quad\text{and}\quad ||\hat{\boldsymbol{\omega}}|| = || \boldsymbol{\omega}||.\]
Moreover, by construction:
\[|\mathbf{1}^\top\hat{\boldsymbol{\omega}}| = \sum_j \frac{\hat{a}_j}{\sigma_j} = \sum_j \big|\frac{a_j}{\sigma_j}\big| \geq\Big|\sum_j \frac{a_j}{\sigma_j}\Big|=\big|\mathbf{1}^\top\boldsymbol{\omega}\big|\,.\]
Since $\rho \geq 0$, this implies:  
\[ \hat{\boldsymbol{a}}^\top \Sigma^{-1}\hat{\boldsymbol{a}}\leq \boldsymbol{a}^\top \Sigma^{-1}\boldsymbol{a}\,,\]
and hence for all $j$, 
\[ \frac{|\hat{a}_j|}{\sqrt{\hat{\boldsymbol{a}}^\top \Sigma^{-1}\hat{\boldsymbol{a}}}}\geq  \frac{|a_j|}{\sqrt{\boldsymbol{a}^\top \Sigma^{-1}\boldsymbol{a}}}\,.\]
The function $g_j$ is even and increasing over the positive range. Therefore:
\[\mathbb{E}\left[\left(\mu_j-p_j+\frac{\hat{a}_j}{\sqrt{\hat{\boldsymbol{a}}^\top \Sigma^{-1}\hat{\boldsymbol{a}}}}\overline{s}\right)_+\right]\geq \mathbb{E}\left[\left(\mu_j-p_j+\frac{a_j}{\sqrt{\boldsymbol{a}^\top \Sigma^{-1}\boldsymbol{a}}}\overline{s}\right)_+\right]\quad \forall j\,.\]
It follows immediately that 
\[\overline{U}(\hat{\boldsymbol{a}}) \geq \overline{U}(\boldsymbol{a})\,, \]
and hence it is without loss of optimality to focus on $\boldsymbol{a} \in \R^K_+$.

Now, for the second claim, note that if we define
\[\beta_i :=  \frac{a_i}{\sqrt{\boldsymbol{a}^\top \Sigma^{-1}\boldsymbol{a}}}\]
then
\[\boldsymbol{\beta}^\top \Sigma^{-1} \boldsymbol{\beta} =  1\,. \]
Conversely, for any $\boldsymbol{\beta}^\top \Sigma^{-1} \boldsymbol{\beta} = 1$, we can define $\boldsymbol{a}:= \boldsymbol{\beta} /||\boldsymbol{\beta}||$. Then 
\[\frac{a_i}{\sqrt{\boldsymbol{a}^\top \Sigma^{-1} \boldsymbol{a}}} = \frac{\beta_i/||\boldsymbol{\beta}||}{1/||\boldsymbol{\beta}||} = \beta_i\,.\]
Therefore, the buyer's problem is also equivalent to the following: 
\[\max_{\boldsymbol{\beta}^\top \Sigma^{-1}\boldsymbol{\beta} = 1} \sum_i g_i(\beta_i)\,.\]
By the first claim, we know that there exists an optimal solution $\boldsymbol{\beta}^* \in \R^K_+$. Then, at this optimal point, there must exist multiplier $\lambda \in \R$ such that for all $i$, 
\[g'_i(\beta^*_i) -2 \lambda (\Sigma^{-1}\boldsymbol{\beta}^*)_i = 0\,.\]
Let 
\[\delta_i := g'_i(\beta^*_i)\,.\]
Note that $\boldsymbol{\delta} \in \R^K_+$ by our previous observation. Moreover, 
\[2 \lambda = 2\lambda (\boldsymbol{\beta}^*)^\top \Sigma^{-1} \boldsymbol{\beta}^* = (\boldsymbol{\beta}^*)^\top \boldsymbol{\delta}\geq 0\,.\]
First, consider the case $\lambda > 0$. Then, we must have 
\[ \boldsymbol{\beta}^* = \frac{1}{2\lambda}\Sigma \boldsymbol{\delta} \,.\]
Thus, there must exist an optimal solution $\boldsymbol{a}^*$ such that 
\[\boldsymbol{a}^* = \boldsymbol{\beta}^* / || \boldsymbol{\beta}^* || \in \mathcal{A}\,,\]
as desired. 

Now, consider the case $\lambda = 0$. Then $\beta^*_i g'_i(\beta^*_i) = 0$ for all $i$. So for each $i$, either $\beta^*_i = 0$ or $g'_i(\beta^*_i) = 0$. Since $g_i$ is convex and nondecreasing on $[0, \infty)$, in both cases $g_i(\beta^*_i) = g_i(0)$. Hence, $\overline{U}(\boldsymbol{\beta}^*/||\boldsymbol{\beta}^*||) = \sum_i g_i(0) \leq \sum_i g_i(\hat{\beta}_i)$ for any feasible $\hat{\boldsymbol{\beta}}$. But then any feasible $\hat{\boldsymbol{\beta}}$ is optimal, and the claim trivially follows.

\subsubsection{Proof of \Cref{lem:pricerange}}
    For any $\boldsymbol{\varphi} \in [0, \frac{\pi}{2}]^{K-1}$, note that we have 
    \[\boldsymbol{a}(\boldsymbol{\varphi}) \in \mathcal{A}(\rho):= \Big\{\frac{\boldsymbol{a}'}{||\boldsymbol{a}'||}:\boldsymbol{a}' = \sum_i c_i\frac{\Sigma e_i}{||\Sigma e_i||}\text{ for some } \boldsymbol{c} \geq 0 \Big\}\,.\]
    We first show that 
    \[\sup_{\boldsymbol{a} \in \mathcal{A(\rho)}}||\boldsymbol{a} - \boldsymbol{\hat{\sigma}}|| \leq h(1 - \rho)\,,\]
    where $\boldsymbol{\hat{\sigma}}:= \boldsymbol{\sigma}/||\boldsymbol{\sigma}||$ and  $h$ is a continuous function with $h(0) = 0$. Indeed, note that 
    \[u_i(\rho) := \Sigma e_i  = (1 - \rho) \sigma^2_i e_i + \rho \sigma_i \boldsymbol{\sigma}\,,\]
    and hence 
    \[||u_i(\rho)|| = \sigma_i \sqrt{\rho^2 ||\boldsymbol{\sigma}||^2 + (1 - \rho^2) \sigma^2_i }\,,\]
    and hence 
    \[\frac{u_i(\rho)}{||u_i(\rho)||} \cdot \boldsymbol{\hat{\sigma}} = \frac{(1-\rho) \sigma^2_i /||\boldsymbol{\sigma}|| + \rho ||\boldsymbol{\sigma}||}{\sqrt{\rho^2 ||\boldsymbol{\sigma}||^2 + (1 - \rho^2) \sigma^2_i }} \,,\]
    which converges to $1$ as $\rho \rightarrow 1$. 
    Therefore, 
    \[\sup_{i} \Big|\Big| \frac{u_i(\rho)}{||u_i(\rho)||}  - \boldsymbol{\hat{\sigma}}\Big|\Big|^2 = \sup_{i}  2\Big(1 -  \frac{u_i(\rho)}{||u_i(\rho)||}\cdot \boldsymbol{\hat{\sigma}} \Big) =: m(\rho) \,,\]
    where $m(\rho) = O((1-\rho)^2)$ by a Taylor expansion around $\rho = 1$.  Thus, for any $a \in \mathcal{A(\rho)}$, we have 
    \[||\boldsymbol{a} - \boldsymbol{\hat{\sigma}}||^2 = 2 \big( 1 - \boldsymbol{a} \cdot \boldsymbol{\hat{\sigma}}\big) = 2 \Big( 1 - \frac{\boldsymbol{a}' \cdot \boldsymbol{\hat{\sigma}}}{||\boldsymbol{a}'||}\Big) \leq 2 \Big( 1 - \frac{(1 - m(\rho)/2)\sum_i c_i}{\sum_i c_i}\Big) = m(\rho)\,.\]
    Thus, 
    \[\sup_{\boldsymbol{a} \in \mathcal{A}(\rho)}||\boldsymbol{a} - \boldsymbol{\hat{\sigma}}|| \leq \sqrt{m(\rho)}\,,\]
    and hence the claim follows.

    Now, for any $i$, let 
    \[\beta_i(\boldsymbol{a}; \rho) :=  \frac{a_i}{\sqrt{\boldsymbol{a}^\top \Sigma^{-1}\boldsymbol{a}}}\,.\]
   We claim that 
    \[\sup_{\boldsymbol{a} \in \mathcal{A}(\rho)}||\boldsymbol{\beta}(\boldsymbol{a}; \rho) - \boldsymbol{\sigma}|| \rightarrow 0\]
    as $\rho \rightarrow 1$. Indeed, let $\boldsymbol{\omega} := (a_i/\sigma_i)_i$ and $\bar{\omega} := \mathbf{1}^\top\boldsymbol{\omega}/K$. Rearranging the expression for $\boldsymbol{a}^\top \Sigma^{-1}\boldsymbol{a}$ derived in the proof of \Cref{lem:dominance} gives
    \[\boldsymbol{a}^\top \Sigma^{-1}\boldsymbol{a} = \frac{||\boldsymbol{\omega} - \bar{\omega}\mathbf{1}||^2}{1-\rho} + \frac{(\mathbf{1}^\top\boldsymbol{\omega})^2}{K\big(1+\rho(K-1)\big)}\,.\]
    The second term converges to $1/||\boldsymbol{\sigma}||^2$ uniformly over $\boldsymbol{a} \in \mathcal{A}(\rho)$, since $\sup_{\boldsymbol{a} \in \mathcal{A}(\rho)}||\boldsymbol{a} - \boldsymbol{\hat{\sigma}}|| \rightarrow 0$ and $\boldsymbol{\omega} = \mathbf{1}/||\boldsymbol{\sigma}||$ at $\boldsymbol{a} = \boldsymbol{\hat{\sigma}}$. For the first term, note that $\boldsymbol{\omega} - \bar{\omega}\mathbf{1}$ is the projection of $\boldsymbol{\omega}$ onto the orthogonal complement of $\mathbf{1}$, which vanishes at $\boldsymbol{a} = \boldsymbol{\hat{\sigma}}$, so that $||\boldsymbol{\omega} - \bar{\omega}\mathbf{1}||^2 \leq ||\boldsymbol{a} - \boldsymbol{\hat{\sigma}}||^2/\min_i \sigma_i^2 \leq m(\rho)/\min_i\sigma_i^2 = O((1-\rho)^2)$. Hence the first term is $O(1-\rho)$ uniformly over $\mathcal{A}(\rho)$. Therefore, $\sup_{\boldsymbol{a} \in \mathcal{A}(\rho)}\big|\boldsymbol{a}^\top \Sigma^{-1}\boldsymbol{a} - 1/||\boldsymbol{\sigma}||^2\big| \rightarrow 0$, and the claim follows since $\boldsymbol{\beta}(\boldsymbol{a};\rho) = \boldsymbol{a}/\sqrt{\boldsymbol{a}^\top \Sigma^{-1}\boldsymbol{a}}$ and $\boldsymbol{a} \rightarrow \boldsymbol{\hat{\sigma}}$ uniformly. 
        
    Now, note that by \Cref{lem:regular}, the set of optimal monopoly prices $p^*_i$ against the distribution of $\theta_i$ generated by strategy $\boldsymbol{\alpha}=\Sigma^{-1}\boldsymbol{a}$ is single-valued, and moreover, it is only a function of $\boldsymbol{\beta}(\boldsymbol{a}; \rho)$ (see the proof of \Cref{lem:dominance}). By the uniform convergence of $\boldsymbol{\beta}$ above, for any $\varepsilon > 0$, there exists some $\rho_\varepsilon$ such that for all $\rho > \rho_\varepsilon$, we have $\boldsymbol{\beta} \in \boldsymbol{\sigma} + \mathbf{b}_\varepsilon$, where $\mathbf{b}_\varepsilon$ is the $\varepsilon$-radius ball in $\R^K$. Let $\boldsymbol{p}^*(\boldsymbol{\beta})$ be the optimal monopoly prices across the goods, given $\boldsymbol{\beta}$. Fix some $\varepsilon$ such that $\boldsymbol{\sigma} + \mathbf{b}_\varepsilon \geq \boldsymbol{\sigma}/2$. By Berge's theorem, $\boldsymbol{p}^*$ is continuous in $\boldsymbol{\beta}$. By the Heine–Cantor theorem, it follows that $\boldsymbol{p}^*$ is uniformly continuous on the ball $\boldsymbol{\sigma} + \mathbf{b}_\varepsilon$. In particular, for any $\varepsilon_2 > 0$, there exists some $\delta > 0$ such that for any $\boldsymbol{\beta}, \boldsymbol{\beta}' \in \boldsymbol{\sigma} + \mathbf{b}_\varepsilon$ where $||\boldsymbol{\beta}- \boldsymbol{\beta}'|| < \delta$, we have 
    $||\boldsymbol{p}^*(\boldsymbol{\beta}) - \boldsymbol{p}^*(\boldsymbol{\beta}')|| < \varepsilon_2$. Let $\boldsymbol{\beta}' = \boldsymbol{\sigma}$. By \Cref{lem:regular}, $p^*_i(\boldsymbol{\sigma}) > \underline{\theta}_i$ for all $i$. Let 
    \[\varepsilon_2 := \min_{i} \big\{p^*_i(\boldsymbol{\sigma}) - \underline{\theta}_i\}/2 > 0\,.\]
    Let $\delta > 0$ be the one given by the uniform continuity of $\boldsymbol{p}^*$. Then, consider any $\rho > \rho_\delta$. It follows that for any $\boldsymbol{\varphi} \in [0, \frac{\pi}{2}]^{K-1}$, the induced $\boldsymbol{\beta}(\boldsymbol{a}(\boldsymbol{\varphi}); \rho) \in \boldsymbol{\sigma} + \mathbf{b}_\delta$ by construction. Therefore, $|| \boldsymbol{\beta}(\boldsymbol{a}(\boldsymbol{\varphi}); \rho) - \boldsymbol{\sigma}|| < \delta$ by construction, and hence 
    \[||\boldsymbol{p}^*(\boldsymbol{\beta}(\boldsymbol{a}(\boldsymbol{\varphi}); \rho)) - \boldsymbol{p}^*(\boldsymbol{\sigma})|| < \varepsilon_2\,.\]
     This implies that for all $i$
     \[|p^*_i(\boldsymbol{\beta}(\boldsymbol{a}(\boldsymbol{\varphi}); \rho)) - p^*_i(\boldsymbol{\sigma})|< \varepsilon_2 \leq (p^*_i(\boldsymbol{\sigma}) - \underline{\theta}_i)/2\,, \]
     and hence 
     \[p^*_i(\boldsymbol{\beta}(\boldsymbol{a}(\boldsymbol{\varphi}); \rho)) > p^*_i(\boldsymbol{\sigma}) -(p^*_i(\boldsymbol{\sigma}) - \underline{\theta}_i)/2 = \underline{\theta}_i + (p^*_i(\boldsymbol{\sigma}) - \underline{\theta}_i)/2 \geq \underline{\theta}_i + \varepsilon_2\,.\]
     The result follows immediately by letting $\varepsilon := \varepsilon_2$ and $\underline{\rho} := \rho_{\delta}$. 

\subsubsection{Proof of \Cref{lem:quasiconcave}}

Using the same notation as in the proof of \Cref{lem:dominance}, we can rewrite the buyer's payoff from any learning weights $\boldsymbol{\alpha}(\boldsymbol{\varphi})$ as 
\[U(\boldsymbol{\alpha}(\boldsymbol{\varphi})) = \overline{U}(\boldsymbol{a}(\boldsymbol{\varphi})) = \sum_i \mathbb{E}\left[\left(\mu_i-p_i+\frac{a_i(\boldsymbol{\varphi})}{\sqrt{\boldsymbol{a}(\boldsymbol{\varphi})^\top\Sigma^{-1}\boldsymbol{a}(\boldsymbol{\varphi})}}\overline{s}\right)_+\right],\]
where $\boldsymbol{a}(\boldsymbol{\varphi}) = \Sigma \boldsymbol{\alpha}(\boldsymbol{\varphi})$ and $\overline{s}$ is an elliptical random variable whose law does not depend on $\boldsymbol{\varphi}$. 

We want to show $\overline{U}(\boldsymbol{a}(\,\cdot\,))$ has a unique maximizer over $\boldsymbol{\varphi}\in [0, \frac{\pi}{2}]^{K-1}$ for $\rho$ high enough. As argued before, it is equivalent to show that there exists some $\underline{\rho}'< 1$ such that for all $\rho > \underline{\rho}'$, and all $\boldsymbol{p} \in \prod_{i} [\underline{\theta}_i + \varepsilon', \mu_i]$, we have that $\overline{U}(\boldsymbol{a}; \rho, \boldsymbol{p})$  has a unique maximizer in the following set 
\[\mathcal{A}(\rho):=\left\{\frac{\boldsymbol{a}'}{||\boldsymbol{a}'||}:\boldsymbol{a}' = \sum_i c_i\frac{\Sigma e_i}{||\Sigma e_i||}\text{ for some } \boldsymbol{c} \geq 0\right\}\,.\]

The remainder of the proof proceeds in several steps. First, we derive bounds on elements of $\overline{U}$ and their derivative as $\rho$ goes to one. Second, we show that along any short geodesic $\boldsymbol{a}(t)\subset\mathcal{A}$, the function $t\rightarrow\overline{U}(\boldsymbol{a}(t))$ is strictly unimodal. Finally, we show that $\overline{U}$ attains a unique maximizer $\boldsymbol{a}^*\in \mathcal{A}$.

Let $Q(\boldsymbol{a}; \rho):=\frac{1}{\sqrt{\boldsymbol{a}^\top\Sigma^{-1}\boldsymbol{a}}}$. 

\vspace{-3mm}
\paragraph{Limits of $\boldsymbol{a}$ and $Q$.}\hspace{-2mm}As shown in the proof of \Cref{lem:pricerange}, 
\[\sup_{\boldsymbol{a} \in \mathcal{A(\rho)}}||\boldsymbol{a} - \boldsymbol{\hat{\sigma}}|| \rightarrow 0\]
as $\rho \rightarrow 1$, where $\boldsymbol{\hat{\sigma}} = \boldsymbol{\sigma}/ ||\boldsymbol{\sigma}||>0$. Moreover, note that $\boldsymbol{\hat{\sigma}}$ is the principal eigenvector of $\Sigma(\rho)$ when $\rho = 1$. Also, as shown in the proof of \Cref{lem:pricerange}, we have 
    \[\sup_{\boldsymbol{a} \in \mathcal{A}(\rho)}||\boldsymbol{\beta}(\boldsymbol{a}; \rho) - \boldsymbol{\sigma}|| \rightarrow 0\]
as $\rho \rightarrow 1$, where 
\[\boldsymbol{\beta}(\boldsymbol{a}; \rho) = \boldsymbol{a} Q(\boldsymbol{a}; \rho)\,.\]
By the same argument as in the proof of \Cref{lem:pricerange}, it also follows that 
\[\sup_{\boldsymbol{a} \in \mathcal{A}(\rho)}\Big|Q(\boldsymbol{a}; \rho) - ||\boldsymbol{\sigma}|| \Big| \rightarrow 0\,.\]
Moreover, note that $||\boldsymbol{\sigma}||^2$ is the only non-zero eigenvalue of $\Sigma(1)$.   

\vspace{-3mm}
\paragraph{Limits of radial derivatives of $Q$.}\hspace{-2mm}Let $\lambda_1\geq \lambda_2\geq \dots \geq \lambda_K>0$ be the eigenvalues of $\Sigma(\rho)$, which depend on $\rho$ and are strictly positive for $\rho<1$. Let $\{\boldsymbol{\nu}_i\}_i$ be the associated eigenvectors. Recall that $\Sigma^{-1}$ shares the same eigenvectors as $\Sigma$ but with eigenvalues $\{1/\lambda_i\}_i$. The product $\boldsymbol{a}^\top\Sigma^{-1}\boldsymbol{a}$ is the Rayleigh quotient of $\Sigma^{-1}$ and $\boldsymbol{a}$. Given that $||\boldsymbol{a}||=1$, it can be written as
\[\boldsymbol{a}^\top\Sigma^{-1}\boldsymbol{a}=\sum_i\frac{(\boldsymbol{\nu}_i^\top\,\boldsymbol{a})^2}{\lambda_i}.\]
The eigenvectors of $\Sigma$ are orthonormal and can be used as a basis. We can thus write $\boldsymbol{a}=\boldsymbol{\nu}_1\cos(\psi)+\sin(\psi)\sum_{i>1}c_i\boldsymbol{\nu}_i$ with $\psi:=\angle(\boldsymbol{\nu}_1,\boldsymbol{a})$ and $\sum_{i>1}c_i^2=1$. Then, 
\[\boldsymbol{a}^\top\Sigma^{-1}\boldsymbol{a}=\frac{\cos^2(\psi(\boldsymbol{a}))}{\lambda_1}+\sin^2(\psi(\boldsymbol{a}))\sum_{i>1}\frac{c_i^2}{\lambda_i}.\]
Furthermore, for any $\boldsymbol{a}\in\mathcal{A}(\rho)$, we have 
\begin{align*}
    |1-\cos \psi(\boldsymbol{a})|\leq \max_i \Big|1-\left(\frac{\Sigma e_i}{||\Sigma e_i||}\right)^\top\boldsymbol{\nu}_1\Big|\,,
\end{align*}
where, for each $i$: 
\[\Big|1-\left(\frac{\Sigma e_i}{||\Sigma e_i||}\right)^\top\boldsymbol{\nu}_1\Big| =\frac{1}{2}|| \frac{\Sigma e_i}{||\Sigma e_i||} - \boldsymbol{\nu}_1||^2 \leq \frac{1}{2}\big(||\boldsymbol{\nu}_1 - \hat{\boldsymbol{\sigma}}|| + ||\hat{\boldsymbol{\sigma}} - \frac{\Sigma e_i}{||\Sigma e_i||}||\big)^2 = O((1 - \rho)^2)\,,\]
where the rate estimate in the last equality follows from the bound given in the proof of \Cref{lem:pricerange}. Therefore, there exists some constant $C_1 > 0$ such that for all $\boldsymbol{a} \in \mathcal{A}(\rho)$, we have 
\[\cos^2\psi(\boldsymbol{a}) \geq 1- C_1 \cdot (1-\rho)^2 \quad\text{ and }\quad \sin^2\psi(\boldsymbol{a}) = 1-\cos^2\psi(\boldsymbol{a}) \leq  C_1 \cdot (1-\rho)^2.\]
Furthermore, by standard arguments, the eigenvalues of $\Sigma$ (other than the largest one) all converge to zero at rate $\lambda_i = \Theta(1-\rho)$ for all $i>1$. 

Fixing $\{c_i\}_{i > 1}$ and differentiating $Q(\boldsymbol{a})$ along the curve $\psi\rightarrow \boldsymbol{\nu}_1\cos(\psi)+\sin(\psi)\sum_{i>1}c_i\boldsymbol{\nu}_i$ gives
\begin{align*}
    \frac{\partial Q}{\partial \psi}= -\frac{1}{2(\boldsymbol{a}^\top\Sigma^{-1}\boldsymbol{a})^{\frac{3}{2}}}\frac{\partial \boldsymbol{a}^\top\Sigma^{-1}\boldsymbol{a}}{\partial \psi}&= -\frac{1}{2(\boldsymbol{a}^\top\Sigma^{-1}\boldsymbol{a})^{\frac{3}{2}}}\frac{\partial }{\partial \psi}\left(\frac{\cos^2(\psi)}{\lambda_1}+\sin^2(\psi)\sum_{i>1}\frac{c_i^2}{\lambda_i}\right)\\
    &= -\frac{1}{2(\boldsymbol{a}^\top\Sigma^{-1}\boldsymbol{a})^{\frac{3}{2}}}\left(\frac{-\sin(2\psi)}{\lambda_1}+\sin(2\psi)\sum_{i>1}\frac{c_i^2}{\lambda_i}\right)\\
    &= \frac{\sin(2\psi)}{2(\boldsymbol{a}^\top\Sigma^{-1}\boldsymbol{a})^{\frac{3}{2}}}\left(\frac{1}{\lambda_1}-\sum_{i>1}\frac{c_i^2}{\lambda_i}\right)\,.
\end{align*}
Moreover, 
\[\frac{1}{\lambda_2} \leq \sum_{i>1}\frac{c_i^2}{\lambda_i}\leq \frac{1}{\lambda_K}\,.\]
Therefore, there exists some constant $C_2$ such that 
\[\sup_{\boldsymbol{a} \in \mathcal{A}(\rho)}\Big|\frac{\partial Q}{\partial \psi}(\boldsymbol{a}; \rho)\Big| \leq \sup_{\boldsymbol{a} \in \mathcal{A}(\rho)}\Big|\frac{1}{2} Q^{3}(\boldsymbol{a};\rho)\Big| \cdot  \sup_{\boldsymbol{a} \in \mathcal{A}(\rho)}\Big|\sin(2\psi(\boldsymbol{a}))\Big|\cdot  \Big|\frac{1}{\lambda_1} - \frac{1}{\lambda_K}\Big| \rightarrow C_2\]
as $\rho \rightarrow 1$, since the first term converges to a constant, the second term is bounded above by $C'(1 - \rho)$, and the third term is bounded above by $C'' \frac{1}{1 - \rho}$ for some constants $C', C''$.  

Similarly, we also look at the second derivative: Fixing $\{c_i\}_{i > 1}$, differentiating $Q(\boldsymbol{a})$ along $\psi$ twice gives: 
\begin{align*}
    \frac{\partial^2 Q}{\partial \psi^2}= \cos(2\psi)Q^3(\boldsymbol{a}; \rho) \left(\frac{1}{\lambda_1}-\sum_{i>1}\frac{c_i^2}{\lambda_i}\right) + \frac{3}{4}Q^5(\boldsymbol{a}; \rho) \sin^2(2\psi)\left(\frac{1}{\lambda_1}-\sum_{i>1}\frac{c_i^2}{\lambda_i}\right)^2\,.
\end{align*}
By the same argument, the absolute value of the second term is bounded from above by
\[\sup_{\boldsymbol{a}\in \mathcal{A}(\rho)}\frac{3}{4}Q^5(\boldsymbol{a}; \rho) \sin^2(2\psi(\boldsymbol{a}))\Big|\frac{1}{\lambda_1} - \frac{1}{\lambda_K}\Big|^2 \rightarrow C_3\]
as $\rho\rightarrow 1$, where $C_3$ is a constant. The first term, however, satisfies  
\[\cos(2\psi(\boldsymbol{a}))Q^3(\boldsymbol{a}; \rho) \left(\frac{1}{\lambda_1}-\sum_{i>1}\frac{c_i^2}{\lambda_i}\right) \leq\cos(2\psi(\boldsymbol{a}))Q^3(\boldsymbol{a}; \rho) \left(\frac{1}{\lambda_1}-\frac{1}{\lambda_2}\right)\,.\]
Moreover, 
\begin{align*}
    \sup_{\boldsymbol{a}\in \mathcal{A}(\rho)} \Big\{\cos(2\psi(\boldsymbol{a}))Q^3(\boldsymbol{a}; \rho) \left(\frac{1}{\lambda_1}-\frac{1}{\lambda_2}\right)\Big\}&\leq \left(\frac{1}{\lambda_1}-\frac{1}{\lambda_2}\right) \underbrace{\inf_{\boldsymbol{a}\in \mathcal{A}(\rho)} \Big\{\cos(2\psi(\boldsymbol{a}))Q^3(\boldsymbol{a}; \rho)\Big\}}_{=:w(\rho)}\,. 
\end{align*}
Note that, as $\rho\rightarrow 1$, $\frac{1}{\lambda_1}$ converges to $1/||\boldsymbol{\sigma}||^2$, $\frac{1}{\lambda_2}$ diverges to $\infty$ at a rate $\Theta(\frac{1}{1 - \rho})$, and $w(\rho)$ converges to a constant. Therefore, 
\[ \sup_{\boldsymbol{a}\in \mathcal{A}(\rho)} \Big\{\cos(2\psi(\boldsymbol{a}))Q^3(\boldsymbol{a}; \rho) \left(\frac{1}{\lambda_1}-\frac{1}{\lambda_2}\right)\Big\} \rightarrow - \infty\,,\]
as $\rho\rightarrow 1$. It follows immediately that 
\[\sup_{\boldsymbol{a} \in \mathcal{A}(\rho)}\frac{\partial^2 Q}{\partial \psi^2}(\boldsymbol{a}; \rho) \rightarrow - \infty\]
as $\rho \rightarrow 1$. 

\vspace{-3mm}
\paragraph{Limits of derivatives of $\overline{U}$.}\hspace{-2mm}For any $i$, let 
    \[\beta_i(\boldsymbol{a}; \rho) :=  \frac{a_i}{\sqrt{\boldsymbol{a}^\top \Sigma^{-1}\boldsymbol{a}}}\,.\]
As shown in the proof of \Cref{lem:dominance}, 
\[\frac{\partial \overline{U}}{\partial \beta_i} = \mathbb{E}\Big[\overline{s}\mathbbm{1}\big\{\mu_i - p_i +\beta_i \overline{s}\geq 0\big\}\Big]\geq \mathbb{E}\Big[\overline{s}\mathbbm{1}\big\{\mu_i - (\underline{\theta}_i + \varepsilon') +\beta_i \overline{s}\geq 0\big\}\Big]\,,\]
where the inequality is due to $\underline{\theta}_i + \varepsilon' \leq p_i \leq \mu_i$ and $\beta_i \geq 0$. Therefore, 
\begin{align*}
\inf_{\boldsymbol{a} \in \mathcal{A}(\rho)} \frac{\partial \overline{U}}{\partial \beta_i}(\boldsymbol{a}; \rho) &\geq \inf_{\boldsymbol{a} \in \mathcal{A}(\rho)} \mathbb{E}\Big[\overline{s}\mathbbm{1}\big\{\mu_i - (\underline{\theta}_i + \varepsilon') +\beta_i(\boldsymbol{a};\rho) \overline{s}\geq 0\big\}\Big] \\
&= \mathbb{E}\Big[\overline{s}\mathbbm{1}\big\{\mu_i - (\underline{\theta}_i + \varepsilon') + \overline{s} \cdot \inf_{\boldsymbol{a} \in \mathcal{A}(\rho)}\beta_i(\boldsymbol{a};\rho) \geq 0\big\}\Big] \,,
\end{align*}
where the last equality uses the convexity of $\overline{U}$ in $\beta_i$. Let $\mathcal{P}:=\prod_{i}[\underline{\theta}_i + \varepsilon', \mu_i]$. 

By the proof of \Cref{lem:pricerange}, we know that $\displaystyle \inf_{\boldsymbol{a} \in \mathcal{A}(\rho)}\beta_i(\boldsymbol{a};\rho)\rightarrow \sigma_i$, as $\rho \rightarrow 1$, and hence 
\[\lim_{\rho\rightarrow 1}\inf_{\boldsymbol{a} \in \mathcal{A}(\rho), \, \boldsymbol{p} \in \mathcal{P}} \frac{\partial \overline{U}}{\partial \beta_i}(\boldsymbol{a}; \rho) \geq \mathbb{E}\Big[\overline{s}\mathbbm{1}\big\{\mu_i - (\underline{\theta}_i + \varepsilon') + \sigma_i \overline{s} \geq 0\big\}\Big] > 0\,,\]
where the strict inequality is due to
\[\Pr\Big(\mu_i + \sigma_i \overline{s} <  \underline{\theta}_i + \varepsilon' \Big) > 0\,,\]
which holds by construction given $\varepsilon' > 0$ and $\underline{\theta}_i:= \displaystyle\lim_{\rho \rightarrow 1} \inf_{v\in V} v_i$. By a similar argument, we also have 
\[\lim_{\rho\rightarrow 1}\sup_{\boldsymbol{a} \in \mathcal{A}(\rho), \, \boldsymbol{p} \in \mathcal{P}} \frac{\partial \overline{U}}{\partial \beta_i}(\boldsymbol{a}; \rho) \leq  \mathbb{E}\Big[\overline{s}\mathbbm{1}\big\{ \overline{s} \geq 0\big\}\Big] < \infty \,. \]
Also, we have that 
\[\sup_{\boldsymbol{a} \in \mathcal{A}(\rho),\, \boldsymbol{p}\in \mathcal{P} }\Big|\frac{\partial^2 \overline{U}}{\partial \beta^2_i}(\boldsymbol{a}; \rho) \Big| \leq \sup_{\boldsymbol{a} \in \mathcal{A}(\rho)} \frac{\mu_i^2}{\beta^3_i(\boldsymbol{a}; \rho)}\cdot  f_{\overline{s}}(0) \rightarrow \frac{\mu^2_i}{\sigma^3_i}f_{\overline{s}}(0)\,,\]
as $\rho \rightarrow 1$.

\vspace{-3mm}
\paragraph{Concavity of $\overline{U}$ on a particular geodesic.}\hspace{-2mm}Fixing $\{c_i\}_{i > 1}$ and differentiating $\overline{U}(\boldsymbol{a})$ along the curve $\boldsymbol{a}: \psi\rightarrow \boldsymbol{\nu}_1\cos(\psi)+\sin(\psi)\sum_{i>1}c_i\boldsymbol{\nu}_i$ with respect to $\psi$ gives: 
\begin{align*}
    \frac{\partial \overline{U}}{\partial \psi} &= \sum_i \frac{\partial \overline{U}}{\partial \beta_i}\Big(a_i(\psi)Q(\boldsymbol{a}(\psi)) \Big)\times\left[a_i'(\psi)\times Q(\boldsymbol{a}(\psi)) + a_i(\psi)\times\frac{\partial Q}{\partial\psi}(\boldsymbol{a}(\psi))\right]\\
    &= Q(\boldsymbol{a}(\psi))\sum_i a_i'(\psi)\frac{\partial \overline{U}}{\partial \beta_i}(a_i(\psi)Q(\boldsymbol{a}(\psi)))\\
    &\hspace{5cm}+ \frac{\partial Q}{\partial\psi}(\boldsymbol{a}(\psi))\sum_i a_i(\psi)\frac{\partial \overline{U}}{\partial \beta_i}(a_i(\psi)Q(\boldsymbol{a}(\psi))).
\end{align*}
Let $h_i(\beta_i) := \frac{\partial \overline{U}}{\partial \beta_i}$. Also abuse the notation to write $Q' := \frac{\partial Q}{\partial\psi}$, $a_i' := \frac{\partial a_i}{\partial\psi}$, and $\beta'_i:= \frac{\partial \beta_i}{\partial \psi}$. The above can be written as 
\[  \frac{\partial \overline{U}}{\partial \psi} = \sum_i h_i(\beta_i) \beta'_i =  Q\sum_i a'_i h_i(\beta_i) + Q'\sum_i a_i h_i(\beta_i)\,.\]
Therefore, we also have 
\[  \frac{\partial^2 \overline{U}}{\partial \psi^2} = \sum_i h'_i(\beta_i) (\beta'_i)^2 + \sum_i h_i(\beta_i) \beta''_i\,,\]
where 
\[\beta'_i = a'_i Q + a_i Q' \quad \text{ and }\quad \beta''_i = a''_i Q + 2 a'_i Q' + a_i Q''\,.\]
Equivalently, we can write 
\[\frac{\partial^2 \overline{U}}{\partial \psi^2} = \underbrace{Q'' \sum_i a_i h_i(\beta_i)}_{\text{I}} + \underbrace{2Q' \sum_i a'_i h_i(\beta_i)}_{\text{II}} + \underbrace{Q \sum_i a''_i h_i(\beta_i)}_{_{\text{III}}} + \underbrace{\sum_i h'_i(\beta_i)(a'_i Q + a_i Q')^2}_{_{\text{IV}}}\,. \]
Note that 
\[\boldsymbol{a}'(\psi) = -\boldsymbol{\nu}_1\sin(\psi)+\cos(\psi)\sum_{i>1}c_i\boldsymbol{\nu}_i \text{ and }  \boldsymbol{a}''(\psi) = -\boldsymbol{\nu}_1\cos(\psi)-\sin(\psi)\sum_{i>1}c_i\boldsymbol{\nu}_i\,,\]
and hence 
\[||\boldsymbol{a}'(\psi) || = ||\boldsymbol{a}''(\psi) || = 1\,.\]
Thus, $|a'_i|, |a''_i| \leq 1$ uniformly. Now, we show that the parts II, III, IV in the above expression are uniformly bounded. First, note that 
\[\lim_{\rho \rightarrow 1}\sup_{\boldsymbol{a}\in \mathcal{A}(\rho),\,\boldsymbol{p} \in \mathcal{P}} |\text{II}(\boldsymbol{a};\boldsymbol{p}, \rho)| \leq 2 \lim_{\rho \rightarrow 1}  \Big(\sup_{\boldsymbol{a}\in \mathcal{A}(\rho)} |Q'| \Big)\cdot \Big(\sum_i \sup_{\boldsymbol{a}\in \mathcal{A}(\rho),\,\boldsymbol{p} \in \mathcal{P}} |h_i(\beta_i)| \Big) \leq C_{\text{II}}\,,  \]
for some constant $C_{\text{II}}$. Second, note that 
\[\lim_{\rho \rightarrow 1}\sup_{\boldsymbol{a}\in \mathcal{A}(\rho),\,\boldsymbol{p} \in \mathcal{P}} |\text{III}(\boldsymbol{a};\boldsymbol{p}, \rho)| \leq  \lim_{\rho \rightarrow 1}  \Big(\sup_{\boldsymbol{a}\in \mathcal{A}(\rho)} |Q| \Big)\cdot \Big(\sum_i \sup_{\boldsymbol{a}\in \mathcal{A}(\rho),\,\boldsymbol{p} \in \mathcal{P}} |h_i(\beta_i)| \Big) \leq C_{\text{III}}\,,  \]
for some constant $C_{\text{III}}$. Third, note that 
\[\lim_{\rho \rightarrow 1}\sup_{\boldsymbol{a}\in \mathcal{A}(\rho),\,\boldsymbol{p} \in \mathcal{P}} |\text{IV}(\boldsymbol{a};\boldsymbol{p}, \rho)| \leq  \lim_{\rho \rightarrow 1}  \sum_i \sup_{\boldsymbol{a}\in \mathcal{A}(\rho),\,\boldsymbol{p} \in \mathcal{P}} |h'_i(\beta_i)| \Big(\sup_{\boldsymbol{a} \in \mathcal{A}(\rho)} |Q| + \sup_{\boldsymbol{a} \in \mathcal{A}(\rho)} |Q'| \Big)^2 \leq C_{\text{IV}}\,,  \]
for some constant $C_{\text{IV}}$. Now, we show that part I in the previous expression diverges to $-\infty$ uniformly: 
\[\lim_{\rho \rightarrow 1}\sup_{\boldsymbol{a}\in \mathcal{A}(\rho),\,\boldsymbol{p} \in \mathcal{P}} \text{I}(\boldsymbol{a};\boldsymbol{p}, \rho) \leq  \lim_{\rho \rightarrow 1}  \Big(\sup_{\boldsymbol{a}\in \mathcal{A}(\rho)} Q''\Big) \cdot \lim_{\rho \rightarrow 1}   \sum_i \inf_{\boldsymbol{a}\in \mathcal{A}(\rho),\,\boldsymbol{p} \in \mathcal{P}} \Big(a_i h_i(\beta_i)\Big)\,. \]
Moreover, as we have shown, 
\[\lim_{\rho \rightarrow 1}  \Big(\sup_{\boldsymbol{a}\in \mathcal{A}(\rho)} Q''\Big) = -\infty\,,\]
and 
\begin{align*}
\lim_{\rho \rightarrow 1}  \sum_i \inf_{\boldsymbol{a}\in \mathcal{A}(\rho),\,\boldsymbol{p} \in \mathcal{P}} \Big(a_i h_i(\beta_i)\Big)  &\geq \lim_{\rho \rightarrow 1}   \sum_i \inf_{\boldsymbol{a}\in \mathcal{A}(\rho)} \big(a_i\big) \inf_{\boldsymbol{a}\in \mathcal{A}(\rho),\,\boldsymbol{p} \in \mathcal{P}}\Big(h_i(\beta_i)\Big) \\
&= \sum_i \hat{\sigma}_i  \lim_{\rho \rightarrow 1}  \inf_{\boldsymbol{a}\in \mathcal{A}(\rho),\,\boldsymbol{p} \in \mathcal{P}}\Big(h_i(\beta_i)\Big) > 0\,,
\end{align*}
where the strict inequality is due to $\hat{\sigma}_i > 0$ and  $\displaystyle \inf_{\boldsymbol{a}\in \mathcal{A}(\rho),\,\boldsymbol{p} \in \mathcal{P}}h_i(\beta_i) > 0$ for each $i$, as we have shown. Together, these imply that 
\[\lim_{\rho \rightarrow 1}\sup_{\boldsymbol{a}\in \mathcal{A}(\rho),\,\boldsymbol{p} \in \mathcal{P}} \text{I}(\boldsymbol{a};\boldsymbol{p}, \rho) = -\infty\,.\]
Fix some $\delta > 0$. By the above arguments, there exists some $\rho_0$ such that for all $\rho > \rho_0$, we have 
\[\sup_{\boldsymbol{a}\in \mathcal{A}(\rho),\,\boldsymbol{p} \in \mathcal{P}} |\text{II}(\boldsymbol{a};\boldsymbol{p}, \rho)| +  |\text{III}(\boldsymbol{a};\boldsymbol{p}, \rho)| + |\text{IV}(\boldsymbol{a};\boldsymbol{p}, \rho)| \leq C_{\text{II}} + C_{\text{III}} +  C_{\text{IV}} + \delta \,.\]
Moreover, there also exists some $\rho_1$ such that for all $\rho > \rho_1$, we have 
\[\sup_{\boldsymbol{a}\in \mathcal{A}(\rho),\,\boldsymbol{p} \in \mathcal{P}} \text{I}(\boldsymbol{a};\boldsymbol{p}, \rho) < -2\cdot \Big(C_{\text{II}} + C_{\text{III}} +  C_{\text{IV}} + \delta\Big)\,.\]
Let $\underline{\rho}' := \max\{\rho_0, \rho_1\}$. It follows immediately that for all $\rho > \underline{\rho}'$, we have 
\[\frac{\partial^2 \overline{U}}{\partial \psi^2}(\boldsymbol{a}; \boldsymbol{p}, \rho) < - \delta \,.\]
for all $\boldsymbol{a} \in \mathcal{A}(\rho)$ and  all $\boldsymbol{p} \in \mathcal{P}$.

\vspace{-3mm}
\paragraph{Generalize to arbitrary geodesics in $\mathcal A(\rho)$.}\hspace{-2mm}We now generalize the above argument to cover any geodesics in $\mathcal{A}(\rho)$. Let $\boldsymbol{a}(\psi)$ be a geodesic in $\mathcal{A}(\rho)$ connecting any two points $\boldsymbol{a}_0,\boldsymbol{a}_1\in\mathcal A(\rho)$, and let $P$ be the $2$-D plane containing $\boldsymbol{a}(\psi)$. 

Define
\[
u_\star \;:=\; \frac{\mathrm{Proj}_P(\boldsymbol{\nu}_1)}{\big\|\mathrm{Proj}_P(\boldsymbol{\nu}_1)\big\|}\,,\qquad
w_\star\in P,\ \ \|w_\star\|=1,\ \ w_\star\perp u_\star.
\]
The geodesic can be written as
\[
\boldsymbol{a}(\psi)\;=\;\cos\psi\,u_\star+\sin\psi\,w_\star\,.
\]
Let $m_{ab}:=e_a^\top \Sigma^{-1} e_b$ with $a, b\in\{1,2\}$, $e_1=u_\star$, $e_2=w_\star$. Then
\[
q(\psi):=\boldsymbol{a}(\psi)^\top \Sigma^{-1}\,\boldsymbol{a}(\psi)
= m_{11}\cos^2\psi + 2m_{12}\sin\psi\cos\psi + m_{22}\sin^2\psi,
\]
and hence
\[
Q'(\psi)= -\tfrac12\, q(\psi)^{-3/2}\, q'(\psi),\qquad
Q''(\psi)= -\tfrac12\, q(\psi)^{-3/2}\, q''(\psi)+\tfrac34\, q(\psi)^{-5/2}\,\big(q'(\psi)\big)^2\,.
\]
By the same argument as before, we only need to consider $|\sin(2\psi)| \leq O(1-\rho)$ and $\cos(2\psi) \geq  1-O((1-\rho)^2)$. Moreover, as noted before, $\frac{1}{\lambda_1}\rightarrow \frac{1}{||\boldsymbol{\sigma}||^2}$, while $\frac{1}{\lambda_i}=\Theta(\frac{1}{1 - \rho})$ for all $i > 1$, as $\rho \rightarrow 1$. We decompose $u_\star$ and $w_\star$ in the $\{\boldsymbol{\nu}_i\}$ basis: 
\[u_\star = c_1 \boldsymbol{\nu}_1 + \sum_{i > 1} c_i \boldsymbol{\nu}_i\,\quad\,w_\star = \sum_{i > 1} d_i \boldsymbol{\nu}_i\,,\]
where we have also used that $w_\star\perp \boldsymbol{\nu}_1$.  Note that by construction 
\[1 - c^2_1 = \sum_{i > 1} c^2_i = O((1 - \rho)^2)\,,\qquad  \sum_{i > 1} d^2_i = 1 \,.\]
Since $\Sigma^{-1}\boldsymbol{\nu}_i = \frac{1}{\lambda_i} \boldsymbol{\nu}_i$, we have 
\begin{align*}
    m_{11} = u_\star^\top \Sigma^{-1} u_\star = c^2_1\frac{1}{\lambda_1} + \sum_{i> 1}\frac{ c^2_i}{\lambda_i} &=  c^2_1\frac{1}{\lambda_1} + O((1 - \rho)^2) \cdot \Theta(\frac{1}{1 - \rho}) \\
    &= \frac{1 - O((1 - \rho)^2)}{\lambda_1} + O(1 - \rho) \rightarrow \frac{1}{||\boldsymbol{\sigma}||^2}
\end{align*}
as $\rho \rightarrow 1$. Similarly, note that 
\[m_{22} = w_\star^\top \Sigma^{-1} w_\star = \sum_{i > 1}  \frac{d^2_i}{\lambda_i} \geq \frac{1}{\lambda_2} = \Theta(\frac{1}{1 - \rho})\,.\]
Moreover, 
\[|m_{12}| = |u^\top_\star \Sigma^{-1} w_\star| =\big| \sum_{i > 1} \frac{c_i d_i}{\lambda_i} \big| \leq \sqrt{\sum_{i > 1} c^2_i} \sqrt{\sum_{i > 1} \frac{d^2_i}{\lambda^2_i}} = O(1 - \rho) \cdot O(\frac{1}{1 - \rho}) = O(1)\,,\]
and hence we can bound $|m_{12}|$ uniformly by some constant. 

Now, we start bounding. First, 
\begin{align*}
  q(\psi) &= m_{11}\cos^2\psi + 2m_{12}\sin\psi\cos\psi + m_{22}\sin^2\psi  \\
  &= m_{11}(1 - O((1-\rho)^2)) + O(1) \cdot O(1 - \rho) \cdot (1 - O((1 - \rho)^2)) + O(\frac{1}{1 - \rho}) \cdot O((1 - \rho)^2) \\
  &\rightarrow \frac{1}{||\boldsymbol{\sigma}||^2}
\end{align*}
as $\rho \rightarrow 1$ (uniformly over all such planes $P$). The same uniform convergence holds for $q^{-1/2}$, $q^{-3/2}$ and $q^{-5/2}$. 

Second, 
\begin{align*}
|q'(\psi)| &= \big|\sin(2\psi) (m_{22} - m_{11}) + 2m_{12} \cos(2\psi) \big| \\
&\leq |\sin(2\psi)|(m_{22} + m_{11}) + 2 | m_{12}| \\
&\leq O(1 - \rho) O(\frac{1}{1 - \rho}) + O(1-\rho) + O(1) = O(1)\,,
\end{align*}
as $\rho \rightarrow 1$. 

Third, 
\begin{align*}
    q''(\psi)&= 2\cos(2\psi) (m_{22} - m_{11}) - 4m_{12} \sin(2\psi) \\
    &\geq 2(1 - O((1-\rho)^2))(\Theta(\frac{1}{1 - \rho}) - O(1)) - O(1)\cdot O(1 - \rho)
    \geq \Theta(\frac{1}{1 - \rho})\,,
\end{align*}
as $\rho \rightarrow 1$. 

Let $\Psi(\boldsymbol{a}_0, \boldsymbol{a}_1)$ be the feasible set of $\psi$ such that it stays in $\mathcal{A}(\rho)$. Combining all the above together, we have that 
\[\sup_{\boldsymbol{a}_0, \boldsymbol{a}_1 \in \mathcal{A}(\rho) ; \psi \in \Psi(\boldsymbol{a}_0, \boldsymbol{a}_1)}|Q'(\psi;\boldsymbol{a}_0, \boldsymbol{a}_1)| \leq C_1(\rho) \rightarrow C_1\]
as $\rho \rightarrow 1$, and  
\[\sup_{\boldsymbol{a}_0, \boldsymbol{a}_1 \in \mathcal{A}(\rho) ; \psi \in \Psi(\boldsymbol{a}_0, \boldsymbol{a}_1)} Q''(\psi;\boldsymbol{a}_0, \boldsymbol{a}_1) \leq C_2(\rho) \rightarrow -\infty\]
as $\rho \rightarrow 1$. Recall the decomposition we had: 
\[\frac{\partial^2 \overline{U}}{\partial \psi^2} = \underbrace{Q'' \sum_i a_i h_i(\beta_i)}_{\text{I}} + \underbrace{2Q' \sum_i a'_i h_i(\beta_i)}_{\text{II}} + \underbrace{Q \sum_i a''_i h_i(\beta_i)}_{_{\text{III}}} + \underbrace{\sum_i h'_i(\beta_i)(a'_i Q + a_i Q')^2}_{_{\text{IV}}}\,. \]
Note that, as before, we have 
\[||\boldsymbol{a}'(\psi)|| = 1\,, \text{ and } ||\boldsymbol{a}''(\psi)|| = 1\,.\]
Moreover, the bounds we had for $h_i$, $h'_i$, and $Q$ continue to hold here. It follows immediately by our previous argument that parts II, III, and IV are uniformly bounded by a constant for sufficiently high $\rho$, while part I uniformly diverges to $-\infty$ for sufficiently high $\rho$. By the same argument as before, for any $\delta > 0$, there exists some $\underline{\rho}'< 1$ such that for all $\rho > \underline{\rho}'$, we have 
\[\frac{\partial^2 \overline{U}}{\partial \psi^2}(\boldsymbol{a}; \boldsymbol{a}_0, \boldsymbol{a}_1, \boldsymbol{p}, \rho) < - \delta \,\]
for any $\boldsymbol{a}_0, \boldsymbol{a}_1 \in \mathcal{A}(\rho)$, any $\boldsymbol{a}$ in the geodesic that connects $\boldsymbol{a}_0$ and $\boldsymbol{a}_1$, and any $\boldsymbol{p} \in \mathcal{P}$. 

\paragraph{Completion of the proof.}\hspace{-2mm}Now we complete the proof by showing that for any $\rho > \underline{\rho}'$, where $\underline{\rho}'$ is given in the previous step, for any $\boldsymbol{p} \in \mathcal{P}$, we have that 
\[\argmax_{\boldsymbol{a} \in \mathcal{A}(\rho)} \overline{U}(\boldsymbol{a}; \boldsymbol{p}, \rho)\]
is single-valued. Suppose for contradiction that there exist some  $\rho > \underline{\rho}'$ and some $\boldsymbol{p} \in \mathcal{P}$ such that 
\[|\argmax_{\boldsymbol{a} \in \mathcal{A}(\rho)} \overline{U}(\boldsymbol{a}; \boldsymbol{p}, \rho)| > 1\,.\]
Fix any two optimizers $\boldsymbol{a}, \boldsymbol{\hat{a}} \in \mathcal{A}(\rho)$. Note that there must be a geodesic connecting $\boldsymbol{a}$ and $\boldsymbol{\hat{a}}$. It follows immediately by the strict concavity proved in the previous step that there exists some $\tilde{\boldsymbol{a}}$ such that 
\[\overline{U}(\tilde{\boldsymbol{a}}; \boldsymbol{p}, \rho) > \overline{U}(\boldsymbol{a}; \boldsymbol{p}, \rho)\,,\]
a contradiction. 

\subsection{Proof of \Cref{prop:no-separate-sales}}

Let $\rho<0$.  Toward a contradiction, suppose that there exists an equilibrium $(\boldsymbol{\alpha}^*, \mathcal{M}^*)$ in which the seller offers a separate sales mechanism. That is, mechanism $\mathcal{M}^*$ is a menu that offers each good $i$ at price $p_i$ and each bundle $B\subseteq\{1,\dots, K\}$ at price $\sum_{i\in B} p_i$. Recall that optimal prices always lie below the mean since the virtual value function crosses zero only once at some $t_i^*\leq 0.5$ (\Cref{lem:regular}). So $p_i\leq \mu_i$ for all $i$.  By \Cref{thm:main}, the buyer's learning must be vertical. However, we show that against separate sales, any vertical learning strategy is strictly dominated by a horizontal learning strategy when $\rho<0$. Thus, $(\boldsymbol{\alpha}^*, \mathcal{M}^*)$ cannot be an equilibrium. 

\vspace{-3mm}
\paragraph{Rewriting of the buyer's payoff.}\hspace{-2mm}We follow the same reformulation as in the proof of \Cref{lem:dominance}.  Under separate sales, the buyer's payoff is separable across goods: 
\[U(\boldsymbol{\alpha}) = \mathbb{E}\left[\sum_i\left(\theta_i(s; \boldsymbol{\alpha})-p_i\right)_+\right]\,.\]
Instead of optimizing over $\boldsymbol{\alpha}$, it is equivalent to optimizing over $\boldsymbol{a}:= \Sigma \boldsymbol{\alpha}$. As in the proof of \Cref{lem:dominance}, for any choice $\boldsymbol{a}$, the buyer's expected payoff equals: 
\[\overline{U}(\boldsymbol{a}):=\sum_i\mathbb{E}\left[\left(\theta_i(s; \boldsymbol{\alpha}(\boldsymbol{a}))-p_i\right)_+\right]= \sum_i\mathbb{E}\left[\left(\mu_i-p_i+\frac{a_i}{\sqrt{\boldsymbol{a}^\top \Sigma^{-1}\boldsymbol{a}}}\overline{s}\right)_+\right]\,,\]
where 
\[\overline{s}:= \frac{\boldsymbol{a}^\top \Sigma^{-1} \mathbf{v} - \boldsymbol{a}^\top \Sigma^{-1}\boldsymbol{\mu}}{\sqrt{\boldsymbol{a}^\top \Sigma^{-1} \boldsymbol{a}}}\]
is a standardized signal, whose law does not depend on $\boldsymbol{a}$. 

Consider the function
\[g_i(l):=\mathbb{E}\left[\left(\mu_i-p_i+l\overline{s}\right)_+\right]\,.\]
As in the proof of \Cref{lem:dominance}, note that 
\[g_i'(l) = \mathbb{E}\big[\overline{s}\mathbbm{1}\{\mu_i-p_i+l\overline{s}\geq 0\}\big]=\mathbb{E}\big[\overline{s}\mathbbm{1}\{\overline{s}\geq -(\mu_i-p_i)/l\}\big]\geq \mathbb{E}\big[\overline{s}\big]=0\,,\]
where the last inequality comes from $l>0$. The inequality is strict whenever $-(\mu_i-p_i)/l$ is strictly greater than the lowest possible realization of $\overline{s}$. Also note that the distribution of $\overline{s}$ is symmetric around zero, and hence the function $g_i$ is even: $g_i(l) = g_i(-l)$ for all $l$, $i$. 

Moreover, as in the proof of \Cref{lem:dominance}, write 
\[\Sigma = \text{diag}(\sigma_i)\, R\, \text{diag}(\sigma_i)\quad\text{where}\quad R = (1-\rho)I+\rho\boldsymbol{\iota}\boldsymbol{\iota}^\top, \]
where $\iota$ is the constant $\mathbf{1}$ vector. Letting $\boldsymbol{\omega}:=(a_i/\sigma_i)_i$, we have 
\begin{align*}
    \boldsymbol{a}^\top \Sigma^{-1}\boldsymbol{a}=\frac{1}{1-\rho} \, \left(||\boldsymbol{\omega}||^2-\frac{\rho}{1+\rho (K-1)}(\boldsymbol{\iota}^\top\boldsymbol{\omega})^2\right)\,.
\end{align*}
Note that by construction we also have $1+\rho (K-1) > 0$. 

\vspace{-3mm}
\paragraph{Vertical learning is dominated.}\hspace{-2mm}Fix any vertical learning strategy $\boldsymbol{\alpha}$. Let $\boldsymbol{a} = \Sigma\boldsymbol{\alpha}$, so we have $\overline{U}(\boldsymbol{a}) = U(\boldsymbol{\alpha})$. Note that $a_i=\text{Cov}(v_i, \boldsymbol{\alpha}^\top\mathbf{v})$ for all $i$, by construction. By definition of vertical learning, $\text{Cov}(v_i, \boldsymbol{\alpha}^\top\mathbf{v})\geq 0$ for all $i$, and hence $a_i\geq 0$ for all $i$. 

Furthermore, it must be that $a_i>0$ for some $i$. If not, then it implies that $\boldsymbol{\alpha}=0$. However, such a learning strategy cannot be sustained in equilibrium (indeed, if $\boldsymbol{\alpha}=0$, the seller's optimal separate sales prices equal $(\mu_1, \mu_2)$, and learning any $v_i$ yields the buyer $\mathbb E[(v_i-\mu_i)_+]>0$).

Now, we consider two cases.

\textbf{Case (A):} Suppose that there exist at least two goods with $a_j > 0$. We construct an alternative learning strategy $\hat{\boldsymbol{\alpha}}$ that strictly improves over $\boldsymbol{\alpha}$. Pick any good $i$ such that 
\[i \in \argmin\big\{a_j /\sigma_j : a_j > 0\big\}\,.\]
Set $\hat{a}_i=-a_i$, $\hat{\boldsymbol{a}}_{-i} = \boldsymbol{a}_{-i}$. That is, $\hat{\boldsymbol{a}}$ simply flips the sign of $a_i$ but keeps everything else constant. Finally, let $\hat{\boldsymbol{\alpha}}=\Sigma^{-1}\hat{\boldsymbol{a}}$. Note that $U(\hat{\boldsymbol{\alpha}}) = \overline{U}(\hat{\boldsymbol{a}})$.

Using the above notation, let $\boldsymbol{\omega}=(a_j/\sigma_j)_j$ and $\hat{\boldsymbol{\omega}}=(\hat{a}_j/\sigma_j)_j$. By construction:
\[|\hat{a}_j| = |a_j|\quad \forall j\quad\text{and}\quad ||\hat{\boldsymbol{\omega}}|| = || \boldsymbol{\omega}||\,.\]
Furthermore, by construction:
\[0 \leq \boldsymbol{\iota}^\top\hat{\boldsymbol{\omega}} = \sum_j \frac{\hat{a}_j}{\sigma_j}<\sum_j \frac{a_j}{\sigma_j}=\boldsymbol{\iota}^\top\boldsymbol{\omega}\,.\]
Since $\rho<0$, this implies:  
\[ \hat{\boldsymbol{a}}^\top \Sigma^{-1}\hat{\boldsymbol{a}}< \boldsymbol{a}^\top \Sigma^{-1}\boldsymbol{a}\,,\]
and 
\[ \frac{|\hat{a}_j|}{\sqrt{\hat{\boldsymbol{a}}^\top \Sigma^{-1}\hat{\boldsymbol{a}}}}\geq  \frac{|a_j|}{\sqrt{\boldsymbol{a}^\top \Sigma^{-1}\boldsymbol{a}}}\quad \forall j\,,\]
and strictly so for $j=i$. The function $g_j$ is even and increasing over the positive range. Therefore:
\[\mathbb{E}\left[\left(\mu_j-p_j+\frac{\hat{a}_j}{\sqrt{\hat{\boldsymbol{a}}^\top \Sigma^{-1}\hat{\boldsymbol{a}}}}\overline{s}\right)_+\right]\geq \mathbb{E}\left[\left(\mu_j-p_j+\frac{a_j}{\sqrt{\boldsymbol{a}^\top \Sigma^{-1}\boldsymbol{a}}}\overline{s}\right)_+\right]\quad \forall j\,.\]
Furthermore, the inequality holds strictly for $j=i$ as long as 
\[\Pr\left(\frac{a_i}{\sqrt{\boldsymbol{a}^\top \Sigma^{-1}\boldsymbol{a}}}\overline{s}<-(\mu_i - p_i)\right)>0\,.\]
Since $a_i>0$, $\text{Var}(\theta_i(s; \boldsymbol{\alpha}))>0$. We know from \Cref{lem:optx} that an optimal mechanism cannot allocate good $i$ to \emph{all} buyer-types, that is: 
\[\Pr\Big(\theta_i(s; \boldsymbol{\alpha})<p_i\Big)>0\implies \Pr\left(\mu_i +\frac{a_i}{\sqrt{\boldsymbol{a}^\top \Sigma^{-1}\boldsymbol{a}}}\overline{s} <p_i\right)>0\,.\]
Thus, the inequality holds strictly for good $i$. Summing over all goods $j$, we get 
\[\overline{U}(\hat{\boldsymbol{a}})>\overline{U}(\boldsymbol{a})\implies U(\hat{\boldsymbol{\alpha}}) >U(\boldsymbol{\alpha})\,.\]
Therefore, against the separate sales mechanism $\mathcal{M}^*$, the buyer has a strictly profitable deviation under any vertical learning strategy. But by \Cref{thm:main}, every equilibrium must have vertical learning. Thus, $(\boldsymbol{\alpha}^*, \mathcal{M}^*)$ cannot be an equilibrium.

\textbf{Case (B):} Suppose that there exists only one good $i$ such that $a_i > 0$ (and hence $a_j = 0$ for all $j \neq i$). By optimality of $\mathcal{M}^*$, it must be that all equilibrium types $t$ consume goods $j \neq i$ (with prices being $\mu_j$). Moreover, as noted before, there must exist a positive measure of equilibrium types $t$ that do not consume good $i$. Therefore, 
\[U(\boldsymbol{\alpha}^*) = \mathbb{E}\left[\sum_j\left(\theta_j(s; \boldsymbol{\alpha}^*)-p_j\right)_+\right] = \E\Bigg[\big(\theta_i(s; \boldsymbol{\alpha}^*)-p_i\big)_+\Bigg]\,.\]
However, consider the learning strategy $\hat{\alpha}_i = 1$ and $\hat{\alpha}_j = 0$ for all $j \neq i$. Since $\rho < 0$, note that this is a horizontal learning strategy (since $a_j < 0$ for all $j \neq i$). Therefore, $\boldsymbol{\hat{\alpha}}$ is a different signal from $\boldsymbol{\alpha}^*$ and induces a strict mean-preserving spread of $\theta_i$. By \Cref{lem:strictJensen}, we have 
\[U(\boldsymbol{\hat{\alpha}}) = \mathbb{E}\left[\sum_j\left(\theta_j(s; \boldsymbol{\hat{\alpha}})-p_j\right)_+\right] \geq  \E\Bigg[\big(\theta_i(s; \boldsymbol{\hat{\alpha}})-p_i\big)_+\Bigg] > U(\boldsymbol{\alpha}^*) \,,\]
and hence $\boldsymbol{\hat{\alpha}}$ is a strict improvement for the buyer, contradicting $(\boldsymbol{\alpha}^*, \mathcal{M}^*)$  being an equilibrium.

\subsection{Proof of \Cref{prop:lineartypes}}

\paragraph{Vertical Learning.}\hspace{-2mm}Suppose for contradiction that an equilibrium signal $s \in \mathcal{S}$ induces a horizontal demand system. Reversing the parametrization and relabeling the goods if necessary, we can write 
\[ \theta_1(t)=a_1t+b_1,\qquad \theta_2(t)=a_2t+b_2, \]
where $a_1 > 0 > a_2$, $ a_1\geq |a_2|$, $a_1+a_2 \geq 0$, and $t \in [0, 1]$. Let $r := -a_2/a_1 \in (0,1]$. Since $\theta_2(1) \geq 0$, we have $b_2 \geq -a_2 > 0$. 

Consider an optimal direct mechanism $(x(t),p(t))$ and write
\[U(t):=\boldsymbol\theta(t)\cdot x(t)-p(t),\qquad
q(t):=a_1x_1(t)+a_2x_2(t) \,.\]
We first record the implication of the mechanism characterization in the proof of \Cref{thm:main}. The auxiliary program becomes
\[
\max_{x\in[0,1]^2}\ b_1x_1+b_2x_2 \qquad\text{subject to}\qquad a_1x_1 + a_2x_2 = 0\,.
\]
Note that its unique solution is $X^*=\{(r,1)\}$. Moreover,
\[
\lambda:=-\frac{b_1}{a_1}
\]
is an optimal multiplier: $b_1 + a_1\lambda = 0$ and $b_2 + a_2\lambda = b_2 + rb_1  > 0$. Thus, good $2$ is the unique negative good, good $1$ is the unique positive balancing good, and there are no positive non-balancing goods.

Let $F$ denote the distribution of $t$ and, following the proof of \Cref{lem:saddlepoint}, let
\[
g_F(t_0):=\overline{\Phi}(t_0;t_0).
\]
The regularity in \Cref{lem:regular} is used there to ensure that $g_F(0)<\lambda$ and to obtain strict single-crossing for the positive non-balancing goods. The latter role is vacuous here. If $g_F(0)<\lambda$, continuity of $g_F$ and $g_F(1)>0\geq\lambda$ yield an interior $t_0^*$ such that $g_F(t_0^*)=\lambda$. The saddle-point argument then applies directly. The initial $\lambda$-ironing interval as in \Cref{lem:saddlepoint} contains $t_0^*$ and every lower type. Ironing consistency and the fact that $t_0^*$ is a worst-off type then imply that $q(t) = 0$ on this interval. Moreover, as in \Cref{lem:optx}, the negative good must be allocated with full probability for almost all types (the negative-good argument there does not use \Cref{lem:regular}). 

If instead $g_F(0) \geq \lambda$, then the same construction with the candidate worst-off type $t_0^* = 0$ gives a saddle point. Indeed, by monotonicity of $\overline{\Phi}(\,\cdot\,; 0)$, for the positive balancing good, we have that for all $t$: 
\[a_1 \overline{\Phi}(t; 0) + b_1 \geq a_1 \overline{\Phi}(0; 0) + b_1 \geq a_1 \lambda + b_1 = 0\,.\]
Moreover, for the negative good, we have that for all $t$:
\[a_2 \overline{\Phi}(t; 0) + b_2 \geq a_2 \overline{\Phi}(1; 0) + b_2 = a_2 + b_2 \geq 0\,,\]
where the first inequality is strict for all $t < 1$ since $t = 1$ cannot be in an ironing interval. Therefore, letting $x^*_1 = r \mathbbm{1}\{a_1 \overline{\Phi}(t; 0) + b_1 = 0\} + \mathbbm{1}\{a_1 \overline{\Phi}(t; 0) + b_1 > 0\}$ and $x^*_2 = \mathbbm{1}\{t \in [0, 1]\}$ gives an allocation rule $x^*$ that maximizes the ironed objective pointwise. Moreover, clearly $\sum_i a_i x^*_i(\,\cdot\,)$ is monotone and $x^*$ is consistent with ironing. Also, since $\sum_i a_i x^*_i(\,\cdot\,) \geq 0$, $t^*_0 = 0$ is indeed a worst-off type. Thus, we have found a saddle point as in \Cref{lem:saddlepoint}. The rectangularity property of saddle points then implies that every optimal mechanism forms a saddle point with $t = 0$. Then, the above argument implies that good $2$ is allocated with full probability for almost all types under every optimal mechanism. 

Combining the two cases, every optimal mechanism therefore satisfies
\begin{equation}\label{eq:horizontal-seller-properties}
x_2(t)=1\,\,\text{ and }\,\,0\leq q(t)\leq a_1+a_2 \,,
\end{equation}
for almost all $t$ and that the lowest type is a worst-off type with $U(0) = 0$. Let $\mathcal{M}$ be the equilibrium menu and $(x(t), p(t))_t$ be the equilibrium outcomes selected by the types. Let 
\[
G(z) := \sup_{(x,p)\in\mathcal{M}}\{zx_1+\mu_2x_2-p\} \,.
\]
This is the payoff of a hypothetical type $(z, \mu_2)$ given the equilibrium menu. In particular, $G$ is convex and continuous. By \eqref{eq:horizontal-seller-properties}, we have
\[G\big(\theta_1(t)\big) \geq \theta_1(t) x_1(t) + \mu_2 x_2(t) - p(t) = U(t)  + \mu_2 x_2(t) - \theta_2(t) x_2(t) = U(t) + \mu_2 - \theta_2(t)\,. \]
Moreover, since good $2$ is the negative good, we also have $U(0) + \mu_2 - \theta_2(0) < 0$, while $G(\theta_1(0)) \geq 0$ given the menu has the outside option. Thus, by continuity, the above inequality is strict for a positive measure of types. Thus, 
\begin{equation}\E\big[G\big(\theta_1(t)\big)\big] > \E[U(t) + \mu_2 - \theta_2(t)] = \E[U(t)]\,.  \label{eq:horizontal-strict-gain}
\end{equation}

Note that learning $v_1$ is feasible by assumption, and mutual mean independence implies that it induces posterior types $(v_1, \mu_2)$. Since $\theta_1 = \mathbb{E}[v_1 \mid s]$, by \eqref{eq:horizontal-strict-gain} and Jensen's inequality, we have 
\[ \mathbb{E}[G(v_1)] \geq \mathbb{E}[G\big(\E[v_1 \mid s]\big)] = \mathbb{E}[G(\theta_1)] > \mathbb{E}[U(t)]\,.\]
Thus, learning $v_1$ is a strictly profitable deviation---a contradiction.

\vspace{-3mm}
\paragraph{Equilibrium Existence.}\hspace{-2mm}Suppose now that the values are exchangeable.  By exchangeability, we have 
\[
\mathbb{E}[v_1\mid v_1+v_2] = \mathbb{E}[v_2\mid v_1+v_2] = \frac{1}{2}(v_1+v_2)\,.
\]
Therefore, learning $v_1+v_2$ induces posterior types $(\frac{v_1 + v_2}{2}, \frac{v_1 + v_2}{2})$. Now, let
\[
k^* \in \argmax_k \Big\{k \cdot \Pr\Big( \frac{1}{2}(v_1+v_2) \geq k\Big)\Big\}\,.
\]
Against diagonal types, only the total quantity $x_1+x_2\in[0,2]$ matters. Any incentive-compatible total allocation is a mixture of cutoff rules, so its expected revenue is at most $2\max_k \{k \Pr\big(\frac{1}{2}(v_1+v_2) \geq k\big)\}$. Pure bundling at price $2k^*$ attains this bound. Moreover, learning about the bundle $v_1 + v_2$ resolves all payoff-relevant uncertainty when the menu only contains $\{1,2\}$ and $\varnothing$. Hence, learning $v_1+v_2$ and pure bundling at price $2k^*$ form a deterministic-menu equilibrium.

\vspace{-3mm}
\paragraph{Nested Bundling.}\hspace{-2mm}Fix a deterministic-menu equilibrium $(s, \mathcal{M})$. By the first part, we know that the equilibrium has vertical learning. Thus, we can write $\theta_i(t) = a_i t+ b_i$ with $a_i \geq 0$ for all $i$. Suppose first that $a_1 b_2 \neq a_2 b_1$. Let $q(t):=a_1 x_1(t) + a_2 x_2(t)$. A type $t$ that reports $\hat{t}$ obtains $U(\hat{t}) + (t-\hat{t}) q(\hat{t})$. Therefore, note that holding $(q,U)$ fixed, seller optimality requires
\[
x(t)\in \argmax_{x\in[0,1]^2} \big\{ b_1x_1 + b_2x_2 : a_1x_1+a_2x_2 = q(t) \big\}\,.
\]
If $a_1, a_2>0$, then we have $b_1/a_1 \neq b_2/a_2$. Without loss of generality, say $b_1/a_1 > b_2/a_2$. Then, for all $t$, we must have $x_2(t) > 0 \implies x_1(t) = 1$. Then, since the on-path assignment is deterministic by assumption, the on-path assignments can only be in $\big\{\varnothing, \{1\},  \{1, 2\}\big\}$.  The same conclusion is immediate if $a_i = 0$ for some $i$, since that good must be included in every non-outside allocation. After removing options that are not chosen by a positive measure of types, we then have a nested-bundling equilibrium: Indeed, as before, the original signal remains optimal because the new menu is a subset of $\mathcal{M}$.

It remains to consider the case where $a_1b_2 = a_2b_1$, which implies that the posterior mean line can be extended to pass the origin $(0, 0)$. Since exchangeability implies $\mu_1 = \mu_2$ and the posterior-mean line contains the prior mean, the line must be the diagonal. Let $Z := \E\big[v_1 \mid s\big] = \E\big[v_2 \mid s\big]$. Then the posterior type is $(Z, Z)$, which also satisfies $Z = \E[\frac{v_1 + v_2}{2} \mid s]$. Since the menu is deterministic, its on-path total quantity $x_1 + x_2$ is nondecreasing and takes values in $\{0, 1, 2\}$. Therefore, there exist cutoffs $k_1 \leq k_2$ such that its on-path indirect utility is
\[
g(z) = (z-k_1)_+ + (z-k_2)_+\,.
\]
Let $R_Z(k):=k \Pr(Z \geq k)$. The seller's revenue is $R_Z(k_1)+R_Z(k_2)$. Since selling both goods at a common cutoff is feasible, seller optimality implies
\begin{equation}\label{eq:radial-monopoly-prices}
k_1,\, k_2\in\argmax_k R_Z(k)\,.
\end{equation}

The signal $W := v_1 + v_2$ induces posterior types $(T, T) := (\frac{v_1 + v_2}{2}, \frac{v_1 + v_2}{2})$ and gives the buyer at least $\mathbb{E}[g(T)]$ against the original menu. Since $Z = \mathbb{E}[T \mid s]$, by Jensen's inequality, we have
\[
\mathbb{E}[g(T)] \geq \mathbb{E}[g(Z)]\,.
\]
Moreover, since $s$ is optimal for the buyer, the reverse inequality must also hold. Thus, the above holds with equality. Moreover, since $g$ is the sum of two convex functions, this implies that we must have 
\begin{equation}\label{eq:radial-cutoff-sufficiency}
\mathbb{E}[(T-k_1)_+] = \mathbb{E}[(Z-k_1)_+]\,.
\end{equation}

Now, replace $\mathcal{M}$ with pure bundling at price $2 k_1$. By \eqref{eq:radial-monopoly-prices}, this gives the seller the same revenue. Moreover, against this menu, any signal $s'$ gives the buyer at most
\[
\mathbb{E}[(W - 2k_1)_+]=2 \cdot \mathbb{E}[(T-k_1)_+]
=2 \cdot \mathbb{E}[(Z-k_1)_+]\,,
\]
where the last equality follows from \eqref{eq:radial-cutoff-sufficiency}. The original signal $s$ attains this bound, so it remains a best response. Therefore, we have found a nested-bundling equilibrium. Note that the buyer's expected utility changes from
\[
\mathbb{E}[(Z-k_1)_+]+\mathbb{E}[(Z-k_2)_+]
\]
to $2 \cdot \mathbb{E}[(Z-k_1)_+]$, which is weakly higher since $k_1 \leq k_2$. Thus, the resulting nested-bundling equilibrium weakly Pareto dominates the original equilibrium, concluding the proof.

\subsection{Proof of \Cref{prop:pure}}
We construct a pure bundling equilibrium. Consider learning the grand bundle $s = \sum_i v_i$, or in the linear signal notation $\boldsymbol{\alpha} = \boldsymbol{1}$. Since $\mathbf{v}$ is elliptical, such a learning strategy leads to the following mapping between signal realizations and types: for all $i$, 
\[\theta_i(s; \boldsymbol{\alpha}) := \mu_i+\frac{\sigma_i^2+\rho\sigma_i\sum_{j\neq i}\sigma_j}{\sum_k\Big(\sigma_k^2+\rho\sum_{l\neq k}\sigma_l\sigma_k\Big)}\left(s-\sum_j\mu_j\right)\,,\]
with $s = \sum_iv_i$.  By the proof of \Cref{thm:main}, we know that pure bundling is optimal in response to such a comonotonic type distribution if $t_i^*=t_j^*$ for all $i$, $j$, where $t_i^*$, $t_j^*$ are as defined in \Cref{lem:regular}. Using the notation from the proof of \Cref{thm:main}, this is the case if $a_i/b_i = a_j/b_j$ for all $i$, $j$, or equivalently, if the segment on which types are supported can be extended to cross the origin---i.e., there exists some $s_0\in \mathbb{R}$ such that $\theta_i(s_0; \boldsymbol{\alpha})=0$ for all $i$. We can substitute out $s_0$ and rewrite $\theta_i$ as a function of any $\theta_j$ as follows: 
\[\theta_i(s; \boldsymbol{\alpha}) = \mu_i+\frac{\sigma_i^2+\rho\sigma_i\sum_{k\neq i}\sigma_k}{\sigma_j^2+\rho\sigma_j\sum_{k\neq j}\sigma_k}[\theta_j(s; \boldsymbol{\alpha})-\mu_j]\,.\]
Thus, there exists $s_0$ such that $\theta_i(s_0; \boldsymbol{\alpha})=0$ for all $i$ if and only if we have the following condition: 
\[\mu_i = \frac{\sigma_i^2+\rho\sigma_i\sum_{k\neq i}\sigma_k}{\sigma_j^2+\rho\sigma_j\sum_{k\neq j}\sigma_k}\mu_j\quad\forall i\neq j\,.\]
Now, under this condition, since the extended posterior mean line must connect $\boldsymbol{\mu}$ and $\boldsymbol{0}$ and $\boldsymbol{\mu} > 0$, the strategy $\boldsymbol{\alpha} = \boldsymbol{1}$ must be a vertical learning strategy. Hence, the seller finds it optimal to offer only the grand bundle at some price. Then, the buyer, of course, finds it optimal to learn fully the grand bundle value and nothing else,  irrespective of how many signals he can acquire.

\subsection{Proof of \Cref{prop:cost}}

Note that with $\rho = 0$, the elliptical distribution satisfies the mutual mean independence condition in \Cref{prop:lineartypes}. Now, suppose for contradiction that there exists a purely horizontal learning equilibrium. Since $\{v_1, v_2\} \subseteq \mathcal{S}$, by the proof of \Cref{prop:lineartypes}, note that the buyer has a strictly profitable deviation against the seller's equilibrium menu, even if information costs are not taken into account. Now, given that the deviations also carry a weakly lower information cost by assumption, any such profitable deviation must continue to be strictly profitable for the buyer---a contradiction.

\subsection{Proof of \Cref{cor:bound}}

\paragraph{Classifying All Equilibria.}\hspace{-2mm}First, the buyer must acquire at least one informative signal. Otherwise, the seller faces the deterministic type $\boldsymbol{\mu}$ and extracts the entire surplus by offering the grand bundle at price $\mu_1 + \mu_2$. Because the first signal is free, the buyer can profitably deviate by learning $v_1 + v_2$ and conditioning his purchase decision on the realized bundle value.

Now suppose the buyer acquires exactly one signal. Any one-dimensional linear signal is free under the assumed cost structure. With $\rho=0$, purely horizontal learning corresponds to a linear signal whose two weights have strictly opposite signs. Learning either individual value is also feasible and free. Thus, \Cref{prop:cost} rules out purely horizontal learning, so every one-signal equilibrium has vertical learning. Moreover, because every alternative one-signal deviation is available at the same zero cost, such an equilibrium is also an equilibrium of our baseline one-signal game. By \Cref{thm:main}, it is therefore outcome-equivalent to a nested-bundling equilibrium.

Finally, suppose that the buyer acquires two signals,
\[
s^A=\boldsymbol{\alpha}^A \cdot \mathbf v \qquad\text{and}\qquad s^B=\boldsymbol{\alpha}^B \cdot \mathbf v\,.
\]
The vectors $\boldsymbol{\alpha}^A$ and $\boldsymbol{\alpha}^B$ cannot be collinear: otherwise, the second signal would reveal no information beyond the first, and the buyer could drop it while saving the strictly positive cost $c$. Hence, the two vectors are linearly independent. Since there are two goods, the buyer can invert the resulting system of two equations and recover $(v_1, v_2)$ exactly. Thus, every two-signal equilibrium involves full learning, and the seller's equilibrium mechanism is optimal against the original value distribution, as desired.

\vspace{-3mm}
\paragraph{Robustness of One-Signal Equilibrium.}\hspace{-2mm}First, we record the following lemma: 

\begin{lemma}\label{lemma:pricebelowmean}
    Let $F$ be a symmetric unimodal distribution supported on $[\underline{v}, \overline{v}]$ where $\underline{v}\geq 0$. Then, there exists a unique optimal monopoly price $p_F \leq \mathbb{E}[v]$.
\end{lemma}
 
\begin{proof}
This follows by the proof of \Cref{lem:regular}.  
\end{proof}

Now, fix any nested bundling equilibrium of our main model. If the equilibrium involves pure bundling, then the equilibrium signal must be $v_1+v_2$, in which case the buyer clearly does not have incentive to acquire a second signal at cost $c > 0$. Otherwise, without loss of generality, suppose that good 1 is the base good, such that the equilibrium menu sells good 1 at price $p_1$ and the bundle at price $p_1+p_\delta$. Let $U^*$ be the buyer's equilibrium payoff, i.e., his payoff when he acquires only one signal. If the buyer acquires a second signal, he can become fully informed and achieve his full information payoff, which we denote by $U_{FI}^{NB}$. Thus, if the associated cost $c\geq U_{FI}^{NB}-U^*$, then the buyer finds it optimal not to acquire the second signal. 

We first bound the buyer's equilibrium payoff $U^*$ from below. A strategy available to the buyer is to fully learn his value for the upgrade good $v_2$ and to learn nothing about the base. Since $p_1\leq \mu_1$ by \Cref{lemma:pricebelowmean} (recall that by our characterization, $p_1$ must be a monopoly price for good $1$ when good $1$ is sold alone), doing so leads the buyer to always buy good 1 and to buy the upgrade if and only if $v_2\geq p_\delta$. This learning strategy then yields an expected payoff of $U_2^{NB}=\mathbb{E}[\max\{v_2-p_\delta,0\}]+\mu_1-p_1$. His equilibrium payoff must then be weakly higher:  $U^*\geq U_2^{NB}$. 

We now bound the full information payoff $U_{FI}^{NB}$ from above. Note that the payoff from learning good 2 fully is unchanged under the separate sales mechanism that sells good 1 at price $p_1$ and good 2 at price $p_\delta$ since $p_1\leq \mu_1$. So $U_2^{NB}=U_2^{SS}$. The full information payoff is always weakly greater under the separate sales mechanism than under the nested bundling mechanism since the separate sales mechanism expands the choice set of the buyer: $U_{FI}^{SS}\geq U_{FI}^{NB}$.    

Combining the above bounds, we have 
\[U_{FI}^{NB}-U^*\leq U_{FI}^{NB}-U_2^{NB}=U_{FI}^{NB}-U_2^{SS}\leq U_{FI}^{SS}-U_2^{SS} = \mathbb{E}\big[\max\{v_1-p_1,0\}\big]-(\mu_1-p_1)\,.\]
Thus, for any  
\[c \geq \mathbb{E}\big[\max\{v_1-p_1,0\}\big]-(\mu_1-p_1) =  \mathbb{E}\big[\max\{v_1-p_1,0\}\big]-\max\{\mu_1-p_1, 0\} \,,\]
we have that the buyer prefers not to acquire the second signal under the one-signal nested bundling equilibrium, proving the result. 

\section{Additional Results and Proofs}\label{app:extensions}

\Cref{subsec:nonadditive,subsec:productioncosts,subsec:weak} provide the extensions discussed in the main text; \Cref{subsec:nonadditive-proof,subsec:costs-proof,subsec:weak-proof} provide the proofs of these additional results. 

\subsection{Non-Additive Values}\label{subsec:nonadditive}

Suppose that we have two goods and the value for the bundle $\{1, 2\}$ is
\[\gamma \big(v_1 + v_2\big)\,,\]
where $\gamma > 1$ when the two goods are complements, $\gamma < 1$ when the two goods are substitutes, and $\gamma = 1$ when the two goods are additive. We assume that a larger bundle gives a strictly higher value: for all $\boldsymbol{v} \in V$, 
\[\gamma \big(v_1 + v_2\big) > \max\{v_1, v_2\}\,.\]

As in the main model, we allow the buyer to learn any $\alpha_1 v_1 + \alpha_2 v_2$, which in this case is equivalent to learning any linear combination of various bundle values. We say that an equilibrium has \textit{\textbf{vertical learning}} if the posterior means for all bundle values are comonotonic, and \textit{\textbf{horizontal learning}} otherwise. Perhaps surprisingly, our main result continues to hold in this setting regardless of whether the goods are complements or substitutes: 

\begin{proposition}\label{prop:non-additive} 
With two goods and uncorrelated values, for any $\gamma$, every equilibrium has vertical learning and is outcome-equivalent to a nested bundling equilibrium.
\end{proposition}

The proof follows the same logic as the proof of \Cref{thm:main}. For the vertical learning part, it shows that against any horizontal learning strategy, the seller's optimal mechanism turns out to have the same structure as identified in \Cref{sec:proof} regardless of complementarity and substitutability, which then implies a profitable deviation by the buyer. For the nested bundling part, given that equilibrium types must be comonotonic, the proof leverages the nesting condition in \citet{yang2023nested}, which does not require additive values.

\subsection{Production Costs}\label{subsec:productioncosts}

Our main model assumes that the buyer's value for each good $k$ is always weakly higher than the seller's cost. It is then always efficient to allocate all the goods and the only reason the seller might refrain from doing so is to extract more surplus. However, when buyer types are horizontally differentiated, the seller only needs to distort the allocation of \emph{some} goods to maximize revenue. The remaining goods are allocated to all buyer types, who then have no incentive to learn how much they value them. This is a key step in the proof of \Cref{thm:main}, but it does rely on the seller's production costs being lower than any buyer's realized value. We now investigate the robustness of \Cref{thm:main} when this assumption is relaxed.

Suppose that the seller has a constant marginal cost  $c_k$ of producing good $k$ and that $c_k<\mu_k$ for all $k$. Thus, the buyer may have a value for good $k$ below its production cost under some signal realization, but the expected value for good $k$ is still above its cost. 

\begin{proposition}\label{prop:costs}
    Suppose that there are two goods with uncorrelated and log-concave value distributions and that $c_k < \mu_k$ for both goods $k$. Then, every equilibrium has vertical learning, and is outcome-equivalent to a nested bundling equilibrium. 
\end{proposition}

The proof follows the same logic as the proof of \Cref{thm:main}. In particular, the characterization of optimal mechanisms against distributions supported on any line segment in \Cref{sec:proof} can be generalized to incorporate constant marginal costs. If the marginal costs are not too high, then the optimal mechanism against any horizontal learning strategy continues to involve enough bundling responses by the seller (in the form of mixed bundling) such that the buyer finds it optimal to deviate to a vertical learning strategy.

\subsection{A Weakening of Nash Equilibrium}\label{subsec:weak}

This section introduces a solution concept weaker than Nash equilibrium under which our main results in \Cref{sec:main} continue to hold, and equilibrium existence is always guaranteed. The solution concept weakens the requirement that the buyer chooses a fully optimal learning strategy, and only requires an appropriate sense of Blackwell undominance.   

Given some mechanism $\mathcal{M}$, we say a learning strategy $\boldsymbol{\alpha}$ is $\mathcal{M}$-\emph{\textbf{Blackwell dominated}} by another strategy $\boldsymbol{\alpha}'$ if there exists $M^*\subseteq M$ such that \textit{(i)} for any signal realization $s$ in the support of $\boldsymbol{\alpha} \cdot \mathbf{v}$, 
\[M^*\cap \argmax_{m\in M} \Big\{\boldsymbol{\theta}(s;\boldsymbol{\alpha})\cdot \boldsymbol{x}(m)  -p(m)\Big\}\neq \varnothing\]
and \textit{(ii)} $\boldsymbol{\alpha}'$ is strictly Blackwell more informative than $\boldsymbol{\alpha}$ about the relevant normalized payoffs, in the sense that for some $m_0\in M^*$, we have 
\[\Big(\boldsymbol{\theta}(s; \boldsymbol{\alpha})\cdot \big(\boldsymbol{x}(m) - \boldsymbol{x}(m_0) \big)\Big)_{m\in M^*\setminus \{m_0\}} \preceq_{\text{cx}} \Big(\boldsymbol{\theta}(s; \boldsymbol{\alpha}')\cdot \big(\boldsymbol{x}(m) - \boldsymbol{x}(m_0) \big)\Big)_{m\in M^*\setminus \{m_0\}} \,,\]
where $\preceq_{\text{cx}}$ is the usual convex order and these two random vectors are not the same. In words, the set $M^*$ includes an optimal option for each type in the support of the posterior mean distribution under signal $\boldsymbol{\alpha}$, and signal $\boldsymbol{\alpha}'$ is strictly more informative about the options in $M^*$ (since the buyer's decision problem is mean-measurable, by Blackwell's theorem, it suffices to consider the convex order over the posterior mean distributions).\footnote{Moreover, what matters for decision making is \emph{relative} payoffs across options in the menu, hence the normalization. For example, if the buyer either buys good 1 or buys good 2 (but always buys something), then what matters for his decision is not the whole vector $(v_1, v_2)$ but only $v_1-v_2$.} 

We say that a learning strategy $\boldsymbol{\alpha}$ is $\mathcal{M}$-\emph{\textbf{Blackwell undominated}} given some mechanism $\mathcal{M}$ if it is not $\mathcal{M}$-Blackwell dominated by any other strategy. A strategy profile $(\boldsymbol{\alpha}, \mathcal{M})$ forms a \emph{\textbf{weak equilibrium}} if the buyer's learning strategy  $\boldsymbol{\alpha}$ is $ \mathcal{M}$-Blackwell undominated,  and mechanism $\mathcal{M}$ is profit-maximizing against $\boldsymbol{\alpha}$. Unlike Nash equilibrium, this solution concept does not require that the buyer fully best-responds to the seller's mechanism, but only that he chooses a signal that is not Blackwell dominated anticipating the options offered by the seller. Note that the weak equilibria form a superset of the Nash equilibria.\footnote{Indeed, for any Nash equilibrium, if the buyer's strategy were to be Blackwell dominated by another strategy $\boldsymbol{\alpha}'$, then there would exist some $M^*$ and some $m_0$ such that the normalized posterior mean distribution supported on a line segment in $\mathbb{R}^{|M^*|-1}$ under $\boldsymbol{\alpha}'$ would be a strict mean-preserving spread. However, by the same arguments as in the proofs of \Cref{lem:infovaluable} and \Cref{lem:learnzero}, the buyer would then have a strictly profitable deviation---a contradiction.
}

\begin{proposition}\label{prop:weak}
There exists a weak equilibrium. Moreover: 
\begin{itemize}
    \item[(i)] Every weak equilibrium has vertical learning,  and is outcome-equivalent to a nested bundling weak equilibrium. 
    \item[(ii)] In any nested bundling weak equilibrium, the buyer's log-scale posterior variances $\emph{Var}(\log(\theta_i))$ of different items are ordered according to their tiers.
\end{itemize}
\end{proposition}

\Cref{prop:weak} shows that our main results in \Cref{sec:main} generalize to this concept of weak equilibrium, and such a weak equilibrium is guaranteed to exist. In particular, \Cref{thm:main} does not rely on the buyer being able to fully optimize in response to the seller's mechanism. Fairly weak rationality requirements are sufficient to rule out horizontal learning equilibria.  At the same time, even under this weak rationality requirement by the buyer, the prediction about the buyer's learning strategy as in \Cref{prop:ordering} continues to hold---indeed, as we explained in \Cref{sec:main}, the ordering of the posterior variance is mostly due to the optimization by the \textit{seller}, and hence continues to hold even if the buyer does not fully optimize.

\subsection{Proof of \Cref{prop:non-additive}}\label{subsec:nonadditive-proof}
We use $B \in \big\{\{1\}, \{2\}, \{1,2\}\big\}$ to denote a non-empty bundle. The allocation probabilities are given by $x_B$, which must satisfy $\sum_B x_B \leq 1$. 

Given a learning strategy $\boldsymbol{\alpha}$, the buyer's posterior means over these bundles are given by $a_B t + b_B$, where $t \in [0, 1]$, and 
\[a_{\{1, 2\}} = \gamma \big(a_{1} + a_{2} \big)\,,\, b_{\{1, 2\}} = \gamma \big(b_{1} + b_{2}\big)\,.\]
Since $\gamma (v_1 + v_2) > \max\{v_1, v_2\}$ on the compact set $V$, we have that for any $t \in [0, 1]$, 
\[a_{\{1, 2\}} t +  b_{\{1, 2\}} > \max \Big\{a_{1} t + b_{1}\,, a_{2} t + b_{2} \Big\}\,.\]
In particular, $b_{\{1, 2\}} > \max\{b_1, b_2\}$. 

The proof proceeds in the same way as the proof of \Cref{thm:main}. Suppose for contradiction that there exists an equilibrium where the buyer uses a horizontal learning strategy. We first derive the properties of the optimal mechanism and then construct a deviation by the buyer. 

\paragraph{Optimal Mechanisms.}\hspace{-2mm}As before, under horizontal learning, it must be that one good has strictly positive sign and one good has strictly negative sign. We follow the same sign convention as before: 
\[a_1 + a_2 \geq 0\,.\]
Note that in equilibrium, it cannot be that $a_1 + a_2 =0$, because if so, by the logic in the introduction, the seller's mechanism would be to offer the grand bundle which extracts the full surplus of the buyer, but then the buyer would deviate to learn about $v_1 + v_2$, leading to a contradiction. 

Thus, suppose that $a_1 + a_2 > 0$, which implies that 
\[a_{\{1,2\}} = \gamma \big(a_1 + a_2 \big) > 0\,.\]
Now, consider the following auxiliary program: 
\begin{equation}
\max_{\mathbf{x}\in [0,1]^3;\, \sum_B x_B \leq 1}\sum_B b_Bx_B \quad \text{ subject to }\quad \sum_B a_B x_B=0 \,.\label{eq:auxiliary2}  
\end{equation}
By strong duality, let $\lambda$ be an optimal dual multiplier on the equality constraint in \eqref{eq:auxiliary2}. We claim that $\lambda < 0$. Indeed, by strong duality, we know that every optimal solution to \eqref{eq:auxiliary2} must solve the following problem
\[ \max_{\mathbf{x}\in [0,1]^3;\, \sum_B x_B \leq 1} \sum_B \big(a_B \lambda  + b_B \big) x_B \,,\]
and also satisfy the equality constraint.  However, if $\lambda \geq 0$, since $b_{\{1, 2\}} > \max\{b_1, b_2\}$, then for the negative good $i$, 
\[a_{\{1, 2\}} \lambda + b_{\{1, 2\}} \geq  b_{\{1, 2\}}  > b_i \geq a_i \lambda + b_i\,. \]
Thus, every optimal solution to the dualized problem must assign zero probability to the negative good. But then every optimal solution to the dualized problem must violate the equality constraint $\sum_B a_B x_B = 0$, which is impossible by strong duality. 

Now, since $\lambda < 0$, we claim that every optimal solution to the dualized problem must assign zero probability to the positive good $j$. Clearly, this would be the case if 
\[\lambda a_j + b_j < 0\,,\]
and hence suppose otherwise. Now, if $\gamma \leq 1$, then 
\[a_{\{1, 2\}} \lambda + b_{\{1, 2\}} - \big(a_j \lambda + b_j \big) = \big(a_{\{1, 2\}} - a_j\big)\lambda + \big(b_{\{1, 2\}} - b_j\big) > 0\,, \]
since $b_{\{1, 2\}} - b_j > 0$ and 
\[a_{\{1, 2\}} - a_j = \gamma a_i +\gamma a_j - a_j = \gamma a_i - (1 - \gamma) a_j  \leq 0\,. \]
If $\gamma > 1$, then we also have that 
\[a_{\{1, 2\}} \lambda + b_{\{1, 2\}} - \big(a_j \lambda + b_j \big) \geq (a_1 + a_2) \lambda + (b_1 + b_2)  - \big(a_j \lambda + b_j \big)  = a_i \lambda + b_i >  0\,, \]
where the first inequality is due to the fact that 
\[(a_1 + a_2) \lambda + (b_1 + b_2) = \underbrace{a_j  \lambda + b_j}_{\geq 0}  + \underbrace{a_i \lambda + b_i}_{ > 0} >  0\,.\]
It follows immediately that every optimal solution to the dualized problem cannot assign positive probability to the positive good $j$. However, in order to satisfy the equality constraint, this implies that every optimal solution must assign a strictly positive probability on $\{1, 2\}$ and a strictly positive probability on the negative good $i$. It follows that we must have 
\[a_{\{1, 2\}} \lambda + b_{\{1, 2\}} = a_{i} \lambda + b_i \geq 0\,, \]
for the optimal dual multiplier $\lambda$. 

Following the notation in \Cref{sec:proof}, by the proof of \Cref{lem:saddlepoint}, we know that there exists some $t^*_0 > 0$ such that\footnote{In particular, note that by the same reasoning as before, $a_{\{1, 2\}}\overline{\Phi}(0; 0) + b_{\{1, 2\}}$ must be strictly negative, and hence $\overline{\Phi}(0; 0) < \lambda$.} 
\[0 > \lambda = \overline{\Phi}(t^*_0; t^*_0)\,.\]
Moreover, the ironing interval $\mathcal{I}$ that includes $t^*_0$ must also include $0$. We claim that there exists some $x^*$ such that $(x^*, t^*_0)$ forms a saddle point: 
    \begin{align*}
        \max_{x \in \text{MON}} \min_{t_0\in[0,1]} \mathbb{E}\Bigg[\sum_B \Big(a_B x_B(t) \Phi(t;& t_0) + b_B x_B(t)\Big)\Bigg] \\
        &= 
  \min_{t_0\in[0,1]} \max_{x \in \text{MON}} \mathbb{E}\Bigg[\sum_B \Big(a_B x_B(t) \Phi(t; t_0) + b_B x_B(t)\Big)\Bigg]\,.
\end{align*}
where 
\[\text{MON}:= \Bigg\{x:[0,1] \rightarrow [0, 1]^3 \text{ s.t. }  \sum_Ba_Bx_B(t) \text{ is nondecreasing in $t$ and $\sum_B x_B(t) \leq 1$}\Bigg \}\,.\]

By  \Cref{lem:ironing} and the proof of \Cref{thm:main}, it suffices to show that there exists some $x^* \in \text{MON}$ such that \textit{(i)} it maximizes the following ironed objective pointwise: 
\[\mathbb{E}\Bigg[\sum_B \Big(a_B x_B(t) \overline{\Phi}(t; t^*_0) + b_B x_B(t)\Big)\Bigg]\,,\]
and \textit{(ii)} it is consistent with ironing, and \textit{(iii)} $t^*_0$ is a worst-off type under the induced mechanism. 

For any $t \in \mathcal{I}$, note that the pointwise maximization problem is 
\[\max_{x \in [0, 1]^3;\, \sum_B x_B \leq 1} \sum_B a_B x_B \overline{\Phi}(t; t^*_0) + b_B x_B = \max_{x \in [0, 1]^3;\, \sum_B x_B \leq 1} \sum_B  \big(a_B \lambda + b_B\big) x_B\,.\]
By construction, there exists some solution $x^\dagger$ such that 
\[\sum_B a_B x^\dagger_B = 0\,.\]
Now, for any $t \not\in \mathcal{I}$, note that the pointwise maximization problem is 
\[\max_{x \in [0, 1]^3;\, \sum_B x_B \leq 1} \sum_B a_B x_B \tilde{\lambda}_t + b_B x_B \,,\]
where 
\[1 \geq \tilde{\lambda}_t := \overline{\Phi}(t; t^*_0) > \lambda \,.\]
Note that for the negative good $i$, we have 
\[a_{\{1, 2\}} \tilde{\lambda}_t + b_{\{1, 2\}} > a_{\{1, 2\}} \lambda  + b_{\{1, 2\}} =  a_{i} \lambda  + b_{i} > a_{i} \tilde{\lambda}_t  + b_{i} \,,\]
and hence its ironed virtual value function is everywhere strictly dominated by that of the bundle $\{1, 2\}$ for all types $t \not \in \mathcal{I}$. Now, for the positive good $j$, note that if $\gamma \leq 1$, then since
\[a_{\{1, 2\}} + b_{\{1, 2\}} > a_{j} + b_{j}\,,\]
which is equivalent to 
\[(a_{\{1, 2\}} - a_{j}) + b_{\{1, 2\}} - b_{j} > 0\,,\]
we must have 
\[\tilde{\lambda}_t \underbrace{(a_{\{1, 2\}} - a_{j})}_{=\gamma a_i - (1 - \gamma) a_j \leq 0} + b_{\{1, 2\}} - b_{j} \geq (a_{\{1, 2\}} - a_{j}) + b_{\{1, 2\}} - b_{j} > 0\,,\]
and thus 
\[\tilde{\lambda}_t a_{\{1, 2\}} + b_{\{1, 2\}} >\tilde{\lambda}_t a_{j} + b_{j} \,.  \]
Moreover, if $\gamma > 1$, we also have that 
\[\tilde{\lambda}_t a_{\{1, 2\}} + b_{\{1, 2\}}  = \gamma \big(\tilde{\lambda}_t a_{i} + b_{i} + \tilde{\lambda}_t a_{j} + b_{j} \big) > \tilde{\lambda}_t a_{i} + b_{i} + \tilde{\lambda}_t a_{j} + b_{j} \geq \tilde{\lambda}_t a_{j} + b_{j} \,,\]
where the first inequality is due to the fact that 
\[\tilde{\lambda}_t a_{\{1, 2\}} + b_{\{1, 2\}} > \lambda a_{\{1, 2\}} + b_{\{1, 2\}} \geq 0\,,\]
and the second inequality is due to the fact that 
\[\tilde{\lambda}_t a_{i} + b_{i} \geq a_i + b_i \geq 0\,.\]
Therefore, combining these two cases, we have that the virtual value function of the positive good $j$ is also everywhere strictly dominated by that of the bundle $\{1, 2\}$ for all types $t \not \in \mathcal{I}$. It follows immediately that for all types $t \not \in \mathcal{I}$, the pointwise maximization has a unique solution that puts full probability on $\{1, 2\}$. 

Now, we construct $x^*$ as follows. Let $x^*(t) = x^\dagger$ for all $t \in \mathcal{I}$, and let $x^*$ put full probability on $\{1, 2\}$ for all $t \not \in \mathcal{I}$. By the above arguments, clearly the constructed $x^*$ satisfies properties \textit{(i)} and \textit{(ii)}. Moreover, note that $x^*$ is in MON. It remains to show that $t^*_0$ is a worst-off type under $x^*$ (and the payment rule induced by the Envelope theorem). This follows by the same argument as in \Cref{lem:saddlepoint}. Together, these certify that $(x^*, t^*_0)$ is a saddle point. 

Now, for every optimal mechanism by the seller, it must induce some optimal allocation rule $x'$ such that $(x', t^*_0)$ form a saddle point (by the rectangularity property of saddle points). As in the proof of \Cref{thm:main}, it follows immediately that every optimal mechanism by the seller must maximize the ironed objective pointwise in such a way that it is consistent with the ironing interval $\mathcal{I}$ and induces $t^*_0$ as a worst-off type. These three features together imply that every optimal mechanism must assign almost every type $t \in \mathcal{I}$ some allocation $x(t)$ that solves the auxiliary program. Thus, for almost all types $t \in \mathcal{I}$, we have that the buyer's indirect utility $U(t) = 0$. For all $t \not \in \mathcal{I}$, by the previous arguments, the pointwise solution is unique, and hence every optimal mechanism must assign full probability to the bundle $\{1, 2\}$ for almost all $t \not \in \mathcal{I}$. 

\vspace{-3mm}
\paragraph{Optimal Learning.}\hspace{-2mm}By the previous part, we know that in the equilibrium, the seller must be offering $\{1, 2\}$ with full probability, and any other option in the menu  consumed by some type must yield $U(t) = 0$ to all types $t$.  By the same argument in the proof of \Cref{thm:main}, it must be that in this equilibrium, the buyer learns about $v_1 + v_2$, but that would lead to a comonotonic distribution of $\theta_1$ and $\theta_2$ under uncorrelated values---hence, a contradiction. 

\vspace{-3mm}
\paragraph{Nested Bundling.}\hspace{-2mm}We claim that given vertical learning, there exists a unique optimal direct-revelation mechanism (up to measure zero) that is deterministic and can be represented by a nested menu. 

To prove the claim, one can verify the conditions in \citet{yang2023nested}. For completeness, we prove the claim directly. Under vertical learning, the posterior mean distribution can be written as: for each $B$,  
\[\theta_B = a_B t + b_B\]
where $a_{\{1, 2\}} = \gamma (a_1 + a_2)$, $b_{\{1, 2\}} = \gamma (b_1 + b_2)$, and $a_1 \geq 0$, $a_2 \geq 0$, $b_1 \geq 0$, $b_2 \geq 0$. 
The claim is easy to see if $a_i = 0$ for some good $i$, since then good $i$ must be sold to all types. Thus, suppose $a_i > 0$ for all goods $i$. Moreover, the claim is also easy to see if $b_1 + b_2 = 0$, since then $b_1 = b_2 = 0$, and pure bundling is optimal. Thus, suppose $b_1 + b_2 > 0$. 

Without loss of generality, suppose that 
\[\frac{b_2}{a_2} \geq \frac{b_1}{a_1}\,.\]
It follows that 
\[\frac{b_2}{a_2} \geq \frac{b_{\{1,2\}}}{a_{\{1,2\}}} = \frac{b_1 + b_2}{a_1 + a_2}\geq \frac{b_1}{a_1}\,.\]
We make two observations. First, note that 
\[\gamma (b_1 + b_2) \geq b_2 \]
and hence 
\[\gamma \geq \frac{b_2}{b_1 + b_2} \geq \frac{a_2}{a_1 + a_2} \,.\]
Note that if 
\[\gamma = \frac{a_2}{a_1 + a_2} \,,\]
then 
\[\frac{b_2}{a_2} = \frac{b_1}{a_1} = \frac{b_1 + b_2}{a_1 + a_2}\,,\]
and $a_{\{1, 2\}} = a_2\,, b_{\{1, 2\}} = b_2$, in which case it is easy to see that pure bundling is also optimal. Thus, suppose $\gamma > \frac{a_2}{a_1 + a_2}$. Then, $(a_{\{1, 2\}} t + b_{\{1, 2\}}) - (a_2 t + b_2)$ is strictly increasing in $t$. Moreover, by \Cref{lem:regular}, the virtual value function induced by $(a_{\{1, 2\}} t + b_{\{1, 2\}}) - (a_2 t + b_2)$ is strictly single-crossing. 

Second, consider good $1$ and any $t$ such that $a_1 \Psi(t) + b_1 > 0$. Then 
\begin{align*}
a_1 t + b_1  - \frac{1 - F(t)}{f(t)} a_1 &= \big(a_1 t + b_1\big)\big(1   - \frac{1 - F(t)}{f(t)} \frac{a_1}{a_1 t + b_1}\big) \\
&< \big(a_{\{1, 2\}} t + b_{\{1, 2\}}\big) \big(1   - \frac{1 - F(t)}{f(t)} \frac{a_{\{1, 2\}}}{a_{\{1, 2\}} t + b_{\{1, 2\}}}\big)  \\ 
&= a_{\{1,2\}} \Psi(t) + b_{\{1,2\}}\,.
\end{align*}
As in the proof of \Cref{thm:main}, consider the following relaxed problem: 
\[\max_{x:[0, 1]\rightarrow [0, 1]^3\,; \sum_B x_B \leq 1 } \mathbb{E}\Bigg[\sum_B (a_B \Psi(t) + b_B)x_B(t)\Bigg]\,,\]
where we maximize pointwise the unironed objective. By \Cref{lem:regular}, we have that for all $B$, $a_B \Psi(t) + b_B$ is strictly single-crossing. By our second observation, every optimal solution must assign good $1$ with probability $0$ (almost everywhere). Moreover, the crossing point $t^*_2$ of $a_2 \Psi(t) + b_2$ is weakly less than the crossing point $t^*_{\{1,2\}}$ of $a_{\{1, 2\}} \Psi(t) + b_{\{1, 2\}}$, which implies that $a_{\{1, 2\}} \Psi(t) + b_{\{1, 2\}}$ single-crosses $a_2 \Psi(t) + b_2$ from below at some point $t^\dagger \geq t^*_{\{1,2\}} \geq t^*_{2}$. It follows immediately that the relaxed problem has a unique solution given by assigning $\varnothing$ on $[0, t^*_2)$, assigning good $2$ with full probability on $[t^*_2, t^\dagger)$ (which may be empty), and assigning the bundle $\{1, 2\}$ with full probability on $[t^\dagger, 1]$. The allocation rule is implementable given the first observation that the values for the bundle $\{1, 2\}$ and for good $2$ satisfy increasing differences. 

The rest of the proof is identical to the proof of \Cref{thm:main}. In this equilibrium, the options in the seller's menu that are consumed by some types must be $\{2\}$ and $\{1, 2\}$, but then removing the other options in the menu results in a nested menu, under which the buyer's strategy continues to be optimal, and the seller's menu continues to be optimal. Thus, we have found an outcome-equivalent nested bundling equilibrium. 

\subsection{Proof of \Cref{prop:costs}}\label{subsec:costs-proof}

The proof of \Cref{prop:costs} is similar to that of \Cref{thm:main}. We first show that every equilibrium has vertical learning, and then that it is outcome equivalent to a nested bundling equilibrium. We assume throughout that $K=2$ and $\rho=0$. 

\subsubsection{Vertical Learning}

Toward a contradiction, suppose that there exists an equilibrium $(\boldsymbol{\alpha}, \mathcal{M})$ with horizontal learning, such that $\text{sign}(\alpha_1)\cdot\text{sign}(\alpha_2)<0$ (recall that we assume uncorrelated values here). The proof follows the same steps as that of \Cref{thm:main}. First, we characterize optimal mechanisms against the type distribution induced by $\boldsymbol{\alpha}$. Second, we construct a profitable deviation for the buyer. 

\vspace{-3mm}
\paragraph{Optimal Mechanisms.}\hspace{-2mm}Using the same normalization as in the proof of \Cref{thm:main}, we can index types by some parameter $t\in [0,1]$ such that a type-$t$ buyer has posterior expected value $\theta_i(t; \boldsymbol{\alpha}) = a_it+b_i$ for each good $i$. We can furthermore normalize signs such that $\sum_i a_i\geq 0$. Since $\boldsymbol{\alpha}$ is a horizontal learning strategy, we must have $a_i>0$ for one good $i$ and $a_j<0$ for the other. Let good 1 be the positive good. Note that the sign normalization is equivalent to $a_1\geq -a_2>0$. 

With production costs, what matters for the seller are effective types:
\[\tilde{\theta}_i(t;\boldsymbol{\alpha}) = a_it+b_i-c_i=:\tilde{a}_it+\tilde{b}_i.\]
To characterize optimal mechanisms, we can then use the same arguments as in the proof of \Cref{thm:main}, replacing the buyer's types by effective types. The main difference is that effective types can be negative. However, we know that $\tilde{b}_2> 0$. Indeed, $\tilde{b}_2 >0.5\tilde{a}_2+\tilde{b}_2=\mu_2-c_2>0$. Furthermore, $\tilde{a}_1+\tilde{b}_1>0$. Indeed, $\tilde{a}_1+\tilde{b}_1>0.5\tilde{a}_1+\tilde{b}_1=\mu_1-c_1>0$. Moreover, we have assumed in this proposition that the distribution $\mathbf{v}$ is log-concave, which implies that $\theta_i$ must be log-concave by the linear projection property of elliptical distributions and Prékopa's Theorem, and hence $a_i \Psi(t) + b_i$ is strictly increasing for $a_i > 0$ and strictly decreasing for $a_i < 0$. 

We start by solving the following auxiliary program: 
\begin{align*}
    \max_{\mathbf{x}\in [0,1]^K}&\sum_i \tilde{b}_ix_i \tag{Auxiliary Program} \\ 
    \text{ subject to } &\sum_i\tilde{a}_ix_i=0\,.
\end{align*}
Letting $\lambda$ denote an optimal multiplier on the equality constraint, any solution $x^*$ to the auxiliary program must solve
\[x_i^*\in \argmax_{x_i\in [0,1]} (\tilde{b}_i+\lambda\tilde{a}_i) x_i\,.\]
We show that $(-a_2/a_1,1)=:\bar{x}$ is the unique solution to this problem. 

First, we argue that any solution must have $x_2^*>0$. If not, then the only candidate that satisfies the equality constraint is $x=(0,0)$. But this is strictly worse than $\bar{x}$ since
\begin{align*}
    \sum_i\tilde{b}_i\bar{x}_i = -\frac{a_2}{a_1}\tilde{b}_1+\tilde{b}_2 &= -\frac{a_2}{a_1}[b_1-c_1]+b_2-c_2 \\
    &= -\frac{a_2}{a_1}[\mu_1 - 0.5a_1-c_1]+\mu_2-0.5a_2 - c_2\\
    &= -\frac{a_2}{a_1}[\mu_1-c_1]+\mu_2 - c_2>0\,.
\end{align*}
Thus, any solution has $x_2^*>0$ and the equality constraint pins down $x_1^* = -(a_2/a_1) x_2^*$. But since $-\frac{a_2}{a_1}\tilde{b}_1+\tilde{b}_2>0$, any optimal solution must set $x_2^*=1$. 

The following facts are worth noting. First, since $\bar{x}$ allocates a positive amount of both goods, it must be that $\tilde{a}_i\lambda +\tilde{b}_i\geq 0$ for $i=1,2$. Furthermore, if $-a_2<a_1$, then good 1 is rationed. Optimality then requires $\tilde{a}_1\lambda +\tilde{b}_1=0$ and $\tilde{a}_2\lambda +\tilde{b}_2=(-a_2/a_1)\times\tilde{b}_1+\tilde{b}_2>0$.  Finally, we can set $\lambda\leq 0.5$. There are two cases. Either $-a_2<a_1$, in which case:
\[\lambda = -\frac{\tilde{b}_1}{\tilde{a}_1} = \frac{-\mu_1+0.5a_1+c_1}{a_1} = 0.5 - \frac{\mu_1-c_1}{a_1}<0.5.\]
If $-a_2=a_1$, then there are many optimal multipliers, which only need to satisfy $\tilde{a}_i\lambda +\tilde{b}_i\geq 0$ for $i=1,2$. Setting $\lambda = 0.5$ satisfies both constraints. 

Unlike in our baseline model, the multiplier on the equality constraint can be either positive or negative depending on parameter values. Thus, we distinguish between several cases in our characterization of optimal mechanisms.

\begin{lemma}\label{lem:opt_costs}
    First, let $\lambda\leq 0$. Then, there exist thresholds $0<\bar{t}_0<\bar{t}_1\leq 1$ such that, for any optimal mechanism, up to a measure-zero set of types, we have
    \begin{enumerate}
        \item[(i)] $x(t) = \bar{x}$ and $U(t)=0$ for all $t\in [0,\bar{t}_0]$;
        \item[(ii)] $x(t) = (1,1)$ for all $t\in (\bar{t}_0,\bar{t}_1]$;
        \item[(iii)] $x(t) = (1,0)$ for all $t>\bar{t}_1$.  
    \end{enumerate}
    
    Now, let $\lambda>0$. Then, there exist thresholds $0\leq \bar{t}_0<\bar{t}_1\leq\bar{t}_2\leq 1$ such that, for any optimal mechanism, up to a measure-zero set of types, we have
    \begin{enumerate}
         \item[(i)] $x(t) = (0,1)$ for all $t\in [0,\bar{t}_0]$;
       \item[(ii)] $x(t) = \bar{x}$ and $U(t)=0$ for all $t\in (\bar{t}_0,\bar{t}_1]$;
        \item[(iii)] $x(t) = (1,1)$ for all $t\in (\bar{t}_1,\bar{t}_2]$;
        \item[(iv)] $x(t) = (1,0)$ for all $t>\bar{t}_2$.  
    \end{enumerate}
    Furthermore, $\bar{t}_1=\bar{t}_2$ only if $\bar{x} = (1,1)$.  
\end{lemma}

\begin{proof}
   Following the same argument as in \Cref{lem:optx}, we know that if $(x^*, t_0^*)$ is a saddle point of
   \[\max_{x \in \text{MON}} \min_{t_0}\mathbb{E}\Bigg[\sum_i \Big(\tilde{a}_i x_i(t) \Phi(t; t_0) + \tilde{b}_i x_i(t)\Big)\Bigg]\,,\]
   then $x^*$ is an optimal mechanism and any optimal mechanism $x'$ must also form a saddle point with $t_0^*$. We first construct a saddle point, which then allows us to characterize all optimal mechanisms. 

   \vspace{-3mm}
   \paragraph{Case (A).}\hspace{-2mm}First, consider the case of $\lambda \leq 0$, which is the only case that arises in the proof of \Cref{thm:main}. Following the same argument as in \Cref{lem:saddlepoint}, we know that there exists some $t_0^*$ such that 
   \[\overline{\Phi}(t_0^*; t_0^*)=\lambda\,.\]
   We also know that the ironing interval including $t_0^*$ must also include $0$. That is, there exists $\bar{t}_0\geq t_0^*$ such that $\overline{\Phi}(t; t_0^*)=\lambda$ for all $t\in [0,\bar{t}_0]$ and $\overline{\Phi}(t; t_0^*)>\lambda$ for all $t>\bar{t}_0$. 
   
   We show that $t^*_0$ is part of a saddle point. Fixing the conjectured worst-off type $t_0^*$, consider the maximization problem: 
   \[\max_{x:[0,1]\rightarrow[0,1]^K}\mathbb{E}\Bigg[\sum_i \Big(\tilde{a}_i x_i(t) \overline{\Phi}(t; t_0^*) + \tilde{b}_i x_i(t)\Big)\Bigg]\,.\]

    We can solve this problem pointwise. By construction, for any $t\in [0,\bar{t}_0]$, 
    \[\tilde{a}_i  \overline{\Phi}(t; t_0^*) + \tilde{b}_i=\tilde{a}_i  \lambda + \tilde{b}_i\quad \forall i\,,\]
    which can be maximized by setting $x(t) = \bar{x}$.   Furthermore, for any $t>\bar{t}_0$, $\overline{\Phi}(t; t_0^*)>\lambda$. Thus, $\tilde{a}_1\overline{\Phi}(t; t_0^*)+\tilde{b}_1>0$, and any solution must set $x_1(t)=1$ for all $t>\bar{t}_0$. Finally, $\tilde{a}_2\overline{\Phi}(t; t_0^*)+\tilde{b}_2\geq 0$ at $t=\bar{t}_0$ and strictly decreases in $t$ over $[\bar{t}_0,1]$. Let $\bar{t}_1:=\max\{t\mid \tilde{a}_2\overline{\Phi}(t; t_0^*)+\tilde{b}_2\geq 0\}$. By construction $\bar{t}_1 \geq \bar{t}_0$. Any solution to the above problem must set $x_2(t)=1$ for all $\bar{t}_0 < t <\bar{t}_1$ and $x_2(t)=0$ for $t>\bar{t}_1$.

    Consider the following allocation rule: $x^*(t) =\bar{x}$ for $t \leq \bar{t}_0$, $x^*(t) =(1,1)$ for $t\in (\bar{t}_0, \bar{t}_1]$, and $x^*(t) = (1, 0)$ for $t > \bar{t}_1$. By the above argument, $x^*$ pointwise maximizes the ironed objective given $t^*_0$. Moreover, note that $x^*$ is in MON and is consistent with ironing, and therefore maximizes the MON-constrained original objective given $t^*_0$ by \Cref{lem:ironing}. Moreover, $t^*_0$ is a worst-off type given $x^*$ by the same argument as in \Cref{lem:saddlepoint}. Indeed, since $\sum_i \tilde{a}_i x^*_i(t^*_0) = 0$ and $\sum_i \tilde{a}_i x^*_i(\,\cdot\,)$ is monotone, we have $U(t^*_0) = 0$ while $U(t) \geq 0$ for all $t$. Thus, $(x^*, t_0^*)$ is a saddle point, which implies $x^*$ is an optimal mechanism. Furthermore, any other optimal mechanism $x'$ must also form a saddle point with $t_0^*$. However, up to a measure-zero set of types, $x^*$ is the only mechanism that maximizes the ironed objective given $t_0^*$, is consistent with ironing, and has $t^*_0$ as a worst-off type.

    We have left to prove that $\bar{t}_1>\bar{t}_0$. Since the function $\overline{\Phi}(t; t_0^*)$ is continuous in $t$, if $\bar{t}_1=\bar{t}_0$, then $\tilde{a}_2\overline{\Phi}(\bar{t}_0; t_0^*)+\tilde{b}_2= 0$, which is equivalent to $\tilde{a}_2\lambda+\tilde{b}_2= 0$. Thus, if $\bar{t}_1=\bar{t}_0$, then $\lambda = -\tilde{b}_2/\tilde{a}_2>0$, a contradiction. 

   \vspace{-3mm}
    \paragraph{Case (B).}\hspace{-2mm}Now consider the case of $\lambda>0$. First,  we show that there exists $t^*_0$ such that $\overline{\Phi}(t_0^*; t_0^*)=\lambda$. The function $g(t_0) = \overline{\Phi}(t_0; t_0)$ is continuous in $t_0$ (\Cref{lem:continuous}) and negative at $t_0=0$ (\Cref{lem:regular}). Thus, we only need to show that $g(1)\geq \lambda$.  Recall that at $t_0=1$, we have 
    \[\Phi(t; 1) = t + \frac{F(t)}{f(t)}\,,\]
    which, by log-concavity, is strictly increasing in $t$. Thus, $\overline{\Phi}(1; 1) =\lim_{t \rightarrow 1}\Phi(t;1)\geq 1$. Therefore, $g(1)\geq 1/2 \geq \lambda$, and, by the intermediate value theorem, there exists $t^*_0$ such that $\overline{\Phi}(t_0^*; t_0^*)=\lambda$.

    We construct a saddle point $(x^*, t_0^*)$. Given the conjectured worst-off type $t_0^*$, consider again the maximization problem:
   \[\max_{x:[0,1]\rightarrow[0,1]^K}\mathbb{E}\Bigg[\sum_i \Big(\tilde{a}_i x_i(t) \overline{\Phi}(t; t_0^*) + \tilde{b}_i x_i(t)\Big)\Bigg].\]
    Let $\mathcal{I}\subseteq[0,1] $ be the ironing interval that includes $t_0^*$. Note that an ironing interval containing the highest type $t = 1$ must have level at least $\Psi(1) = 1 > \lambda$; therefore, $1 \notin \mathcal{I}$. Let $\bar{t}_2=\max\{t\mid \tilde{a}_2\overline{\Phi}(t; t_0^*) + \tilde{b}_2\geq 0\}$ and $\bar{t}_0=\min\{t\mid \tilde{a}_1\overline{\Phi}(t; t_0^*) + \tilde{b}_1\geq 0\}$. Since $\overline{\Phi}(t; t_0^*)$ is monotonically increasing and equals $\lambda>0$ for all $t\in \mathcal{I}$, it must be that $\bar{t}_2\geq \max_{t'\in \mathcal{I}}{t'} $ and $\bar{t}_0\leq \min_{t'\in \mathcal{I}}{t'} $. Thus, the allocation 
    \[x^*(t) =\begin{cases}
         \bar{x}\quad \text{if } t\in \mathcal{I}\\
         (\mathbbm{1}_{t\geq \bar{t}_0},\mathbbm{1}_{t\leq \bar{t}_2}),\quad \text{if } t\notin \mathcal{I}\,,
    \end{cases} \]
    maximizes the ironed objective pointwise. Note that $x^*$ is in MON:    If $a_1 = -a_2$, then $\bar{x}=(1,1)$, and clearly by construction $x^*_1(\,\cdot\,)$ is weakly increasing and $x^*_2(\,\cdot\,)$ is weakly decreasing. Otherwise, we have $a_1 > - a_2$. Then, $\tilde{a}_1\lambda + \tilde{b}_1=0$ and hence  $\tilde{a}_1\overline{\Phi}(t; t_0^*) + \tilde{b}_1=0$ for all $t\in \mathcal{I}$, and $\bar{t}_0= \min_{t'\in \mathcal{I}}{t'}$. Consequently, $x^*_1(\,\cdot\,)$ is weakly increasing. Moreover $x^*_2(\,\cdot\,)$ is also weakly decreasing since $\bar{x}_2 = 1$. Together these show that $x^*$ is in MON. Moreover, $x^*$ is consistent with ironing by inspection. Moreover, since $x^*\in\text{MON}$ and $\sum_i \tilde{a}_i x^*_i(\,\cdot\,)=0$ on $\mathcal{I}$, it follows that $t^*_0$ is a worst-off type. Thus, $(x^*, t^*_0)$ is a saddle point. Again, as argued before, any other optimal mechanism must also form a saddle point with $t_0^*$. Note that any such mechanism must coincide with $x^*$ almost everywhere (by pointwise optimization, ironing consistency, and the fact that $t_0^*$ is a worst-off type). 
    
    We have left to prove that $\bar{t}_2=\max_{t'\in \mathcal{I}}{t'} =: \bar{t}_1$ only if $\bar{x}=(1,1)$. Since $\overline{\Phi}(\,\cdot\,,t_0^*)$ is continuous, $ \bar{t}_2=\bar{t}_1$ if and only if $\tilde{a}_2\lambda + \tilde{b}_2=0$. Thus, if $\bar{t}_2=\max_{t'\in \mathcal{I}}{t'}$, we then have $\lambda = -\tilde{b}_2/\tilde{a}_2$ and $\tilde{a}_1\lambda + \tilde{b}_1 = \tilde{b}_1-(a_1/a_2)\tilde{b}_2 >0$ (as shown before). Thus, it must be that $\bar{x}=(1,1)$ and $-a_2=a_1$. 
 \end{proof}

\vspace{-3mm}
\paragraph{Optimal Learning.}\hspace{-2mm}We show that, against any optimal mechanism $\mathcal{M}$, the buyer has a strictly profitable deviation under horizontal learning.  Let 
\[O:=\Big\{\big(x(t), p(t)\big)\Big\}_{t\in [0, 1]}\]
denote the minimal menu that implements the seller's optimal mechanism (which would give the same ex ante payoff to the buyer under strategy $\boldsymbol{\alpha}$). 

We jointly consider \textbf{Cases (A)} and \textbf{(B)} in \Cref{lem:opt_costs}. In both cases, $(\bar{x},\bar{p})\in O$ and $((1,1), p_{12})\in O$, where $\bar{p}$ is the price of the rationing option and $p_{12}$ the price of the grand bundle. If $-a_2=a_1$, then $\bar{x}=(1,1)$ by \Cref{lem:opt_costs}, so the rationing option coincides with the grand bundle; the argument below applies verbatim to this case. 

The set $O$ can include up to two other outcomes: $((1,0),p_1)$ for some $p_1$ and $((0,1), p_2)$ for some $p_2$. Note that if $((1,0),p_1)\in O$, then $p_1>\min_s \theta_1(s; \boldsymbol{\alpha})$. That is, not all types are willing to buy good 1 by itself. If it were not true, then all types but the lowest would get a strictly positive payoff out of option $((1,0),p_1)$. Yet we know that a positive mass of them must get the rationing option and zero payoff under any optimal mechanism. Similarly, if $((0,1),p_2)\in O$, then $p_2>\min_s \theta_2(s; \boldsymbol{\alpha})$. Finally, if both options are included in $O$, then we must have $p_1+p_2>p_{12}$. Indeed, by \Cref{lem:opt_costs}, we know that type $\bar{t}_0$ must be indifferent between $((0,1),p_2)$ and nothing, type $\bar{t}_1$ between $((1,1),p_{12})$ and nothing, and type $\bar{t}_2$ between $((1,0),p_{1})$ and $((1,1),p_{12})$. Thus, we have 
\begin{align*}
    p_1+p_2 - p_{12} = -a_2\bar{t}_2 - b_2 + a_2\bar{t}_0 +b_2= -a_2(\bar{t}_2-\bar{t}_0)>0\,,
\end{align*}
since $a_2<0$ and $\bar{t}_2>\bar{t}_0$. 

We show that, against menu $O$, the buyer strictly prefers the ``flipped'' vertical learning strategy $\hat{\boldsymbol{\alpha}}=(\alpha_1, -\alpha_2)$ to $\boldsymbol{\alpha}$. 

Note that the two learning strategies lead to the same distribution of posterior expected values for good 1. That is, $\theta_1(s; \boldsymbol{\alpha})$ and $\theta_1(\hat{s}; \hat{\boldsymbol{\alpha}})$ follow the same distribution, which we denote by $G$. From now on, we index types under both learning strategies by $\theta_1$. We can then write the value that a type $\theta_1$-buyer has for good 2 as 
\begin{align*}
    &\theta_2(\theta_1) = \mu_2 + \frac{\alpha_2\sigma^2_2}{\alpha_1\sigma^2_1}(\theta_1 - \mu_1)\quad \text{under strategy }\boldsymbol{\alpha}\,,\\
      &\hat{\theta}_2(\theta_1) = \mu_2 - \frac{\alpha_2\sigma^2_2}{\alpha_1\sigma^2_1}(\theta_1 - \mu_1)\quad \text{under strategy }\hat{\boldsymbol{\alpha}}\,.
\end{align*}
Let $U_\mathcal{M}(\boldsymbol{\alpha})$ and $U_\mathcal{M}(\hat{\boldsymbol{\alpha}})$ denote the buyer's expected payoff under mechanism $\mathcal{M}$ when he chooses learning strategy $\boldsymbol{\alpha}$ and $\hat{\boldsymbol{\alpha}}$, respectively. Define $U_O(\boldsymbol{\alpha})$ and $U_O(\hat{\boldsymbol{\alpha}})$ similarly. We want to show that $U_\mathcal{M}(\boldsymbol{\alpha})<U_\mathcal{M}(\hat{\boldsymbol{\alpha}})$. By construction, $U_\mathcal{M}(\boldsymbol{\alpha})=U_O(\boldsymbol{\alpha})$. Furthermore, since $O\subseteq \mathcal{M}$, $U_\mathcal{M}(\hat{\boldsymbol{\alpha}})\geq U_O(\hat{\boldsymbol{\alpha}})$. Thus, it is enough to show that $U_O(\boldsymbol{\alpha})<U_O(\hat{\boldsymbol{\alpha}})$.

Let $O'$ be the menu constructed from removing the rationing option from $O$ if $\bar{x} \neq (1,1)$, and let $O'=O$ otherwise. Because the rationing option yields zero surplus to any type under strategy $\boldsymbol{\alpha}$, it must be that $U_{O'}(\boldsymbol{\alpha})=U_O(\boldsymbol{\alpha})$ while $U_{O'}(\hat{\boldsymbol{\alpha}})\leq U_O(\hat{\boldsymbol{\alpha}})$. Thus, it is enough to show that $U_{O'}(\boldsymbol{\alpha})<U_{O'}(\hat{\boldsymbol{\alpha}})$.

In our proof, we consider a fictitious separate sales mechanism $SS$. If both $((1,0),p_1), ((0,1),p_2)\in O$, then $SS$ simply consists of good 1 at price $p_1$, good 2 at price $p_2$, and the grand bundle at $p_1+p_2$. If there is no $(x,p)\in O$ with $x=(0,1)$, then define $p_2:=\max_{\theta_1} \theta_2(\theta_1)$. By construction, this ensures that if $x=(0,1)$ is not included in $O$, then the constructed $((0,1),p_2)$ is not purchased by any type under $\boldsymbol{\alpha}$. Note that this option is not purchased by any type under $\hat{\boldsymbol{\alpha}}$ either, since $\max_{\theta_1} \hat{\theta}_2(\theta_1)=\max_{\theta_1} \theta_2(\theta_1)$. Similarly, if  there is no $(x,p)\in O$ with $x=(1,0)$, then define $p_1:=p_{12}-\min_{\theta_1} \theta_2(\theta_1)$. As before, the price $p_1$ is chosen so that no type ever purchases this option under both $\boldsymbol{\alpha}$ and $\hat{\boldsymbol{\alpha}}$. 

Let $U_{SS}(\boldsymbol{\alpha}) $ and $U_{SS}(\hat{\boldsymbol{\alpha}})$ denote the buyer's payoffs under separate sales mechanism $SS$.  Note that learning strategies $\boldsymbol{\alpha}$ and $\hat{\boldsymbol{\alpha}}$ yield the same expected payoff to the buyer under any separate sales mechanism. Thus, $U_{SS}(\boldsymbol{\alpha}) = U_{SS}(\hat{\boldsymbol{\alpha}})$. What we want to show ($U_{O'}(\boldsymbol{\alpha})<U_{O'}(\hat{\boldsymbol{\alpha}})$) is then equivalent to: 
\[U_{O'}(\hat{\boldsymbol{\alpha}})-U_{SS}(\hat{\boldsymbol{\alpha}})>U_{O'}(\boldsymbol{\alpha})-U_{SS}(\boldsymbol{\alpha}).\]
For the buyer's payoff, the only difference between menus $O'$ and $SS$ is that the former sells the grand bundle at $p_{12}$ while the latter sells the grand bundle at $p_1+p_2>p_{12}$. Indeed, if both $((1,0),p_1), ((0,1),p_2)\in O'$, this is the only difference between menu $O'$ and menu $SS$. If either option is not included in $O'$, then $SS$ includes it but its price is set such that the buyer never purchases it under either learning strategy. Therefore, it suffices to show that the buyer suffers more from an increase in the price of the grand bundle under the vertical learning strategy $\hat{\boldsymbol{\alpha}}$ than under the horizontal learning strategy $\boldsymbol{\alpha}$. 

\usetikzlibrary{decorations.pathreplacing}
 \begin{figure}[!t]
\begin{center}
\begin{tikzpicture}[scale=0.9, every node/.style={transform shape}]

  \draw[->, thick] (-0.1,0) -- (5.5,0) node[below] {$v_1$};
  \draw[->, thick] (0,-0.1) -- (0,4.5) node[left] {$v_2$};
  \filldraw[black] (2,2) circle (1pt);
     \fill[black!20, opacity=0.4] (2,2) ellipse [x radius=2, y radius=2];
  \draw[thick] (2,0.1) -- (2,-0.1) ;
  \draw[thick] (0.1, 2) -- (-0.1, 2) ;
  \draw[thick, purple] (0.2,1.1) -- (3.8, 2.9) node[above right] {\small$\hat{\boldsymbol{\alpha}}$};
  \draw[thick, purple] (0.2,2.9) -- (3.8, 1.1) node[below right] {\small$\boldsymbol{\alpha}$};
  \draw[thick, blue, dashed] (4, 1.3) -- (2.9,1.3) -- (1.9,2.3) --  (-0.1, 2.3) node[left] {${\color{blue} p_2}$};
  \draw[thick, blue, dashed] (2.9,1.3) -- (2.9,-0.1) node[below] {${\color{blue} p_1}$};
  \draw[thick, blue, dashed] (1.9,2.3) --  (1.9, 4) ;
\node at (1,4) {${\color{blue} O'}$};
  \draw[thick, blue, dashed] (4.2,0.1) -- (4.2,-0.1) node[below] {${\color{blue} p_{12}}$};
  \draw[thick, purple, dotted] (2.4, 1.8) -- (2.4, -0.6);
  \draw[thick, purple, dotted] (3.4, 1.3) -- (3.4, -0.6);
\draw[thick, purple] [decorate,decoration={brace,amplitude=6pt,mirror}] (2.4, -0.6) -- (3.4, -0.6) node[midway,below=8pt] {\small${\color{purple} \Theta_{\{1,2\}}}$};
  \draw[thick, purple, dotted] (2.15,2.1) --(2.15,-1.8);
  \draw[thick, purple, dotted] (3.8, 2.9) --(3.8,-1.8);
\draw[thick, purple] [decorate,decoration={brace,amplitude=6pt,mirror}] (2.15,-1.8) -- (3.8,-1.8) node[midway,below=8pt] {\small${\color{purple} \hat{\Theta}_{\{1,2\}}}$};

  \draw[->, thick] (6.9,0) -- (7+5.5,0) node[below] {$v_1$};
  \draw[->, thick] (7+0,-0.1) -- (7+0,4.5) node[left] {$v_2$};
  \filldraw[black] (7+2,2) circle (1pt);
     \fill[black!20, opacity=0.4] (7+2,2) ellipse [x radius=2, y radius=2];
  \draw[thick] (7+2,0.1) -- (7+2,-0.1) ;
  \draw[thick] (7+0.1, 2) -- (7-0.1, 2) ;
  \draw[thick, purple] (7+0.2,1.1) -- (7+3.8, 2.9) node[above right] {\small$\hat{\boldsymbol{\alpha}}$};
  \draw[thick, purple] (7+0.2,2.9) -- (7+3.8, 1.1) node[below right] {\small$\boldsymbol{\alpha}$};
  \draw[thick, blue, dashed] (11, 1.3) -- (9.4, 1.3) -- (8,2.7) --  (7-0.1, 2.7) node[left] {${\color{blue} p_2}$};
  \draw[thick, blue, dashed] (9.4, 1.3) -- (9.4,-0.1) node[below] {${\color{blue} p_1}$};
  \draw[thick, blue, dashed] (8,2.7) -- (8,4);
  \draw[thick, purple, dotted] (8.45, 2.3) -- (8.45, -0.6);
  \draw[thick, purple, dotted] (10.4,1.3) -- (10.4, -0.6);
\draw[thick, purple] [decorate,decoration={brace,amplitude=6pt,mirror}] (8.45, -0.6) -- (10.4, -0.6) node[midway,below=8pt] {\small${\color{purple} \Theta_{\{1,2\}}}$};
  \draw[thick, purple, dotted] (8.8, 1.8) -- (8.8, -1.8);
  \draw[thick, purple, dotted] (10.8, 2.9) -- (10.8, -1.8);
\draw[thick, purple] [decorate,decoration={brace,amplitude=6pt,mirror}] (8.8, -1.8) -- (10.8, -1.8) node[midway,below=8pt] {\small${\color{purple} \hat{\Theta}_{\{1,2\}}}$};
\end{tikzpicture}
\end{center}
\caption{Illustration of the two cases for the comparison of $\boldsymbol{\alpha}$ and $\boldsymbol{\hat{\alpha}}$.}\label{fig:costs_proof}
\end{figure}
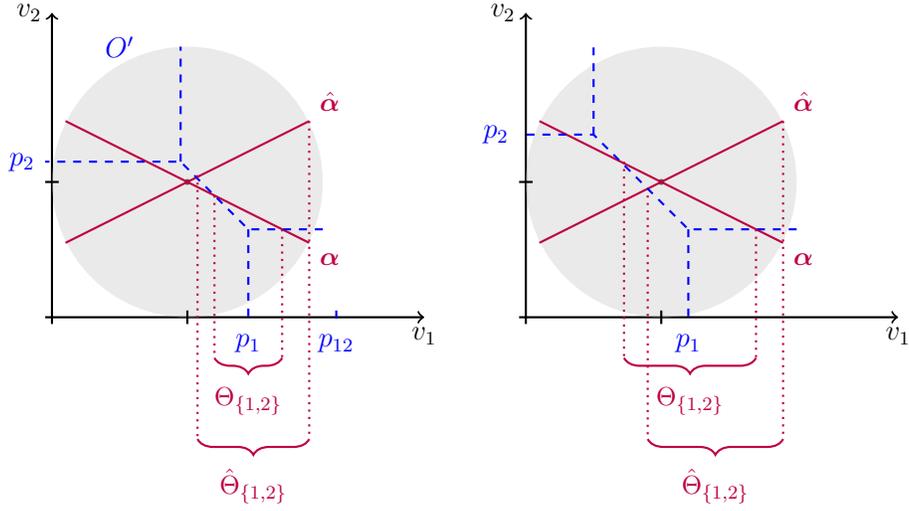

Let $\Theta_{\{1,2\}}:=\Big\{\theta_1: \theta_1+\theta_2(\theta_1)-p_{12}\geq \max\{0, \theta_1 - p_1, \theta_2(\theta_1)-p_2\}\Big\}$ be the set of types who purchase the grand bundle under $\boldsymbol{\alpha}$.  Only these types are potentially affected by an increase in the price of the grand bundle. Thus,
\begin{align*}
    U_{O'}(\boldsymbol{\alpha})-U_{SS}(\boldsymbol{\alpha}) = \int_{\theta_1\in \Theta_{\{1,2\}}}\Delta(\theta_1)dG(\theta_1)\,,
\end{align*}
where
\[\Delta(\theta_1) := \theta_1+\theta_2(\theta_1)-p_{12} - (\theta_1 - p_1)_+-(\theta_2(\theta_1)-p_2)_+\geq 0\]
is the effect on a buyer with realized type $\theta_1\in \Theta_{\{1,2\}}$. Set $\Delta(\theta_1)=0$ for all $\theta_1\notin \Theta_{\{1,2\}}$ since any such type is unaffected by an increase in the grand bundle price. Define $\hat{\Theta}_{\{1,2\}}$ and $\hat{\Delta}(\,\cdot\,)$ similarly.

There are two cases, depending on whether the lowest type who purchases the grand bundle under $\boldsymbol{\alpha}$ has a value for good 1 above or below the mean (\Cref{fig:costs_proof} illustrates both). First, consider what happens when $\min\{\theta_1:\theta_1\in \Theta_{\{1,2\}}\} \geq \mu_1$, that is, when the lowest type who purchases the grand bundle under $\boldsymbol{\alpha}$ has a value for good 1 weakly greater than the mean. Note that this implies $\hat{\theta}_2(\theta_1) \geq \theta_2(\theta_1)$ for all $\theta_1\in \Theta_{\{1,2\}}$, strictly whenever $\theta_1>\mu_1$. We show that \textit{(i)} $\Theta_{\{1,2\}}\subset \hat{\Theta}_{\{1,2\}}$, and that \textit{(ii)} $\hat{\Delta}(\theta_1)\geq \Delta(\theta_1)$ for all $\theta_1\in \Theta_{\{1,2\}}$, strictly for some. Together, these imply that $ U_{O'}(\hat{\boldsymbol{\alpha}})-U_{SS}(\hat{\boldsymbol{\alpha}})>U_{O'}(\boldsymbol{\alpha})-U_{SS}(\boldsymbol{\alpha})$. 

To establish \textit{(i)}, note that if a buyer who values good 1 at $\theta_1$ and good 2 at $\theta_2(\theta_1)$ finds it optimal to purchase the grand bundle, then so does a buyer who values good 1 at $\theta_1$ and good 2 at $\hat{\theta}_2(\theta_1)>\theta_2(\theta_1)$. For all $\theta_1\in \Theta_{\{1,2\}}$, we thus have:
\begin{align*}
    \hat{\Delta}(\theta_1)- \Delta(\theta_1) = &\theta_1+\hat{\theta}_2(\theta_1)-p_{12} - (\theta_1 - p_1)_+-(\hat{\theta}_2(\theta_1)-p_2)_+\\
    &\qquad \quad-\theta_1-\theta_2(\theta_1)+p_{12} + (\theta_1 - p_1)_++(\theta_2(\theta_1)-p_2)_+\\
    =&\hat{\theta}_2(\theta_1)-\theta_2(\theta_1)+(\theta_2(\theta_1)-p_2)_+-(\hat{\theta}_2(\theta_1)-p_2)_+\geq 0\,,
\end{align*}
where the inequality is strict if $\theta_2(\theta_1)<p_2$ and $\theta_1>\mu_1$. Note that both must hold for a positive measure of $\theta_1 \in \Theta_{\{1, 2\}}$ since $\boldsymbol{\alpha}$ is a horizontal learning strategy and $\Theta_{\{1,2\}}$ has positive measure. This establishes \textit{(ii)}.

Now consider what happens when $\min\{\theta_1:\theta_1\in \Theta_{\{1,2\}}\} <  \mu_1$. This case is slightly more involved as $\hat{\theta}_2(\theta_1)<\theta_2(\theta_1)$ for some $\theta_1\in \Theta_{\{1,2\}}$, such that the above argument is no longer sufficient. We can, however, leverage the symmetry of the type distribution around the mean: any type $\theta_1<\mu_1$ has the same probability (under both $\boldsymbol{\alpha}$ and $\hat{\boldsymbol{\alpha}}$) as type $2\mu_1-\theta_1$. We show that $\hat{\Delta}(\theta_1)+\hat{\Delta}(2\mu_1-\theta_1)\geq \Delta(\theta_1)+\Delta(2\mu_1-\theta_1)$ for all $\theta_1\in \Theta_{\{1,2\}}$ where $\theta_1<\mu_1$, and strictly so for a positive measure of these types. This then implies that $ U_{O'}(\hat{\boldsymbol{\alpha}})-U_{SS}(\hat{\boldsymbol{\alpha}})>U_{O'}(\boldsymbol{\alpha})-U_{SS}(\boldsymbol{\alpha})$. In particular, this argument covers all $\theta_1 \in  \Theta_{\{1,2\}}$ if for any $\theta'_1 > \mu_1$ where $\theta'_1 \in \Theta_{\{1,2\}}$, we have $2\mu_1 - \theta'_1 \in  \Theta_{\{1,2\}}$. Otherwise, for any ``unmatched type'' $\theta'_1 > \mu_1$ where $\theta'_1 \in \Theta_{\{1,2\}}$, it is easy to see that $\hat{\Delta}(\theta'_1) \geq \Delta(\theta'_1)$ by the previous argument since $\hat{\theta}_2(\theta'_1)>\theta_2(\theta'_1)$. 

Now, there are two subcases: either $\theta_1\in \hat{\Theta}_{\{1,2\}}$ or $\theta_1\notin \hat{\Theta}_{\{1,2\}}$. Consider the first subcase first, such that type $\theta_1$ purchases the grand bundle under both the vertical and horizontal learning strategy. Then, as above: 
\begin{align*}
    \hat{\Delta}(\theta_1)- \Delta(\theta_1) = \hat{\theta}_2(\theta_1)-\theta_2(\theta_1)+(\theta_2(\theta_1)-p_2)_+-(\hat{\theta}_2(\theta_1)-p_2)_+\,.
\end{align*}
Note that, by construction, $\hat{\theta}_2(2\mu_1-\theta_1)=\theta_2(\theta_1)$ and $\theta_2(2\mu_1-\theta_1)=\hat{\theta}_2(\theta_1)$. Thus, $2\mu_1-\theta_1\in  \hat{\Theta}_{\{1,2\}}$ and  $2\mu_1-\theta_1\in  \Theta_{\{1,2\}}$, and, as above: 
 \begin{align*}
    \hat{\Delta}(2\mu_1-\theta_1)- \Delta(2\mu_1-\theta_1)     &=\theta_2(\theta_1)-\hat{\theta}_2(\theta_1)+(\hat{\theta}_2(\theta_1)-p_2)_+-(\theta_2(\theta_1)-p_2)_+\\
    & =  \Delta(\theta_1)- \hat{\Delta}(\theta_1)\,.
\end{align*}
Thus, $\hat{\Delta}(\theta_1)+\hat{\Delta}(2\mu_1-\theta_1)= \Delta(\theta_1)+\Delta(2\mu_1-\theta_1)$. 

Now consider the latter subcase where type $\theta_1$ does not purchase the grand bundle under the vertical learning strategy $\hat{\boldsymbol{\alpha}}$. Note that there must exist a positive mass of such types. Indeed, let $\underline{\theta}_1 := \min\{\theta_1 : \theta_1 \in \Theta_{\{1,2\}}\} < \mu_1$. Under $\boldsymbol{\alpha}$, type $\underline{\theta}_1$ is indifferent between the grand bundle and nothing, so $\underline{\theta}_1 + \theta_2(\underline{\theta}_1) = p_{12}$. Since $\hat{\theta}_2(\underline{\theta}_1) < \theta_2(\underline{\theta}_1)$, we have $\underline{\theta}_1 + \hat{\theta}_2(\underline{\theta}_1) < p_{12}$. By continuity, all $\theta_1 \in \Theta_{\{1,2\}}$ sufficiently close to $\underline{\theta}_1$ are thus not in $\hat{\Theta}_{\{1,2\}}$. Also, note that for these types, we must have either $\theta_1 + \hat{\theta}_2(\theta_1)<p_{12}$ or $\hat{\theta}_2(\theta_1)<p_{12}-p_1$ (the remaining possibility, $\theta_1 < p_{12} - p_2$, is ruled out by $\theta_1 \in \Theta_{\{1,2\}}$).

For all such types, we have $\hat{\Delta}(\theta_1)=0$ and 
\begin{align*}
    \Delta(\theta_1)- \hat{\Delta}(\theta_1) =  \theta_1+\theta_2(\theta_1)-p_{12} - (\theta_1 - p_1)_+-(\theta_2(\theta_1)-p_2)_+\,.
\end{align*}
As before, $2\mu_1-\theta_1\in  \hat{\Theta}_{\{1,2\}}$. If $2\mu_1-\theta_1\notin  \Theta_{\{1,2\}}$, then $ \Delta(2\mu_1-\theta_1)=0$, and 
\begin{align*} \hat{\Delta}(2\mu_1-\theta_1)- &\Delta(2\mu_1-\theta_1) \\
&= 2\mu_1-\theta_1 + \theta_2(\theta_1) -p_{12} - (2\mu_1-\theta_1 - p_1)_+-(\theta_2(\theta_1)-p_2)_+\\
&=  \Delta(\theta_1)- \hat{\Delta}(\theta_1) +2(\mu_1-\theta_1) +(\theta_1-p_1)_+ -  (2\mu_1-\theta_1 - p_1)_+\\
&\geq  \Delta(\theta_1)- \hat{\Delta}(\theta_1)\,,
\end{align*}
where the inequality is due to that $2\mu_1 - \theta_1 > \theta_1$, and is strict if $\theta_1<p_1$. Moreover, note that there exists a positive measure of such $\theta_1$ with $\theta_1<p_1$ since, under $\boldsymbol{\alpha}$, there exists a type who is indifferent between consuming nothing and consuming the bundle and there exists a positive measure of types consuming the bundle. Now, if $2\mu_1-\theta_1\in  \Theta_{\{1,2\}}$,  then  
\begin{align*} 
\hat{\Delta}(2\mu_1-\theta_1)- &\Delta(2\mu_1-\theta_1) =\theta_2(\theta_1)-\hat{\theta}_2(\theta_1)+(\hat{\theta}_2(\theta_1)-p_2)_+-(\theta_2(\theta_1)-p_2)_+\\
 &=  \Delta(\theta_1)- \hat{\Delta}(\theta_1) +p_{12}-\theta_1-\hat{\theta}_2(\theta_1)+(\hat{\theta}_2(\theta_1)-p_2)_++(\theta_1-p_1)_+\\
 &>  \Delta(\theta_1)- \hat{\Delta}(\theta_1)\,,
\end{align*}
since either $\theta_1 + \hat{\theta}_2(\theta_1)<p_{12}$ or $\hat{\theta}_2(\theta_1)<p_{12}-p_1$.

\subsubsection{Nested Bundling}
By the previous parts, we know that every equilibrium must have vertical learning. Now, fix any equilibrium. Then, the effective types (after adjusting for costs) can be written as: for each $i$,  
\[\tilde{\theta}_i = \tilde{a}_i t + \tilde{b}_i\]
where $\tilde{a}_i \geq 0$, and $t \in [0, 1]$. The only difference compared to \Cref{sec:proof} is that $\tilde{b}_i$ may be negative. The proof of the nested bundling part of \Cref{thm:main} uses the fact that $a_i \Psi(t) + b_i$ is strictly single-crossing under positive $a_i, b_i$ (\Cref{lem:regular}). However, as noted before, since we have assumed the distribution $\mathbf{v}$ is log-concave, we have that $\theta_i$ is log-concave by the linear projection property of elliptical distributions and Prékopa's Theorem, and hence $\tilde{a}_i \Psi(t) + \tilde{b}_i$ is strictly increasing under positive $\tilde{a}_i$ and hence strictly single-crossing. The rest of the proof is identical.

\subsection{Proof of \Cref{prop:weak}}\label{subsec:weak-proof}

\paragraph{Existence.}\hspace{-2mm}We construct a weak equilibrium. For any learning strategy $\boldsymbol{\alpha}$, normalize types as in the proof of \Cref{thm:main} such that type $t$ values good $i$ at $a_it+b_i$, and let $\Psi(t)$ be the associated virtual value function as defined in \Cref{lem:regular}. Fix any vertical learning strategy $\boldsymbol{\alpha}$ such that $a_i>0$ for all $i$, and $b_i/a_i\neq b_j/a_j$ for all $i \neq j$. There must exist such an $\boldsymbol{\alpha}$ since the set of vertical learning strategies is a non-empty convex $(K-1)$-dimensional set, and learning strategies that fail to satisfy these conditions have measure zero in that set. 

Consider the following direct revelation mechanism $\mathcal{M}$: 
\begin{align*}
    x_i(t)  = \mathbbm{1}\{a_i\Psi(t) + b_i\geq 0\}\\
    p(t) = \sum_i [a_it+b_i]x_i(t) - \int_0^t\sum_i a_i x_i(s)ds\,.
\end{align*}
We know from the proof of \Cref{thm:main} that this mechanism is optimal against $\boldsymbol{\alpha}$. We now argue that $\boldsymbol{\alpha}$ is $\mathcal{M}$-Blackwell undominated.

Let $t_i:=\min\{t:x_i(t) = 1\}$ be the lowest type who is allocated good $i$, and label goods such that $t_1\leq t_2\leq \dots \leq t_K$. We show that these inequalities can be strengthened to $t_1<t_2<\dots <t_K<1$. By \Cref{lem:regular}, we know that $\Psi(t)>0$ for any $t>0.5$, which means $x_i(t)= 1$ for any $t>0.5$, and thus $t_K<1$. \Cref{lem:regular} also shows that $\Psi(t) \rightarrow-\infty$ as $t\rightarrow 0$. Thus, it can only be optimal to allocate some good $i$ to the lowest type if $a_i=0$. However, by construction, $a_i>0$ for all $i$. Similarly,  $t_i = t_j$ if and only if $b_i/a_i=b_j/a_j$, which is precluded by construction. Thus, the optimal mechanism constructed above allocates all the following bundles with positive probability: $\varnothing$, $\{1\}$, $\{1, 2\}$, $\{1, 2, 3\}$, $\dots$, $\{1, 2, \dots, K\}$. Furthermore, any type $t\in (t_l, t_{l+1})$ finds it \emph{strictly} optimal to buy bundle $\{1, 2, \dots, l\}$. Thus, any selection of optimal reports $M^*$ must include all the above bundles, and $(U_m)_{m\in M^*} = (0, v_1, v_1+v_2, v_1+v_2+v_3, \dots, \sum_k v_k)$. Then, for another strategy $\boldsymbol{\alpha}'$ to $\mathcal{M}$-Blackwell dominate $\boldsymbol{\alpha}$, it must be strictly Blackwell more informative than $\boldsymbol{\alpha}$ about $(U_m)_{m\in M^*} $, in the sense that 
\[\Big(\sum_{l=1}^k \theta_l(s; \boldsymbol{\alpha})\Big)_{k=1\dots K} \preceq_{\text{cx}} \Big(\sum_{l=1}^k \theta_l(s; \boldsymbol{\alpha}')\Big)_{k=1\dots K} \,.\]
This implies that, for any weights $(\lambda_1,\dots, \lambda_K) \in \R^K$, we have 
\[\boldsymbol{\lambda}\cdot\boldsymbol{\theta}(s; \boldsymbol{\alpha}) \preceq_{\text{cx}} \boldsymbol{\lambda}\cdot\boldsymbol{\theta}(s; \boldsymbol{\alpha}')  \,,\]
since the convex order implies the linear convex order. However, we know that signal $\boldsymbol{\alpha}\neq \boldsymbol{0}$ induces the most dispersed distribution of $\boldsymbol{\theta}$ along some line in $\mathbb{R}^K$. In particular, note that for $\boldsymbol{\lambda}^* = \boldsymbol{\alpha}$, we must have 
\[\boldsymbol{\lambda}^*\cdot\boldsymbol{\theta}(s; \boldsymbol{\alpha}') \preceq_{\text{cx}} \boldsymbol{\lambda}^*\cdot\boldsymbol{\theta}(s; \boldsymbol{\alpha})\quad \text{in addition to}\quad  \boldsymbol{\lambda}^*\cdot\boldsymbol{\theta}(s; \boldsymbol{\alpha}) \preceq_{\text{cx}} \boldsymbol{\lambda}^*\cdot\boldsymbol{\theta}(s; \boldsymbol{\alpha}')\,,\]
since the signal $\boldsymbol{\alpha} \cdot \mathbf{v}$ fully reveals the state $\boldsymbol{\lambda}^* \cdot \mathbf{v}$. The above can only be possible if signals $\boldsymbol{\alpha}$ and $\boldsymbol{\alpha}'$ are identical, in the sense that $\boldsymbol{\alpha}'=c\cdot\boldsymbol{\alpha}$ for some constant $c$. But then $\boldsymbol{\alpha}'$ cannot $\mathcal{M}$-Blackwell dominate $\boldsymbol{\alpha}$.

Thus, strategy $\boldsymbol{\alpha}$  is $\mathcal{M}$-Blackwell undominated and $\mathcal{M}$ is revenue-maximizing given $\boldsymbol{\alpha}$. The strategy profile forms a weak equilibrium. 

\vspace{-3mm}
\paragraph{Vertical learning and nested bundling.}\hspace{-2mm}We first show that every weak equilibrium has vertical learning and is outcome-equivalent to a nested bundling equilibrium. The proof is virtually identical to that of \Cref{thm:main}. By contradiction, suppose that there exists a weak equilibrium with horizontal learning $(\boldsymbol{\alpha}, \mathcal{M})$. The characterization of the seller's best response is identical since weak equilibrium imposes the same restriction on the seller's behavior as Nash equilibrium. Thus, \Cref{lem:optx} holds. Using the same arguments as in the proof of \Cref{thm:main}, we can then show a contradiction. Indeed, \Cref{lem:learnzero}, \Cref{lem:learnsame}, \Cref{lem:infovaluable}, and \Cref{lem:learnfinal} only rely on the buyer not choosing a signal that is Blackwell dominated by some other signal in the relevant payoff subspace, which is precisely what the notion of weak equilibrium requires. In particular, by the proof of \Cref{lem:infovaluable}, information must be strictly valuable in any weak equilibrium. By the proofs of \Cref{lem:optx} and \Cref{lem:learnzero}, any $\boldsymbol{\alpha}$ such that $\alpha_i \neq 0$ for some negative good $i$ is $\mathcal{M}$-Blackwell dominated (by selecting $M^*$ to be the induced outcomes in the equilibrium). By the proofs of \Cref{lem:optx} and \Cref{lem:learnsame}, any $\boldsymbol{\alpha}$ such that $\alpha_i \neq \alpha_j$ for some negative good $i$ and some positive balancing good $j$ is $\mathcal{M}$-Blackwell dominated (by selecting $M^*$ to be the induced outcomes in the equilibrium except replacing the rationing options that yield $0$ payoff to all buyer types with the empty set $\varnothing$). Therefore, in the weak equilibrium, we must have $\alpha_i = 0$ for all negative goods and all positive balancing goods $i$, which leads to a sign contradiction as before. Thus, every weak equilibrium has vertical learning. 

The proof of nested bundling part is also unaffected by the weaker solution concept: for any weak equilibrium, removing options in the seller's menu that are not chosen by any equilibrium type continues to sustain a weak equilibrium where the buyer chooses the original signal. 

\vspace{-3mm}
\paragraph{Ordering of log-scale posterior variance.}\hspace{-2mm}Fix any weak equilibrium $(\boldsymbol{\alpha}, \mathcal{M})$. We know from the previous part that $\boldsymbol{\alpha}$ must be a vertical learning strategy. The proof of \Cref{prop:ordering} applies verbatim as the argument only leverages the fact that $\boldsymbol{\alpha}$ is a vertical learning strategy and that $\mathcal{M}$ is revenue-maximizing against $\boldsymbol{\alpha}$. Since the concept of weak equilibrium still requires that the seller chooses a revenue-maximizing mechanism, the result follows. 

\end{document}